\documentclass[reprint,amsmath,amssymb,aps, pra, nolinenumbers, floatfix,longbibliography,superscriptaddress,nofootinbib]{revtex4-2}

\usepackage{graphicx}
\usepackage{dcolumn}
\usepackage{bm}
\usepackage[dvipsnames]{xcolor}
\usepackage{physics}
\usepackage{multirow}
\usepackage{booktabs}
\usepackage{amssymb}
\usepackage{tikz}
\usetikzlibrary{shapes.geometric}
\usepackage{hyperref}
\usepackage{nicematrix}
\NiceMatrixOptions{cell-space-limits = 0.8pt}

\hypersetup{
    unicode=true, 
    plainpages=false,
    colorlinks=true,
    linkcolor=blue,
    citecolor=blue,
    filecolor=blue,
    urlcolor=blue
}
\usepackage{txfonts} 
\allowdisplaybreaks

\renewcommand{\v}[1]{\mathbf{#1}}
\newcommand{\SO}{\ensuremath{\mathrm{SO}}}

\newcommand{\FMtwo}{\ensuremath{\mathrm{FM}_2}}
\newcommand{\FMone}{\ensuremath{\mathrm{FM}_1}}
\newcommand{\nematic}{\ensuremath{\hat{\mathbf{d}}}}

\begin{document}

\title{Topological interfaces crossed by defects and textures of continuous and discrete point group symmetries in spin-2 Bose-Einstein condensates}

\author{Giuseppe Baio}
\affiliation{Physics, Faculty of Science, University of East Anglia, Norwich Research Park, Norwich, NR4 7TJ, UK}
\affiliation{Centre for Photonics and Quantum Science, University of East Anglia, Norwich Research Park, Norwich, NR4 7TJ, UK}
\author{Matthew T. Wheeler}
\affiliation{Physics, Faculty of Science, University of East Anglia, Norwich Research Park, Norwich, NR4 7TJ, UK}
\affiliation{Centre for Photonics and Quantum Science, University of East Anglia, Norwich Research Park, Norwich, NR4 7TJ, UK}
\author{David S. Hall}
\affiliation{Department of Physics and Astronomy, Amherst College, Amherst, Massachusetts 01002-5000, USA}
\author{Janne Ruostekoski}
\affiliation{Department of Physics, Lancaster University, Lancaster, LA1 4YB, UK}
\author{Magnus O. Borgh}
\affiliation{Physics, Faculty of Science, University of East Anglia, Norwich Research Park, Norwich, NR4 7TJ, UK}
\affiliation{Centre for Photonics and Quantum Science, University of East Anglia, Norwich Research Park, Norwich, NR4 7TJ, UK}

\begin{abstract}
We systematically and analytically construct a set of spinor wave functions representing defects and textures that continuously penetrate interfaces between coexisting, topologically distinct magnetic phases in a spin-2 Bose-Einstein condensate. These include singular and nonsingular vortices carrying mass or spin circulation that connect across interfaces between biaxial- and uniaxial nematic, cyclic and ferromagnetic phases, as well as vortices terminating as monopoles on the interface (``boojums'').
The biaxial-nematic and cyclic phases exhibit discrete polytope symmetries featuring non-Abelian vortices and we investigate a pair of non-commuting line defects within the context of a topological interface.
By numerical simulations, we characterize the emergence of non-trivial defect core structures, including the formation of composite defects. Our results demonstrate the potential of spin-2 Bose-Einstein condensates as experimentally accessible platforms for exploring interface physics, offering a wealth of combinations of continuous and discrete symmetries.
\end{abstract}

\maketitle
\section{Introduction}
When topologically distinct phases coexist in a continuous and coherent ordered medium, a topological interface may form at the phase boundary where the different broken symmetries of their order-parameters connect smoothly.
Such interfaces were first discussed in the context of domain walls in the early universe~\cite{zeldovich1975,Kibble1976,Kibble1980}, where they may form the termination points of cosmic strings~\cite{Vilenkin1985},
and later in brane models in superstring theory~\cite{Dvali1999,Sarangi2002,Gudnason2015}.
They appear
universally across many areas of physics, from the $A$--$B$ phase boundary in superfluid liquid $^3$He~\cite{Osheroff1977,Yip1986,Salomaa1987,Finne2006,Bradley2008,Volovik2009}, via atomic Bose-Einstein condensates (BECs)~\cite{Takeuchi2006,Kasamatsu2010,Borgh2012,Borgh2013,Borgh2014,Kaneda2014}, 
to quantum chromodynamics~\cite{Alford2001,Cipriani2012,Eto2014}. 
The different order-parameter symmetries imply that the bulk medium on either side of the interface supports different families of topological defects and textures, which therefore cannot cross the interface unchanged.
Consequently, defects and textures penetrating through the interface must either terminate at the interface or continuously transform into different defects and textures of the corresponding phases. 
Due to the ubiquitous nature of topological interfaces, their study in controlled experiments becomes of general importance, inspiring the use of laboratory systems as emulators for interface physics in contexts otherwise not amenable to experimental observations, such as the simulation of brane-collision processes~\cite{Bradley2008}.

Topological-interface physics becomes especially intriguing when the medium on either or both sides of the interface exhibits discrete polytope point-group order-parameter symmetry~\cite{Xiao2022}, leading to defects whose charges depend on the presence of other defects in the system and whose dynamics is highly constrained~\cite{Poenaru1977,Mermin1979}. Such order-parameter symmetries arise in particular phases of spin-2 and spin-3 BECs~\cite{Barnett2006,Santos2006,Diener2006,Barnett2007,Semenoff2007,Makela2007,Yip2007,Kawaguchi2011,Kobayashi2009,Borgh2016b,Borgh2017,Xiao2022}, which have consequently been proposed, e.g., as candidates for quantum-computation applications using non-Abelian vortex braiding~\cite{Mawson2019}. 

Optically trapped spinor BECs~\cite{Kawaguchi2012,StamperKurn2013}, where the internal spin-degrees of freedom are not frozen out by strong magnetic fields~\cite{StamperKurn1998}, provide an ideal
testing ground for investigating interface physics with different bulk regions exhibiting different magnetic phases and defects~\cite{Borgh2012,Borgh2013,Borgh2014}. Topological interfaces can also form at vortex cores in spinor BECs when the singularity of the bulk order parameter with one symmetry is accommodated by filling the defect core with atoms in a different magnetic phase and symmetry~\cite{Lovegrove2016},
as experimentally realized in spin-1~\cite{Weiss2019,Xiao2021} and spin-2~\cite{Xiao2022} BECs. Here we consider engineering of spatially extended topological interfaces between coexisting bulk regions of distinct spin-2 magnetic phases that are analogous to topological bulk interfaces studied in high-energy physics, superfluid liquid $^3$He, and in spin-1 BECs. Both sides of the interface may then harbor defects and textures with continuous and discrete polytope symmetries that terminate at or cross the interface.

The phase diagrams of spinor BECs~\cite{Kawaguchi2012} exhibit a rich variety of magnetic phases with different order-parameter symmetries, supporting different families of topological defects~\cite{Ueda2014}. These include singular vortices carrying both integer~\cite{Yip1999,Isoshima2002,MizushimaPRA2002,
Sadler2006,Semenoff2007,Lovegrove2012,Lovegrove2016,Borgh2016b,Weiss2019,Xiao2021,Xiao2022} and fractional~\cite{Leonhardt2000,Zhou2003,Ji2008,Seo2015,Semenoff2007,Kobayashi2009,Lovegrove2012,Lovegrove2016,Borgh2016b,Borgh2017,Xiao2021,Xiao2022} charges, as well as  nonsingular vortices (2D Skyrmions)~\cite{Ohmi1998, Ho1998, MizushimaPRL2002,
Martikainen2002, Leanhardt2003, Mizushima2004, Choi2012, Choi2012b, Lovegrove2014,Weiss2019}---whose corresponding objects in magnetic solid-state systems have attracted recent interest~\cite{Muhlbauer2009,Nagaosa2013}---wall-vortex complexes~\cite{Kang2019,Takeuchi2021}, and monopoles~\cite{Stoof2001,Savage2003b,Ruostekoski2003,Pietila2009,Ray2014,Ray2015,Ollikainen2017,Mithun2022,Blinova2023}. In addition, spinor BECs support 3D Skyrmions~\cite{Al_Khawaja2001,Ruostekoski2001,Battye2002,Savage2003a,Ruostekoski2004,Kawakami2012,Tiurev2018,Lee2018}---topologically non-trivial textures first proposed in nuclear physics~\cite{Skyrme1961}---and knotted solitons~\cite{Kawaguchi2008,Hall2016}
with parallels in classical field theories~\cite{Faddeev1997,Battye1998,Hietarinta1999}
and even magnetic materials~\cite{Sutcliffe2017}. It has recently been theoretically proposed that these complex topological objects can be generated in atomic systems through optical excitation~\cite{Parmee2022}. 

Here we analytically construct spinor wave functions representing continuous connections of defects and textures across topological interfaces in spin-2 BECs and numerically simulate their core structure for illustrative examples. Radically different symmetry properties of the magnetic phases mean that their defects and textures exhibit distinct and generally incommensurate topologies that can inhibit connections across the interfaces. We systematically identify and explicitly construct a set of allowed connections for interfaces between biaxial nematic (BN), uniaxial nematic (UN), cyclic (C), and ferromagnetic (FM) phases. These include singular and nonsingular vortices carrying mass circulation as well as spin vortices. Numerical simulation reveals the appearance of complex defect-core structures, including non-axisymmetric cores at the UN-BN interface and composite cores~\cite{Lovegrove2014} of C-BN spin vortices.
Defects may also terminate at the interface, either with a vortex-free state on the opposite side, or as a monopole on the interface, similar to ``boojums'' on the $A$-$B$ phase boundary in superfluid liquid $^3$He~\cite{Blaauwgeers2002,Volovik2009}. We demonstrate that monopole solutions exist on C-BN, C-FM, and UN-BN interfaces as the termination point of singular vortices, and numerically show the formation of half-quantum Alice rings~\cite{Ruostekoski2003,Blinova2023} from singular point defects due to dissipation.
With techniques for controlled creation of vortices in several phases with different internal symmetries having recently been developed~\cite{Xiao2022}, our analytical and numerical results highlight how spinor-BEC systems are poised as immediate candidates for the realization of topological interfaces. Spin-2 topological interfaces offer the potential even for non-Abelian defect physics, which we numerically simulate here by constructing a pair of singular non-commuting fractional BN vortices that terminate at an interface.

This article is organized as follows: Section~\ref{sec: mean-field}, provides a brief overview of the mean-field theory and magnetic phases and order-parameter symmetries of spin-2 BECs, as well as establishes defect nomenclature. In Sec.~\ref{sec: theory}, we then present continuously interpolating spinors across topological interfaces and construct interface-crossing defects and textures. Numerical results are in Sec.~\ref{sec: numerics} before concluding remarks and experimental discussion in Sec.~\ref{sec: conclusions}.

\section{Spin-2 BECs mean-field theory\label{sec: mean-field}}
In the Gross-Pitaevskii mean-field theory, the spin-2 BEC is described by a five-component wave function  $\Psi=\sqrt{n}(\zeta_2, \zeta_1, \zeta_0, \zeta_{-1}, \zeta_{-2})^\text{T}$, where $n(\v{r})$ is the condensate density, which together with the normalized spinor $\zeta(\v{r})$ 
gives the field in the $m=+2, +1, 0, -1, -2$ magnetic sublevels.
The Hamiltonian density reads~\cite{Ueda2002,Kawaguchi2012}
\begin{equation}
    \mathcal{H} = \mathcal{H}_0
    + \frac{c_0}{2}n^2 + \frac{c_1}{2}n^2|\langle\hat{\vb{F}}\rangle|^2 + \frac{c_2}{2}n^2|A_{20}|^2,
    \label{eq: energy-functional}
\end{equation}
with the single-particle contribution
\begin{equation}
    \mathcal{H}_0 = \frac{\hbar^2}{2M}|\nabla\Psi|^2
    + \left(\frac{1}{2}M\omega^2r^2 -p\langle\hat{F}_z\rangle + q\langle\hat{F}_z^2\rangle\right)n,
    \label{eq: single-particle}
\end{equation}
where $M$ is the atomic mass and $\omega$ is the angular frequency of the harmonic  trap, which we, for simplicity, assume to be isotropic.
The term $p = -g\mu_B B$ is a linear Zeeman shift arising from a
uniform magnetic field $B$ oriented along $z$, where $g=1/2$ is the Land\'e factor for $F=2$ and $\mu_B$ is the Bohr magneton. A quadratic Zeeman shift $q$ also arises from such a magnetic field, and its exact form is obtained by means of the Breit-Rabi formula \cite{corney1978atomic}. Besides magnetic fields, the values of $q$ and $p$ can be experimentally controlled by ac Stark shifts induced by microwaves or lasers~\cite{Gerbier2006,Santos2007}.

Interaction terms in Eq.~\eqref{eq: energy-functional} arise from the three possible $s$-wave scattering channels of colliding spin-2 atoms with scattering lengths $a_f$, corresponding to total angular momentum $f=0$, $2$ and $4$, respectively. These combine to form three interactions terms in Eq.~\eqref{eq: energy-functional}: First, a contribution of strength $c_0=4\pi\hbar^2(3a_4+4a_2)/7M$ that depends only on the atomic density. A second interaction term of strength  $c_1 = 4\pi\hbar^2(a_4-a_2)/7M$  also depends on the magnitude of the local condensate spin vector  $\langle\hat{\v{F}}\rangle=\zeta^\dagger\hat{\v{F}}\zeta$, constructed from the vector of spin-2
angular momentum operators $\hat{\v{F}} \equiv (\hat{F}_x, \hat{F}_y, \hat{F}_z)$.
In addition to the density- and spin-dependent interactions, a third interatomic interaction term arises that is proportional to 
\begin{align}
    |A_{20}|^2 = \frac{1}{5}\left|2\zeta_2\zeta_{-2}-2\zeta_1\zeta_{-1}+\zeta_0^2\right|^2,
    \label{eq: a20definition}
\end{align}
where $A_{20}$ is the spin-singlet duo amplitude. The strength of this interaction is $c_2=4\pi\hbar^2(4a_4-10a_2+7a_0)/7M$. 

We consider magnetic phases as stationary solutions to the Gross-Pitaevskii energy functional \eqref{eq: energy-functional} when we ignore the harmonic trapping potential.
The steady-state spinors are found as optimal points of the mean-field density functional. These satisfy the general condition $\delta \mathcal{H}/\delta\zeta_m^\ast =0$, i.e.,
\begin{equation}
\left[-p \hat{F}_z + q \hat{F}^2_z + \tilde{c}_0 \, \zeta^\dagger \zeta + \tilde{c}_1 \, \langle\hat{\v{F}}\rangle\cdot\hat{\v{F}} + \frac{\tilde{c}_2}{5}(\hat{\mathcal{T}}\zeta)^\dagger\, \zeta\hat{\mathcal{T}} -\mu \right] \zeta = 0,
\label{eq: stationary_cond}
\end{equation}
where $\tilde{c}_{0,1,2} = n\,c_{0,1,2}$, $\mu$ is the chemical potential, and the time-reversal operator $\hat{\mathcal{T}}$ is defined by the action $(\hat{\mathcal{T}}\zeta)_m=(-1)^m\zeta^*_{-m}$ on the spinor components. The definition in Eq.~\eqref{eq: a20definition} can then be written $A_{20} = (\hat{\mathcal{T}}\zeta)^\dagger \zeta/\sqrt{5}$~\cite{Kawaguchi2011}.
Equation~\eqref{eq: stationary_cond} generally results in a homogeneous, nonlinear, algebraic system for the unknowns $\zeta_m$ \cite{Ciobanu2000}, which may be solved to find the stationary states. Spinor-BEC experiments~\cite{Ray2015,Hall2016,Xiao2021,Xiao2022,Choi2012b}  
frequently rely on dynamically stable stationary solutions (i.e., robust with respect to small dynamical perturbations) that are long-lived on the experimental time scale. 
A subset of the stationary solutions may also be energetically (meta\nobreakdash-)stable, corresponding to local energetic minima.

In the absence of Zeeman shifts ($p=q=0$), the uniform spin-2 BEC exhibits five distinct magnetic phases, of which four also appear as uniform ground states for different values of the interatomic interactions $c_{0,1,2}$~\cite{Ciobanu2000,Barnett2006,Kawaguchi2012}. 
They are characterized by the symmetries of the corresponding order parameter, which are illustratively visualized using a spherical-harmonics representation of the spinor (see also Ref.~\cite{Xiao2022} for a detailed discussion):
\begin{equation}
    Z(\theta,\varphi) = \sum^2_{m=-2} \zeta_m Y^m_{F=2}(\theta,\varphi),
    \label{eq: spher_harm_repr}
\end{equation}
where $Y^m_{F}(\theta,\varphi)$ denotes the spherical harmonic of degree $F$ and order $m$.
For each phase we state a representative spinor, from which any other may be reached by application of a gauge transformation together with a spin rotation, defined by three Euler angles, such that
\begin{equation}
    \begin{gathered}
        \zeta \to e^{i\tau}\hat{U}(\alpha, \beta, \gamma)\,\zeta, \\
        \hat{U}(\alpha, \beta, \gamma) = \exp(-i\hat{F}_z\alpha)\exp(-i\hat{F}_y\beta)\exp(-i\hat{F}_z\gamma). 
    \end{gathered}
    \label{eq: euler_angles}
\end{equation}
The family of states thus generated forms the order-parameter space $\mathcal{M}$ of each magnetic phase as a subset of the full $G=\text{U}(1)\times\text{SO}(3)$ symmetry group of the Hamiltonian density at zero level shifts.
\begin{figure*}
    \centering
    \hspace*{-.2cm}\includegraphics[width=2\columnwidth]{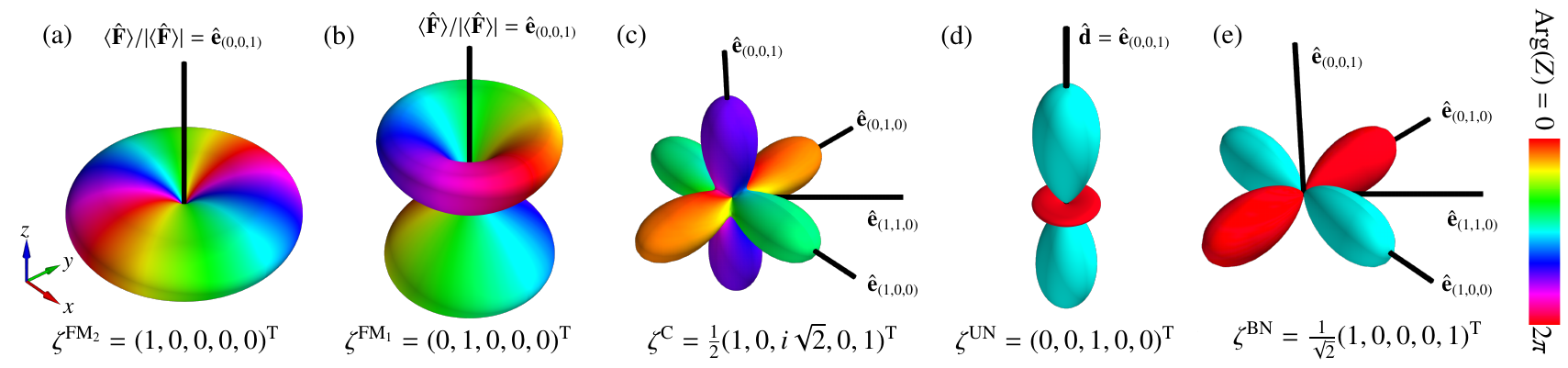}
    \caption{Spherical-harmonic representations [Eq.~\eqref{eq: spher_harm_repr}] of the magnetic phases of a spin-2 BEC. (a) FM$_2$ and (b) FM$_1$ whose order-parameter spaces represent spatial rotations (here $\langle\hat{\vb{F}}\rangle/|\langle\hat{\vb{F}}\rangle|=\hat{\bf{e}}_{(0,0,1)}$). 
    (c) The C phase combines the discrete symmetry of a tetrahedron with the condensate phase.
    (d) The UN phase, whose order parameter is given by an unoriented, nematic axis $\nematic=\hat{\bf{e}}_{(0,0,1)}$, together with the condensate phase. 
    (e) The BN phase combines the discrete symmetry of a square with the phase.} 
    \label{fig: phases}
\end{figure*}

The spin-2 FM ($\FMtwo$) phase, exemplified by 
\begin{align}
    \zeta^{\text{FM}_2}=(1, 0, 0, 0, 0)^\text{T},
    \label{eq: FM2_representative}
\end{align}
is characterized by $|\langle\hat{\v{F}}\rangle|=2$ and $|A_{20}|^2=0$. The order-parameter space $\mathcal{M}_{\text{FM}_2}=\SO(3)/\mathbb{Z}_2$~\cite{Makela2003,Xiao2022} is characterized by a continuous spin-gauge rotation symmetry [Fig.~\ref{fig: phases}(a)]. A further FM phase with $|\langle\hat{\v{F}}\rangle|=1$ (FM$_1$), exemplified by 
\begin{align}
    \zeta^{\text{FM}_1}=(0, 1, 0, 0, 0)^\text{T}
    \label{eq: FM1_representative}
\end{align}
also forms a stationary state with an $\mathcal{M}_{\FMone}=\SO(3)$ order-parameter space~\cite{Makela2003} similar to the FM phase in spin-1 BECs [Fig.~\ref{fig: phases}(b)]~\cite{Ho1998,Ohmi1998,Weiss2019}.

The C phase, exemplified by
\begin{align}
    \zeta^\mathrm{C}=\frac12\left(1, 0, i\sqrt{2}, 0, 1 \right)^\text{T},
    \label{eq: C_representative}
\end{align}
by contrast, has a discrete polytope order-parameter symmetry [Fig.~\ref{fig: phases}(c)] where the tetrahedral subgroup of rotations combined with the condensate phase, collectively $T_{\tau,\mathbf{F}}$, gives rise to the manifold $\mathcal{M}_\text{C}=[\SO(3)\times \mathrm{U}(1)]/T_{\tau,\mathbf{F}}$~\cite{Semenoff2007,Kobayashi2012},
with spinors characterized by $|\langle\hat{\v{F}}\rangle|=0$ as well as $|A_{20}|^2=0$. In the C phase, trios of atoms combine to form a spin-singlet state. The amplitude of singlet-trio formation reads 
\begin{align}
    A_{30} = \frac{3\sqrt{6}}{2}\left(\zeta_1^2\zeta_{-2}+\zeta_{-1}^2\zeta_2\right)
    + \zeta_0\left(\zeta_0^2-3\zeta_1\zeta_{-1}-6\zeta_2\zeta_{-2}\right),
\label{eq: A30_definition}
\end{align}
such that for any C spinor, $|A_{30}|^2=2$~\cite{Ueda2002,Kawaguchi2012}.

The remaining two stationary solutions at zero field are nematic phases with $|\langle\hat{\v{F}}\rangle|=0$ and $|A_{20}|^2=1/5$.
In the UN phase, represented by  
\begin{align}
    \zeta^\mathrm{UN}=(0, 0, 1, 0, 0)^\text{T},
    \label{eq: UN_representative}
\end{align}
the order parameter manifold is $\mathcal{M}_\text{UN} = \text{U}(1)\times (S^2/\mathbb{Z}_2)$, i.e., parametrized by an unoriented vector, the nematic axis $\nematic$ corresponding to a local symmetry axis, and a condensate phase~\cite{Song2007, Kobayashi2012} as illustrated in Fig.~\ref{fig: phases}(d). 
The BN phase exemplified by 
\begin{align}
    \zeta^\mathrm{BN}=\frac{1}{\sqrt{2}}(1, 0, 0, 0, 1)^\text{T},
    \label{eq: BN_representative}
\end{align} 
by contrast, exhibits a discrete, four-fold dihedral symmetry that combines with $\pi$-shifts of the condensate phase [Fig.~\ref{fig: phases}(e)], denoted $(D_4)_{\tau,\mathbf{F}}$. The order-parameter space is thus $\mathcal{M}_{\text{BN}} = [\SO(3)\times \mathrm{U}(1)]/(D_4)_{\tau,\mathbf{F}}$~\cite{Song2007, Kobayashi2012, Borgh2016b}.
In addition to the different order-parameter symmetry, the two nematic states can also be distinguished by the spin-singlet trio amplitude, where
$|A_{30}|=1$ in the UN phase and $|A_{30}|=0$ for the BN. 

The UN and BN phases compete as the likely ground-state phase for spin-2 BECs of $^{87}$Rb~\cite{Ciobanu2000,Klausen2001,Widera2006}, $^{85}$Rb~\cite{Klausen2001}, and  $^{23}$Na~\cite{Ciobanu2000}, though uncertainties overlap with the C phase.
The two nematic phases are energetically degenerate at the mean-field level for $p=q=0$. Beyond mean-field, the degeneracy may be broken by quantum fluctuations through order-by-disorder processes~\cite{Turner2007,Song2007}, but it can also be lifted already at the mean-field level~\cite{Borgh2016b} through a quadratic level shift.

When $p\neq0$ and $q\neq0$ in Eq.~\eqref{eq: single-particle}, the expressions for the magnetic phases as stationary solutions to the Gross-Pitaevskii energy functional \eqref{eq: energy-functional} become considerably more complex. These solutions and their dynamical and energetic stability have been investigated for both spin-1 and spin-2 BECs~\cite{koashi_prl_2000,Ciobanu2000,zhang_njp_2003,murata_pra_2007,ruostekoski_pra_2007}. In particular, the stationary solutions of the magnetic phases no longer follow the straightforward classification and
the Zeeman shifts
can, e.g., cause the condensate to adopt the properties of a FM phase even when nematic or C
phases are favoured by the interactions, and vice versa. The magnetic phases are then described by spinors where each component is a function of the Zeeman shifts and interatomic interactions, continuously interpolating between magnetic phases that would exist for $p=q=0$. 

The symmetries of the order parameter determine the topological properties of defects and textures~\cite{Mermin1979}. The non-trivial elements of the first homotopy group $\pi_1(\mathcal{M})$ represent singular line defects. These include singly and multiply quantized vortices defined by a winding of the condensate phase $\tau$ in Eq.~\eqref{eq: euler_angles}, 
and also vortices that arise purely from spin rotations and therefore carry spin circulation but no superfluid mass current (spin vortices).  Mass and spin circulation may also combine such that a fractional $2 \pi w$ winding of $\tau$ is compensated by a simultaneous spin rotation through $2 \pi \sigma$ (about some symmetry axis of the order parameter). We may thus denote vortex charges by $(w,\sigma)$ whenever these are uniquely defined~\cite{Kawaguchi2012}. For convenience, vortices with integer $w$ defined by winding of the condensate phase alone will in the following be referred to as phase vortices.
Singular vortices with $w=1/2$ are referred to as half-quantum vortices (HQVs) and appear in the BN phase~\cite{Borgh2016b}.  By analogy, spin vortices with $w=0$ (hence carrying no mass circulation) and $\sigma=1/2$ are called spin-HQVs and appear in the UN~\cite{Kawaguchi2012}, BN~\cite{Borgh2016b} and C phases~\cite{Makela2003,Semenoff2007}. The C phase also supports $1/3$- and $2/3$-quantum vortices (i.e., $|w|=1/3$ or $2/3$)~\cite{Makela2003,Semenoff2007,Kobayashi2009}.
Additionally, spin-2 phases support nonsingular vortices, which are textures that carry mass and/or spin circulation. 
Such vortices are trivial in $\pi_1(\mathcal{M})$, but may be topologically classified by the second homotopy group $\pi_2(\mathcal{M})$ when the boundary conditions on the texture are fixed.

The non-trivial elements of $\pi_2(\mathcal{M})$ also correspond to the topological charges of singular point defects (monopoles)~\cite{Mermin1979}. The UN phase is the only example of a spin-2 order parameter manifold where the second homotopy group is non-trivial~\cite{Kobayashi2012}, and thus supports topologically stable monopoles, in which the nematic axis forms a radial hedgehog texture, i.e., $\nematic = (\cos\varphi \sin\theta, \sin\varphi \sin\theta, \cos\theta)$, where $(\theta,\varphi)$ are the spherical coordinates centered on the point singularity. Despite $\pi_2(\mathcal{M})=0$ in the remaining magnetic phases, radial hedgehog textures can still exist, albeit with at least one associated line singularity extending away from the monopole. In the FM phases, these are the generalizations of the spin-1 Dirac monopoles~\cite{Savage2003b,Ray2014}, where the radial hedgehog appears in the condensate spin $\langle\hat{\mathbf{F}}\rangle$, while in the BN and C phases, the monopole can be formed by a chosen order-parameter symmetry axis.

\section{Topological interfaces and defect connections in spin-2 BECs\label{sec: theory}}
In a spinor superfluid, magnetic phases with different order-parameter symmetries can coexist.
For example, this situation arises spontaneously due to energy relaxation of defect cores~\cite{Ruostekoski2003,Lovegrove2012,Lovegrove2016,Borgh2016a,Borgh2016b}, as also observed in detailed experiments~\cite{Weiss2019,Xiao2021,Xiao2022}. The size of the defect core can then be understood from the healing lengths arising from the contributions to the interaction energy. There are  consequently three such length scales in the spin-2 BEC:
\begin{equation}
    \xi_d=\ell \sqrt{\frac{\hbar\omega}{2nc_0}},
    \enskip \xi_F=\ell \sqrt{\frac{\hbar\omega}{2n|c_1|}},
    \enskip \xi_a=\ell \sqrt{\frac{\hbar\omega}{2n|c_2|}},
    \label{eq: healing-lengths}
\end{equation}
where $\ell = (\hbar/M\omega)^{1/2}$ is the harmonic oscillator length. These describe, respectively, the distance over which perturbations of the superfluid density, condensate spin, and singlet duo amplitude heal to the bulk value. Typically, we have $\xi_F,\xi_a > \xi_d$ in current experimental realizations, allowing the core of a singular defect to reduce its energy by expanding and filling with a different superfluid phase~\cite{Ruostekoski2003}. The condensate wave function then smoothly interpolates between the coexisting phases in the bulk and the defect core, establishing a topological interface between them.

An extended topological interface may also be purposefully engineered to create spatially separate bulk regions with different order-parameter symmetries within the same, continuous superfluid. This can be achieved through spatial variation of
interaction strengths in the Hamiltonian~\eqref{eq: energy-functional}, such that different regions exhibit the characteristics of different magnetic phases~\cite{Borgh2012,Borgh2013}. The $s$-wave scattering lengths could be manipulated, e.g., using optical or microwave Feshbach resonances~\cite{Fatemi2000,Papoular2010}. Alternatively, it is possible to exploit stationary solutions in the presence of non-vanishing Zeeman shifts, in which case the BEC with spatially varying $p$ or $q$ can continuously interpolate between magnetic phases that would exist for $p=q=0$~\cite{Borgh2014}. Both approaches can result in stable stationary wave functions that interpolate between the bulk phases of different order-parameter symmetries for defects and textures, separated by a coherent topological interface. Here we use these to explicitly construct wave functions that smoothly connect vortices, other defects or nonsingular textures in the limiting phases.

From a representative spinor interpolating between chosen bulk phases, 
spinor vortices can typically be constructed by defining how the complex argument $\chi_m=\text{Arg}(\zeta_m) = k_m \varphi$, with $k_m \in \mathbb{Z}$, of each component winds as a function of the azimuthal angle $\varphi$ about the vortex line. Moreover,
textures and defects that occupy several spin components are constructed by the symmetry-group transformation of Eq.~\eqref{eq: euler_angles}, where the  Euler angles can be spatially dependent. 
Provided that a single transformation yields well-defined, single-valued states in both sides of the interface, a continuous connection across an interface is generated. Examples of this are shown in Secs.~\ref{sec: UN-BN}--\ref{sec: FM-BN}. Similar solutions can also describe filled vortex cores~\cite{Lovegrove2016} and composite defects~\cite{Lovegrove2014} when the interpolation parameter varies with the radial distance from a singular defect line.

\subsection{Uniaxial to biaxial nematic (UN-BN) \label{sec: UN-BN}}

\begin{figure}
    \centering
    \hspace*{-.15cm}\includegraphics[width=1.04\columnwidth]{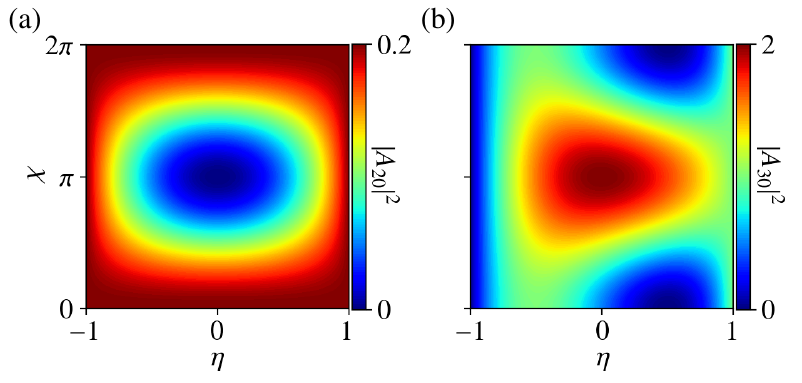}
    \caption{Spin-singlet duo and trio amplitudes $|A_{20}|^2$ (a) and $|A_{30}|^2$ (b), obtained 
    in Eqs.~\eqref{eq: eta_chi_singlets1} and \eqref{eq: eta_chi_singlets2} 
    as functions of the interpolating parameter $\eta$ and the relative phase difference $\chi$. Along $\chi \approx 0$, $\eta$ interpolates between the UN and BN phases. Crossover to the C phase is realized within the regions corresponding to $\pi/2 \lesssim \chi \lesssim 3\pi/2$ and
    $-0.5 \lesssim \eta \lesssim 0.9$.\label{fig: a20-a30-plot}}
\end{figure}

We first focus on stationary spinors satisfying Eq.~\eqref{eq: stationary_cond} and
interpolating between UN and BN phases.
Since we therefore require $\langle\hat{\mathbf{F}}\rangle=0$ and constant $A_{20}$, the five equations in the steady-state system~\eqref{eq: stationary_cond} can be treated as two independent $2\times 2$ linear systems corresponding to $m=\pm2$ and $m=\pm1$, respectively, and a single equation for $m=0$~\cite{Kawaguchi2012}. We choose to work in the three-component limit with $\zeta_{\pm 1}=0$, resulting in the spinor parametrization 
\begin{equation}
    \zeta^{\mathrm{UN-BN}} =
    \frac12
    \left(
        e^{i\chi_2}\sqrt{1-\eta},
        0,
        e^{i\chi_0}\sqrt{2(1+\eta)},
        0,
        e^{i\chi_{-2}}\sqrt{1-\eta}
    \right)^\text{T},
    \label{eq: UN-BN-interpolating-spinor}
\end{equation}
where $\chi_m$ are arbitrary phase coefficients that can assume fixed values or be spatially wound in defects and textures. Crucially, $\eta \in [-1,1]$ now forms an interpolating parameter between UN and BN magnetic phases: for $\eta=1$, only the $\zeta_0$ component is non-zero, representing the familiar UN phase in Eq.~\eqref{eq: UN_representative}, with the nematic axis aligned with the $z$-axis. Similarly, for $\eta=-1$, we retrieve the familiar BN phase in Eq.~\eqref{eq: BN_representative}. However, the interpolating region harbours additional complexity, revealed by the variation of the spin-singlet duo and trio amplitudes as a function of $\eta$ and  $\chi_m$. Calculating these using Eq.~\eqref{eq: UN-BN-interpolating-spinor} gives
\begin{align}
        &|A_{20}|^2 = \frac{1}{10} \bigg{[}\left(1 - \eta^2\right)\cos\left(\chi_2 + \chi_{-2} - 2\chi_0\right)
        + 1 + \eta^2 \bigg{]},
    \label{eq: eta_chi_singlets1}\\[2ex]
    \begin{split}
    &|A_{30}|^2 
    = \frac{1 + \eta}{4} \bigg{[}3 \left(\eta ^2 - 1\right) \cos\left(\chi_2 + \chi_{-2} - 2\chi_0\right)\\
        &\qquad\qquad\quad\;\;\; + \eta\, (5 \eta - 8) + 5 \bigg{]}.
    \end{split}
        \label{eq: eta_chi_singlets2}
\end{align}
Hence, we can study the behavior of Eqs.~\eqref{eq: eta_chi_singlets1} and \eqref{eq: eta_chi_singlets2} in the parameter space $(\chi,\eta)$, where $\chi=\chi_2 + \chi_{-2} - 2\chi_0$,  as shown in Fig.~\ref{fig: a20-a30-plot}. First consider $\chi=0$. We then notice from the variation of $|A_{30}|^2$ in Fig.~\ref{fig: a20-a30-plot}(b) that instead of a monotonic growth from the minimum $|A_{30}|^2=0$ (BN),
to $|A_{30}|^2=1$ (UN), the BN phase reappears around $\eta = 0.5$. If instead $\chi = \pi$, the C phase arises in the vicinity of $\eta = 0$, where $|A_{20}|^2=0$, $|A_{30}|^2=2$ and the spinor~\eqref{eq: UN-BN-interpolating-spinor} coincides with Eq.~\eqref{eq: C_representative}. For all values of $\chi$, the UN and BN limits at $\eta = \pm 1$ remain unchanged.  

The mean-field energy, calculated for a uniform spin-2 BEC in the state \eqref{eq: UN-BN-interpolating-spinor}, reads ($\tilde{c}_{0,1,2} = n c_{0,1,2}$)
\begin{align}
    \mathcal{E}^\text{UN-BN} = \mathcal{H}\left[\Psi^\text{UN-BN}\right]-\frac{\tilde{c}_0}{2} =  2q(1-\eta) + \frac{\tilde{c}_2}{2}|A_{20}|^2,
    \label{eq: un-bn_mfenergy}
\end{align}
where $|A_{20}|^2$ is given in Eq.~\eqref{eq: eta_chi_singlets1}. 
The convexity of $|A_{20}|^2$ as a function of $(\eta,\chi)$ means that for $q=0$, the energy~\eqref{eq: un-bn_mfenergy} is minimized along the $\chi=0,2\pi$ and $\eta=\pm1$ edges in Fig.~\ref{fig: a20-a30-plot}(a) when $\tilde{c}_2<0$ (the case $\tilde{c}_2>0$ will be discussed in Sec.~\ref{sec: C-BN}). This corresponds to the UN-BN degeneracy at zero level shifts.
In the presence of an external field such that $q\neq 0$, the $q$-dependent contribution (linear in $\eta$) in Eq.~\eqref{eq: un-bn_mfenergy} shifts the symmetry axis such that the value of $\eta$ that minimizes Eq.~\eqref{eq: un-bn_mfenergy} now depends on $q$. $\mathcal{E}^\text{UN-BN}$ is minimized in the BN limit ($\eta=-1$) for $q\leq 0$, and in the UN limit ($\eta=1$) for $q\geq 0$. The state~\eqref{eq: UN-BN-interpolating-spinor} thus represents a stationary solution such that one nematic phase is energetically favored over the other depending on the sign of the quadratic level shift, while also providing a smooth interpolation between these limits.

Now letting the interpolating parameter $\eta=\eta(z)$ vary between separated bulk regions along the $z$ axis, the solution provides a smooth UN-BN topological interface. This may be stabilized by varying $q=q(z)$, such that the BN and UN phases are favored on either side of the interface. 
Numerical examples of such interface engineering are provided in Sec.~\ref{sec: un-bn_numeric}. We are now ready to construct vortex states that cross a UN-BN interface, generated by different azimuthal phase windings $\chi_m$. These are summarized in Table~\ref{tab: UN-BN-vortices}.   

\textit{Phase vortices penetrating the interface}---We first consider a phase vortex in both UN and BN limits, obtained from Eqs.~\eqref{eq: UN_representative} and \eqref{eq: BN_representative}, with $k$ winding
\begin{equation}
\zeta^\text{UN}_\text{pv} = \exp(ik\varphi)\zeta^\text{UN},\quad \zeta^\text{BN}_\text{pv} = \exp(ik\varphi)\zeta^\text{BN}.
\end{equation}
These two solutions continuously connect across the interface according to the interpolating spinor in Eq.~\eqref{eq: UN-BN-interpolating-spinor} with $\chi_{0,\pm 2}=k\varphi$~\footnote{This is equivalent to choosing $\tau=k\varphi$ and constant Euler angles in Eq.~(\ref{eq: euler_angles}), acting on a uniform spinor $\zeta^\text{UN-BN}$ on the form Eq.~(\ref{eq: UN-BN-interpolating-spinor}) with $\chi_m=0$.} 
\begin{equation}
    \zeta^\text{UN-BN}_\text{pv} =
    \frac{e^{ik\varphi}}{2}
    \left(
        D_-,
        0,
        \sqrt2 D_+,
        0,
        D_-
    \right)^\text{T}, 
    \label{eq: UN-BN-SQV-SQV}
\end{equation}
representing a singular, interface-penetrating, $k$-quantized phase vortex, carrying mass-circulation. Here we have also introduced the shorthand
\begin{equation}
    D_\pm \equiv \sqrt{1\pm\eta}.
    \label{eq: D_pm}
\end{equation}

\textit{Phase vortices terminating at the interface}---Phase vortices may also terminate at the UN-BN interface. Such states are constructed by choosing $\chi_m$ in Eq.~\eqref{eq: UN-BN-interpolating-spinor} to introduce circulation in one limit of the interface only. For example, the choice $\chi_{\pm 2}=k\varphi,\, \chi_0 = 0$ yields phase vortices in the BN limit terminating to a vortex-free state in the UN while the spinor superfluid remains continuous and coherent everywhere:
\begin{equation}
    \zeta^\text{UN-BN}_\text{vf-pv} =  \frac12
    \left(
        e^{ik\varphi}D_-,
        0,
        \sqrt2 D_+,
        0,
        e^{ik\varphi}D_-
    \right)^\text{T}.
    \label{eq: UN-BN-VF-SQV}
\end{equation}
Conversely, azimuthal winding of the phase $\chi_0$ only, i.e., $\chi_0=k\varphi,\, \chi_{\pm 2} = 0$ results instead in a phase vortex in the UN limit, and a vortex-free state in the BN:
\begin{equation}
    \zeta^\text{UN-BN}_\text{pv-vf} =  \frac12
    \left(
        D_-,
        0,
        e^{ik\varphi}\sqrt2 D_+,
        0,
        D_-
    \right)^\text{T}.
    \label{eq: UN-BN-SQV-VF}
\end{equation}
In addition to describing terminating vortices, Eqs.~\eqref{eq: UN-BN-VF-SQV} and \eqref{eq: UN-BN-SQV-VF} can also parametrize the superfluid UN core of a BN phase vortex and vice versa, where the interface forms part of the vortex core structures~\cite{Xiao2022}. 

The terminating vortices break axisymmetry at the interface. This follows immediately from the $\chi$ dependence of $|A_{30}|^2$ at $\eta=0$ shown in Fig.~\ref{fig: a20-a30-plot}. For Eqs.~\eqref{eq: UN-BN-VF-SQV} and~\eqref{eq: UN-BN-SQV-VF} with $k=1$, $\chi = \pm 2\varphi$, respectively, $\chi$ takes all values $0$ to $\pm 4\pi$ on any closed loop around the vortex. Thus, the termination of a phase vortex implies the appearance of C-phase regions at the interface, where  $D_\pm \approx 1$ in Eq.~\eqref{eq: UN-BN-SQV-SQV} (see also Sec.~\ref{sec: un-bn_numeric}).  

\textit{Connections involving half-quantum vortices}---HQVs form in the BN phase by combining a $\pi$ winding of the condensate phase with a compensating spin rotation to keep the wave function single valued. We can immediately infer from the order-parameter symmetry shown in Fig.~\ref{fig: phases}(e) that an HQV can be formed either through a $\pi/2$ spin rotation about the $\hat{\mathbf{e}}_{(0,0,1)}$ symmetry axis to yield a $(1/2,1/4)$ vortex, or through a $\pi$ spin rotation about $\hat{\mathbf{e}}_{(1,1,0)}$ resulting in a $(1/2,1/2)$ vortex. More generally, vortices with any half-integer quanta of mass circulation can be constructed in a similar way. $(1/2,1/4)$ vortices connecting smoothly to a vortex-free UN limit are obtained from Eq.~\eqref{eq: UN-BN-interpolating-spinor} by choosing $\chi_{2} =\chi_0=0$, $\chi_{-2}=\varphi$ [equivalent to $\gamma=\varphi/4,\,\tau=2\gamma$ in Eq.~\eqref{eq: euler_angles}], to yield
\begin{equation}
    \zeta^\text{UN-BN}_{\text{vf-hqv}} = \frac12
    \left(
        D_-,
        0,
        \sqrt2 D_+,
        0,
        e^{i\varphi}D_-
    \right)^\text{T}.
    \label{eq: UN-BN-VF-HQV}
\end{equation}
The same BN $(1/2,1/4)$ vortex can also smoothly connect to a singly quantized UN phase vortex by instead choosing $\chi_0=\pm\varphi$ as the complex argument of the $\zeta_0$ component in Eq.~\eqref{eq: UN-BN-VF-HQV}.

\begin{table}
    \centering
    \begin{tabular}{cccccc}
        \toprule
        \multicolumn{5}{c}{UN-BN: Vortices from spinor-component phase winding} \\
        \midrule
        UN limit & BN limit &  $\chi_2/\varphi$ & $\chi_0/\varphi$ & $\chi_{-2}/\varphi$  \\
        \midrule
         Phase vortex & Phase vortex & $k$ & $k$ & $k$\\ 
         Vortex-free & Phase vortex & $k$ & 0 & $k$\\
         Phase vortex & Vortex-free & 0 & $k$ & 0\\
         Vortex-free & Spin vortex  & $-k$ & 0 & $k$\\
         Phase vortex & Spin vortex  & $-k$ & $k$ & $k$\\
         Vortex-free & Half-quantum vortex  & 0 & 0 & $1$\\
         Phase vortex & Half-quantum vortex  & 0 & $\pm 1$ & 1\\
        \bottomrule
    \end{tabular}
    \caption{Singular vortex connections across a UN-BN interface, characterized by the phase windings $\chi_m$ in Eq.~\eqref{eq: UN-BN-interpolating-spinor}. Generalizations to multiple quantization are given by $k \in \mathbb{Z}$.}
    \label{tab: UN-BN-vortices}
\end{table}

\textit{Connections involving spin vortices}---Spinor BECs additionally support the non-dissipative flow of spin, which can lead to vortices carrying spin circulation only: spin vortices. 
An example of a singular case is given by the combination $\chi_{\pm 2} = \mp k\varphi, \, \chi_0=0$ in Eq.~\eqref{eq: UN-BN-interpolating-spinor}, where a BN spin vortex terminates at the interface, connecting to a vortex-free UN state. This can equivalently be constructed through the action in Eq.~(\ref{eq: euler_angles}), e.g., by choosing the winding $\alpha=k\varphi/2$. Both constructions result in the spinor
\begin{equation}
    \zeta^\text{UN-BN}_\text{vf-sv} =  \frac12
    \left(
        e^{-ik\varphi}D_-,
        0,
        \sqrt2 D_+,
        0,
        e^{ik\varphi}D_-
    \right)^\text{T}.
    \label{eq: UN-BN-VF-SV}
\end{equation}
Note that for $k=\pm1$, the BN limit of Eq.~\eqref{eq: UN-BN-VF-SV} describes a vortex where the order parameter rotates by $\pi$ around the $\hat{\mathbf{e}}_{(0,0,1)}$ axis as depicted in Fig.~\ref{fig: phases}(e). The vortex is a spin-HQV, analogous to $\pi$-disclinations in nematic liquid crystals.
Spin vortices may also penetrate the interface, connecting to a corresponding spin vortex in the other phase. In the presence of a UN-BN interface, these can be obtained by 
applying Eq.~\eqref{eq: euler_angles} with $\alpha=k\varphi/2$, and constant $\beta=\pi/2$ to the interpolating spinor in Eq.~\eqref{eq: UN-BN-interpolating-spinor}. The resulting state reads
\begin{equation}
    \zeta^\text{UN-BN}_\text{sv-sv} = \frac14 \begin{pNiceMatrix}
        e^{-ik\varphi}(D_- + \sqrt3 D_+) \\
        0 \\
        \sqrt6 D_- - \sqrt2 D_+ \\
        0 \\
        e^{ik\varphi}(D_- + \sqrt3 D_+)
        \label{eq: UN-BN-SV-SV}
    \end{pNiceMatrix}.
\end{equation}
When $k=1$, Eq.~\eqref{eq: UN-BN-SV-SV} contains a spin-HQV in the UN limit, where the nematic director exhibits a radial disgyration in the $x,y$-plane. In the BN limit, the spin-HQV is instead formed by a $\pi$ rotation of the order parameter around the $\hat{\mathbf{e}}_{(1,0,0)}$ axis [see Fig.~\ref{fig: phases}(e)] along any closed loop around the vortex line.

\textit{Nonsingular textures and monopoles}---So far we have considered only singular line defects. However, the nematic spin-2 phases can also form nonsingular spin vortices.  We construct a fountain-like texture of the nematic axis $\nematic$ in the bulk UN phase [cf.\ Fig.~\ref{fig: phases}(d)]. Starting from Eq.~\eqref{eq: UN_representative}, we apply the transformation~\eqref{eq: euler_angles}, where the Euler angles are chosen such that the nematic axis, $\nematic = (\cos\alpha \sin\beta, \sin\alpha \sin\beta, \cos\beta)$, bends away from the vortex line with increasing radial distance $\rho$, i.e., with $\alpha=\varphi$. The resulting spinor reads
\begin{equation}
        \zeta^\text{UN} = \sqrt{\frac38}
    \begin{pNiceMatrix}
        e^{-2i\varphi}\sin^2\beta \\
        -e^{-i\varphi}\sin 2 \beta \\
        \frac{1}{\sqrt{6}}(1 + 3\cos2\beta) \\
        e^{i\varphi} \sin 2 \beta \\
        e^{2i\varphi}\sin^2\beta              \label{eq: UN-ncv}
    \end{pNiceMatrix},
\end{equation}
and the nonsingular texture is obtained by taking $\beta = \beta(\rho)$ to be a monotonically increasing function of the radial distance $\rho$ from the vortex line, with $\beta(0)=0$.

Instead applying the same transformation to the BN spinor~\eqref{eq: BN_representative}, however, results in a singular spin vortex, albeit together with a fountain texture formed by the $\hat{\mathbf{e}}_{(0,0,1)}$ symmetry axis [cf.\ Fig.~\ref{fig: phases}(e)]. The vortex is described by
\begin{equation}
    \zeta^\text{BN} = \frac{1}{\sqrt{8}}
     \begin{pNiceMatrix}
        e^{-2i\varphi}(\cos^2\beta + 1) \\
        e^{-i\varphi}\sin 2 \beta \\
        \sqrt{6}\sin^2\beta \\
        -e^{i\varphi}\sin 2 \beta \\
        e^{2i\varphi}(\cos^2\beta + 1)          \label{eq: BN-sv} \\
    \end{pNiceMatrix}.
\end{equation}
Since Eqs.~\eqref{eq: UN_representative} and~\eqref{eq: BN_representative} are exactly the UN and BN limits, respectively, of the UN-BN interpolating spinor~\eqref{eq: UN-BN-interpolating-spinor} with all $\chi_m=0$, it follows immediately that the nonsingular UN spin vortex in Eq.~\eqref{eq: UN-ncv} can connect smoothly across the interface to the singular spin vortex in Eq.~\eqref{eq: BN-sv} on the BN side. The corresponding wavefunction is given by
\begin{equation}
     \zeta^\text{UN-BN} = \frac{1}{\sqrt{2}}\left( D_+\zeta^\text{UN} + 
                        D_-\zeta^\text{BN}\right), \label{eq: UN-BN-orbit}    
\end{equation}
where $D_\pm(z)$, defined in Eq.~\eqref{eq: D_pm}, parametrize the spatial interpolation between the bulk regions.

On a topological interface, monopoles may form as termination points of singular vortex lines, similar to ‘‘boojums'' in superfluid liquid $^3$He~\cite{Blaauwgeers2002,Volovik2009}. The construction is analogous to that connecting the BN singular spin vortex to the UN nonsingular spin vortex. The spinor is again given by  Eqs.~\eqref{eq: UN-ncv}--\eqref{eq: UN-BN-orbit}, only now taking $\beta=\theta$, independent of radial distance to form the required monopole configuration in the UN limit.
In the BN limit, the spinor still represents a singular spin vortex. As the spinor interpolates across the interface, however, this vortex line now terminates on the UN point defect.
 
It is also possible for a singular UN vortex to terminate as a BN monopole, constructed such that the vortex coincides with the associated line singularity [which always exists since $\pi_2(\mathcal{M}_{\text{BN}}) = 0$]. As in Ref.~\cite{Borgh2016b}, we construct the monopole by applying Eq.~\eqref{eq: euler_angles} with $\alpha = -\gamma =\varphi$ and $\beta = \theta$ to the BN spinor $\zeta^{\text{BN}} = (1,0,\sqrt{6},0,1)^\text{T}/\sqrt8$ [itself obtained applying a $\beta=\pi/2$ rotation to Eq.~\eqref{eq: BN_representative}]. The line singularity is then aligned with the negative $z$-axis. On a topological interface, a BN monopole may then be oriented such that the line defect is ``hidden'' on the opposite side. For example, constructing an interpolating spinor on the form~\eqref{eq: UN-BN-orbit}, where the monopole forms the BN limit for $z>0$ and interpolating to the UN phase for $z<0$, we find 
the limits
\begin{align}
    &\zeta^\text{UN} =\sqrt{\frac38}\begin{pNiceMatrix}
        e^{-2 i \varphi } (\cos \theta \cos \varphi+i \sin \varphi)^2 \\
        2 e^{-i \varphi } \sin \theta \cos \varphi \,(\cos \theta \cos \varphi+i \sin \varphi) \\
        \frac{1}{2 \sqrt{6}}\left(6 \sin ^2\theta \cos 2 \varphi -3 \cos 2 \theta -1 \right) \\
        -2 e^{i \varphi } \sin \theta \cos \varphi \,(\cos \theta \cos \varphi-i \sin \varphi)\\
        e^{2 i \varphi } (\cos \theta \cos \varphi-i \sin \varphi)^2
    \end{pNiceMatrix},
    \label{eq: UN-monopole2}\\[2ex]
    &\zeta^\text{BN} \!=\! \frac{1}{4\!\sqrt 2} \begin{pNiceMatrix}
        2 e^{-4 i \varphi } \sin ^4\frac{\theta }{2}+3 e^{-2 i \varphi } \sin ^2\theta+2 \cos ^4\frac{\theta }{2} \\
        e^{-3 i \varphi } \sin \theta \left[e^{4 i \varphi } (\cos \theta \!-\! 1) \!-\! 6 e^{2 i \varphi } \cos \theta \!+\! \cos\theta \!+\! 1\right] \\
        \sqrt{\frac{3}{2}} \left(2 \sin ^2\theta  \cos 2 \varphi +3 \cos 2 \theta +1\right) \\
        2e^{-i \varphi } \sin \theta  \left[\cos\theta \,(\cos 2\varphi -3) + i\sin 2\varphi \right] \\
        2 e^{4 i \varphi } \sin ^4\frac{\theta }{2}+3 e^{2 i \varphi } \sin ^2\theta+2\cos ^4\frac{\theta }{2}
    \end{pNiceMatrix}.
    \label{eq: BN-monopole2}
\end{align}
The spinor interpolating between these limits then represents a UN singular spin vortex, which terminates as a BN monopole on the interface.
The states obtained in Eqs.~\eqref{eq: UN-BN-SV-SV}--\eqref{eq: BN-monopole2} are summarized in Table~\ref{tab: UN-BN-other}.
Both examples of nematic spin vortices terminating as monopoles on the UN-BN interface are simulated numerically in Sec.~\ref{sec: un-bn_numeric}.
\begin{table}
    \centering
    \begin{tabular}{cccccc}
        \toprule
        \multicolumn{5}{c}{UN-BN: Vortices, textures and monopoles from Euler angles} \\
        \midrule
        UN limit & BN limit &   $\alpha/\varphi$ & $\gamma/\varphi$ & $\beta$ \\
        \midrule
         Spin Half-quantum vortex & Spin Half-quantum vortex & 1/2 & 0 & $\pi/2$ \\
         Nonsingular spin vortex & Spin vortex & 1 & 0 & $\beta(\rho)$ \\ 
         Monopole & Spin vortex & $1$  & 0 & $\theta$\\
         Spin vortex & Monopole & $1$  & $-1$ & $\theta$\\
        \bottomrule
    \end{tabular}
    \caption{Singular and nonsingular spin vortices and monopoles connecting across a UN-BN interface, constructed by azimuthal dependence of Euler angles $\alpha$, $\gamma$ (given as multiples of the azimuthal angle $\varphi$), and $\beta$ (given as a multiple of the polar angle $\theta$ or, for nonsingular vortices, as a monotonically increasing function of the transverse radius $\rho$), with $\tau=0$.}
    \label{tab: UN-BN-other}
\end{table}

\subsection{Cyclic to nematic (C-UN/BN) \label{sec: C-BN}}
As shown in Sec.~\ref{sec: UN-BN}, the steady-state family of phase-mixing spinors in Eq.~\eqref{eq: UN-BN-interpolating-spinor} includes a crossover from both nematic phases to the C phase if we ensure a constant $\pi/2$ phase difference between the components $\zeta_0$ and  $\zeta_{\pm2}$, equivalent to restricting ourselves to the subset of solutions given by
\begin{equation}
    \zeta^{\mathrm{C-UN/BN}} =
    \frac12
    \left(
        e^{i\chi_2}\sqrt{1 \!-\! \eta},
        0,
        ie^{i\chi_0}\sqrt{2(1 \!+\! \eta)},
        0,
        e^{i\chi_{-2}}\sqrt{1 \!-\! \eta}
    \right)^\text{T},
    \label{eq: C-BN-interpolating-spinor}
\end{equation}
where $\chi_2+\chi_{-2}-2\chi_0=0$. Thus, the C spinor in Eq.~\eqref{eq: C_representative}, with tetrahedral symmetry combined with the condensate phase, is obtained for $\eta=0$. Conversely, when $\eta=\pm 1$, we once again retrieve UN or BN states, respectively. The uniform mean-field energy of $\zeta^{\mathrm{C-UN/BN}}$ in Eq.~(\ref{eq: C-BN-interpolating-spinor}) reads 
\begin{equation}
 \mathcal{E}^\text{C-UN/BN}  =  2q(1-\eta) + \frac{\tilde{c}_2}{10}\eta^2,   
\end{equation}
and is minimized by the value $\eta = 10q/\tilde{c}_2 $ when $\tilde{c}_2 > 0$, i.e., we now have a $q$-dependent interpolating parameter. 

We can then construct interface solutions interpolating between C and UN/BN magnetic phases for $|q| \leq \tilde{c}_2/10$, as numerically demonstrated in Sec.~\ref{sec: c-bn_numeric}. We focus here on a C-BN interface only, where the Zeeman shift $q$ varies spatially from
$q=0$ (C) to $q=-\tilde{c}_2 /10$ (BN). 

\textit{Phase and spin vortices penetrating the interface}---The defect connections involving phase and spin vortices discussed in Sec.~\ref{sec: UN-BN} for the UN-BN case are retrieved here with the simple replacement rule $D_+ \rightarrow iD_+$ in Eqs.~\eqref{eq: UN-BN-SQV-SQV}, \eqref{eq: UN-BN-VF-SV} and \eqref{eq: UN-BN-SV-SV}, where the C limit is given by $D_\pm=1$. For example, spin vortices in both C and BN phases connecting across the interface are given by the wave function
\begin{equation}
    \zeta^\text{C-BN}_\text{sv-sv} =  \frac12
    \left(
        e^{-ik\varphi}D_-,
        0,
        i\sqrt2 D_+,
        0,
        e^{ik\varphi}D_-
    \right)^\text{T},
    \label{eq: C-BN-SV-SV}
\end{equation}
obtained from the interpolating spinor in Eq.~\eqref{eq: C-BN-interpolating-spinor} by choosing $\chi_{\pm 2} = \mp k\varphi, \chi_0=0$. For $k=1$, both limits represent spin-HQVs. The vortex connections across a C-BN interface constucted from the phase windings $\chi_m$ are represented in Table~\ref{tab: C-BN-vortices}. Note that, despite the similar forms of Eqs.~\eqref{eq: UN-BN-interpolating-spinor} and \eqref{eq: C-BN-interpolating-spinor}, solutions with $\chi_2 + \chi_{-2} -2\chi_0 \neq 0$ in Eq.~\eqref{eq: C-BN-interpolating-spinor} do not produce well-defined defect states in the C limit. 
\begin{table}
    \centering
    \begin{tabular}{cccccc}
        \toprule
        \multicolumn{5}{c}{C-UN/BN: Vortices from spinor-component phase winding} \\
        \midrule
        C/BN limits & UN limit &  $\chi_2/\varphi$ & $\chi_0/\varphi$ & $\chi_{-2}/\varphi$  \\
        \midrule
         Phase vortex & Phase vortex & $k$ & $k$ & $k$\\ 
         Spin vortex & Vortex free  & $-k$ & 0 & $k$\\
        \bottomrule
    \end{tabular}
    \caption{Vortex connections across a C-UN/BN interface, characterized by the phase windings $\chi_m$ in Eq.~\eqref{eq: C-BN-interpolating-spinor}.}
    \label{tab: C-BN-vortices}
\end{table}

The symmetries of the C and BN order parameters also allow a further connection between quantized spin vortices, distinct from Eq.~\eqref{eq: C-BN-SV-SV}. We now construct the vortex state from spin rotations about the $\hat{\mathbf{e}}_{(1,1,0)}=\left(\hat{\mathbf{e}}_{(1,0,0)}+\hat{\mathbf{e}}_{(0,1,0)}\right)/\sqrt{2}$ in Fig.~\ref{fig: phases}(c) and (e), given by the operator\footnote{The same spin rotation can also be written on the form of Eq.~\eqref{eq: euler_angles} with Euler angles $\alpha=-\gamma =\pi/4,\, \beta=\varphi$. This is, however, less intuitively instructive here than the axis-angle representation.}
\begin{align}
    \hat{U}\left(\hat{\mathbf{e}}_{(1,1,0)}, \delta\right) = 
    \exp\left(-i\frac{\hat{F}_x 
    + \hat{F}_y}{\sqrt{2}}\delta\right)
    \label{eq: C-BN_sv_rot}
\end{align}
in the axis-angle representation. Whenever $\delta$ winds by an integer multiple of $2\pi$ on a closed loop around the vortex line, this defines a spin vortex with integer quantization in both C and BN phases. Applying Eq.~\eqref{eq: C-BN_sv_rot} with $\delta=\varphi$ to the interpolating spinor $\zeta^\text{C-BN}$ [Eq.~\eqref{eq: C-BN-interpolating-spinor} with $\chi_m=0$] leads to a spinor of the form of Eq.~\eqref{eq: UN-BN-orbit} with $D_+ \rightarrow iD_+$ and 
\begin{equation}
    \zeta^\text{UN}_\text{sv} = \sqrt{\frac38}
    \begin{pNiceMatrix}
        -i\sin^2 \varphi \\
        -e^{-\frac{i\pi }{4}} \sin 2\varphi \\
        \frac{1}{\sqrt{6}}\left(1 \!+\! 3\cos 2\varphi \right)\\
        e^{\frac{i \pi }{4}} \sin2 \varphi \\
        i\sin^2 \varphi
    \end{pNiceMatrix}, \quad
    \zeta^\text{BN}_\text{sv} = \frac{1}{\sqrt2}
    \begin{pNiceMatrix}
        \cos \varphi \\
        e^{\frac{i\pi }{4}} \sin \varphi \\
        0\\
        e^{\frac{3i \pi }{4}} \sin \varphi \\
        \cos \varphi
    \end{pNiceMatrix}.
    \label{eq: C-BN-SV2-SV2}
\end{equation}
This state corresponds to singular, singly quantized spin vortices in all C, UN, and BN limits, and thus in particular represent a continuous connection between C and BN phases.

\textit{Nonsingular textures and monopoles}---Line defects that terminate as monopoles or connect to nonsingular nematic textures in the UN limit can similarly be defined through constructions paralleling those discussed in Sec.~\ref{sec: UN-BN}, with the replacement $D_+ \rightarrow iD_+$. The states thus obtained, together with the connecting spin vortices given by Eq.~\eqref{eq: C-BN-SV2-SV2}, are summarized in Table~\ref{tab: C-BN-other}. 
\begin{table}
    \centering
    \begin{tabular}{cccccc}
        \toprule
        \multicolumn{5}{c}{C-UN/BN: Vortices, textures and monopoles from Euler angles} \\
        \midrule
        C/BN limits & UN limit &  $\alpha/\varphi$ & $\gamma/\varphi$ & $\beta$ \\
        \midrule
         Spin vortex & Spin vortex  & $\alpha=\frac{\pi}{4}$ & $\gamma=-\alpha$ & $k\varphi$ \\
         Spin vortex & Nonsingular spin vortex & 1 & 0 & $\beta(\rho)$ \\ 
         Spin vortex & Monopole  & $1$  & 0 & $\theta$\\
        \bottomrule
    \end{tabular}
    \caption{Spin vortices and monopoles connecting across a C-UN/BN interface, constructed by azimuthal dependence of Euler angles $\alpha$, $\gamma$ (given as multiples of the azimuthal angle $\varphi$), and $\beta$ (given as a multiple of the polar angle $\theta$ or, for nonsingular vortices, as a monotonically increasing function of the transverse radius $\rho$), with $\tau=0$.} 
    \label{tab: C-BN-other}
\end{table}

\subsection{Cyclic to ferromagnetic (C-FM) \label{sec: C-FM}}
Thus far, we have considered interfaces that can be induced by a spatially varying quadratic Zeeman shift $q$. Now turning our attention to an interface between C and FM phases, we consider instead engineering of a linear Zeeman shift $p$. When the longitudinal magnetization is allowed to adapt, $p$ can stabilize population imbalances, inducing net longitudinal magnetization, and can act as an interpolation parameter, similarly to the spin-1 case~\cite{Borgh2014}. 
A topological interface can be formed by a spatially inhomogeneous $p$ even when the total magnetization is conserved.
We consider states for which $\langle \hat{\mathbf{F}} \rangle = (0,0,\langle \hat{F}_z \rangle)^\text{T}$ and $\zeta_0 = 0$, resulting in distinct families of interpolating spinors satisfying Eq.~\eqref{eq: stationary_cond} and the following (mutually exclusive) conditions
\begin{gather}
 \zeta_2\zeta_1 = \zeta_{-1}\zeta_{-2} = 0,  \label{eq: C-FM_condition}\\
 \zeta_1 = \zeta_{-1} = 0 \quad \text{or} \quad \zeta_2 = \zeta_{-2} = 0  . \label{eq: FM-BN_condition}
\end{gather} 
We defer Eq.~\eqref{eq: FM-BN_condition} to Sec.~\ref{sec: FM-BN}, and concentrate on spinors satisfying Eq.~\eqref{eq: C-FM_condition}. We then obtain a one-parameter family of steady-state spinors interpolating between C and FM phases:  
\begin{equation}
    \zeta^\text{C-FM} = \frac{1}{\sqrt{3}}
    \left(
        e^{i\chi_2}\sqrt{1 + \eta} ,
        0 ,
        0 ,
        e^{i\chi_{-1}}\sqrt{2 - \eta} ,
        0
    \right)^\text{T},
    \label{eq: C-FM-interpolating-spinor}
\end{equation}
where the interpolating parameter $\eta = \langle \hat{F}_z\rangle$.
For $\eta=0$, Eq.~\eqref{eq: C-BN-interpolating-spinor} reduces to the C limit
\begin{equation}
    \zeta^\text{C} = \frac{1}{\sqrt3}\left(1,0,0,\sqrt{2},0\right)^\text{T},   
\end{equation}
which is related to Eq.~\eqref{eq: C_representative} via a spinor rotation.
For $\eta=-1$ and $\eta=2$, we obtain the $\FMone$ and $\FMtwo$ limits with  $|\langle \hat{\vb{F}} \rangle|=1$ and $|\langle \hat{\vb{F}} \rangle|=2$, respectively. 
Since $|A_{20}|^2=0$ for all $\eta$, the uniform mean-field energy~\eqref{eq: energy-functional} of $\zeta^\text{C-FM}$ depends only on $\eta$ and is independent of the phases $\chi_2,\chi_{-1}$:
\begin{align}
    \mathcal{E}^\text{C-FM} = 2q - (p-q)\,\eta + \frac{\tilde{c}_1}{2} \eta^2 .
    \label{eq: C-FM_energy}
\end{align}
It is minimized when $\eta=(p-q)/\tilde{c}_1$ for $\tilde{c}_1>0$. 
We can then construct a C-FM interface in Eq.~\eqref{eq: C-FM-interpolating-spinor} for $q=0$ by letting $p$ interpolate from $p=-\tilde{c}_1$ (FM$_1$) or $p = 2\tilde{c}_1$ (FM$_2$) to $p=0$ (C).
Alternatively, when $p=q=0$ in Eq.~\eqref{eq: C-FM_energy}, the C phase is favoured for $\tilde{c}_1>0$ and $\FMtwo$ for $\tilde{c}_1<0$, The interface in Eq.~\eqref{eq: C-FM-interpolating-spinor} may therefore also be formed by a spatially varying $\tilde{c}_1$. These are numerically examined in Sec.~\ref{sec: c_fm_numeric}.

\textit{Phase and spin vortices penetrating the interface}---As in Secs.~\ref{sec: UN-BN} and \ref{sec: C-BN}, the vortices connecting C and FM magnetic phases are constructed by windings of $\chi_2$ and $\chi_{-1}$ in Eq.~\eqref{eq: C-FM-interpolating-spinor}, where $\chi_2=\chi_{-1}=k\varphi$ results in phase vortices in both C and FM limits, while $\chi_2=-2k\varphi,\, \chi_{-1}=k\varphi$ yields a spin vortex (in C) connecting to a phase vortex (in FM).

\textit{Connections involving fractional vortices}---Introducing winding of the complex argument in only one of the spinor components in Eq.~\eqref{eq: C-FM-interpolating-spinor} results in the vortex states (for $\chi_2 =\varphi,\,\chi_{-1}=0$, and $\chi_2 =0,\,\chi_{-1}=\varphi$)
\begin{align}
    &\zeta^\text{C-FM}_{\frac13} = \frac{1}{\sqrt{3}}
    \left(
         e^{i\varphi}D_2 ,
         0 ,
         0 ,
         D_{-1},
         0
    \right)^\text{T}, \label{eq: C-FM-third}\\
    &\zeta^\text{C-FM}_{\frac23} = \frac{1}{\sqrt{3}}
    \left(
         D_2 ,
         0 ,
         0 ,
         e^{i\varphi}D_{-1},
         0
    \right)^\text{T}, \label{eq: C-FM-2third}
\end{align}
with $D_{\pm m} \equiv \sqrt{|m - 1|\pm \eta}$. These correspond to $\tau=\varphi/3, \gamma=-\tau$ for Eq.~\eqref{eq: C-FM-third} ($1/3$-vortex) and $\tau=2\varphi/3, \gamma=\tau/2$ for Eq.~\eqref{eq: C-FM-2third} ($2/3$-vortex), when expressed using Eq.~\eqref{eq: euler_angles}~\cite{Semenoff2007, Kobayashi2009, Huhtamaki2009}. 
Since mass circulation is determined by $\tau$, the C limit of Eqs.~\eqref{eq: C-FM-third} and~\eqref{eq: C-FM-2third} represent vortices with $1/3$ and $2/3$ quanta of circulation, respectively. Due to the spin-gauge symmetry of the FM phases, these states connect to phase vortices or vortex-free states in the FM limit, resulting in different mass circulation between the magnetic phases. 
Specifically, Eq.~\eqref{eq: C-FM-third} represents an interpolating solution from a singly quantized phase vortex ($\FMtwo$) to a vortex-free state ($\FMone$) via a $(1/3,-1/3)$ vortex (C), and 
Eq.~\eqref{eq: C-FM-2third} from a vortex-free state
($\FMtwo$) to a singly quantized phase vortex ($\FMone$) via a $(2/3,1/3)$ vortex (C). The C-FM vortex connections are summarized in Table~\ref{tab: C-FM-vortices}.
\begin{table}
    \centering
    \begin{tabular}{cccc}
        \toprule
        \multicolumn{4}{c}{C-FM: Vortices from spinor-component phase winding} \\
        \midrule
        C limit & FM$_2$ limit &  $\chi_2/\varphi$ & $\chi_{-1}/\varphi$  \\
        \midrule
         Phase vortex & Phase vortex & $k$ & $k$\\ 
         Spin vortex & Phase vortex  & $-2k$ & $k$\\
         $1/3$-vortex & Phase vortex & 1 & 0 \\
         $2/3$-vortex & Vortex-free  & 0 & 1 \\
        \bottomrule
        \toprule
        C limit & $\FMone$ limit &  $\chi_2/\varphi$ & $\chi_{-1}/\varphi$  \\
        \midrule
         Phase vortex & Phase vortex & $k$ & $k$\\ 
         Spin vortex & Phase vortex  & $-2k$ & $k$\\
         $1/3$-vortex & Vortex-free  & 1  & 0 \\
         $2/3$-vortex & Phase vortex & 0  & 1 \\
        \bottomrule
    \end{tabular}
    \caption{Singular vortex connections across a C-FM interface, characterized by the phase windings $\chi_m$ in Eq.~\eqref{eq: C-FM-interpolating-spinor}.} 
    \label{tab: C-FM-vortices}
\end{table}

\textit{Nonsingular textures and monopoles}---By considering Eq.~\eqref{eq: euler_angles}, we obtain the interpolating solutions determined by the symmetry group $G$ acting on Eq.~\eqref{eq: C-FM-interpolating-spinor}
\begin{equation}
    \zeta^\text{C-FM} = \frac{1}{\sqrt3}\left(D_2\zeta^{\text{FM}_2^+} + D_{-1}\zeta^{\text{FM}_{1}^-}\right), 
    \label{eq: C-FM-orbit}
\end{equation}
where $\zeta^\text{FM$_{1,2}^\pm$}$ represent the parametrizations of the states with $F_z=\pm 1,2$
\begin{align}  
    &\zeta^{\text{FM}_{2}^+} = 
    e^{i(\tau-2\gamma)}\begin{pNiceMatrix}
        e^{-2i\alpha}\cos^4\frac{\beta}{2}\\
        2e^{-i\alpha}\cos^3\frac{\beta}{2}\sin\frac{\beta}{2}\\
        \sqrt{6}\cos^2\frac{\beta}{2}\sin^2\frac{\beta}{2}\\
        2e^{i\alpha}\cos\frac{\beta}{2}\sin^3\frac{\beta}{2}\\
        e^{2i\alpha}\sin^4\frac{\beta}{2}
    \end{pNiceMatrix}, \label{eq: FM2_orbit}\\[1ex]
    &\zeta^{\FMone^-} = 
    e^{i(\tau+\gamma)}\begin{pNiceMatrix}
        -2e^{-2i\alpha}\cos\frac{\beta}{2}\sin^3\frac{\beta}{2}\\
        e^{-i\alpha}\sin^2\frac{\beta}{2}\left(3\cos^2\frac{\beta}{2} - \sin^2\frac{\beta}{2}\right)\\
        \sqrt{6}\left(\cos\frac{\beta}{2}\sin^3\frac{\beta}{2} - \cos^3\frac{\beta}{2}\sin\frac{\beta}{2}
        \right)\\
        e^{i\alpha}\cos^2\frac{\beta}{2}\left(\cos^2\frac{\beta}{2}-3\sin^2\frac{\beta}{2}\right)\\
        2e^{2i\alpha}\cos^3\frac{\beta}{2}\sin\frac{\beta}{2}
    \end{pNiceMatrix}. \label{eq: FM1_orbit}
\end{align}  

In the FM phases, angular momentum can be carried by
nonsingular (coreless) vortices.
The best known examples exhibit a fountain-like spin texture where the spin density 
aligns with $\hat{\mathbf{z}}$ at $\rho=0$, and  tilts away as $\rho$ increases, corresponding to a monotonically increasing Euler angle $\beta=\beta(\rho)$.
In the FM$_2$ case, the order parameter is kept nonsingular everywhere by any choice of phase and Euler angles such that
\begin{equation}
    \tau-2\gamma=\pm 2\alpha=\pm 2\varphi,
    \label{eq: FM2-coreless-condition}
\end{equation}
whereas the FM$_1$ case requires 
\begin{equation}
    \tau+\gamma=\pm\alpha= \pm\varphi.
    \label{eq: FM1-coreless-condition}
\end{equation}
Substituting these into Eqs.~\eqref{eq: FM2_orbit} and \eqref{eq: FM1_orbit} results in nonsingular FM vortices connecting across the interface to  singular C vortices. 
The latter include phase ($\gamma=0$), spin ($\tau=0$), and fractional ($\tau=\varphi/3$ and $2\varphi/3$) vortices.

Moreover, analogously to spin-1 BECs~\cite{Savage2003b,Borgh2012}, the FM$_{1,2}$ generalizations of the Dirac monopoles can be defined by taking $\beta=\theta$ in Eqs.~\eqref{eq: FM2_orbit} and~\eqref{eq: FM1_orbit}.  The condensate phase and the remaining Euler angles are chosen according to Eq.~\eqref{eq: FM2-coreless-condition} for $\FMtwo$ and Eq.~\eqref{eq: FM1-coreless-condition} for $\FMone$.
This yields the characteristic hedgehog texture of $\langle \hat{\mathbf{F}} \rangle$, embedding a singular vortex line terminating on the monopole. Similarly to nonsingular vortices, $\FMtwo$ Dirac monopoles can connect continuously to phase, spin and fractional vortices in the C limit. 

All interpolating spinors constructed in this way from Eqs.~\eqref{eq: C-FM-orbit}--\eqref{eq: FM1_orbit} are summarized in Table~\ref{tab: C-FM-other} for the C-FM interfaces, with a discussion of the analogous constructions for a FM-BN interface to follow in Sec.~\ref{sec: FM-BN}. 
\begin{table}
    \centering
    \begin{tabular}{cccccc}
        \toprule
        \multicolumn{6}{c}{C-FM: Vortices, textures and monopoles from Euler angles} \\
        \midrule
        C limit & FM$_2$ limit &  $\tau/\varphi$ & $\alpha/\varphi$ & $\gamma/\varphi$ & $\beta$ \\
        \midrule
        Phase vortex & Nonsingular vortex & 2 & 1 & 0 & $\beta(\rho)$ \\ 
        Spin vortex & Nonsingular vortex & 0 & 1 & $\pm 1$ & $\beta(\rho)$ \\ 
        $2/3$-vortex & Nonsingular vortex & 2/3 & 1 & -2/3, 4/3 & $\beta(\rho)$ \\
        Phase vortex & Dirac monopole & 2 & $1$ & 0 & $\theta$ \\ 
        Spin vortex & Dirac monopole & 0 & $ 1$ & $\pm 1$ & $\theta$ \\ 
        $2/3$-vortex & Dirac monopole & 2/3 & $ 1$ & -2/3, 4/3 & $\theta$ \\
        \bottomrule
        \midrule
        C limit & FM$_1$ limit &  $\tau/\varphi$ & $\alpha/\varphi$ & $\gamma/\varphi$ & $\beta$ \\
        \midrule
        Phase vortex & Nonsingular vortex & 1 & 1 & 0 & $\beta(\rho)$ \\ 
        Spin vortex & Nonsingular vortex & 0 & 1 & $\pm 1$ & $\beta(\rho)$ \\ 
        $1/3$-vortex & Nonsingular vortex & 1/3 & 1 & -4/3, 2/3 & $\beta(\rho)$ \\
        Phase vortex & Dirac monopole & 1 & 1 & 0 & $\theta$ \\ 
        Spin vortex & Dirac monopole & 0 & 1 & $\pm 1$ & $\theta$ \\ 
        $1/3$-vortex & Dirac monopole & 1/3 & 1 & -4/3, 2/3 & $\theta$ \\
        \bottomrule
    \end{tabular}
    \caption{Singular and nonsingular vortices and monopoles connecting across a C-FM interface constructed through spatial dependent $\tau$, $\alpha$,
    and $\gamma$ (given as multiples of the azimuthal angle $\varphi$), and $\beta$ (given as a multiple of the polar angle $\theta$ or, for nonsingular vortices, as a monotonically increasing function of the transverse radius $\rho$).}
    \label{tab: C-FM-other}
\end{table}

\subsection{Ferromagnetic to biaxial nematic (FM-BN) \label{sec: FM-BN}}
We now return to the  steady-state spinors with zero transverse magnetization (and arbitrary longitudinal magnetization) that satisfy Eq.~\eqref{eq: FM-BN_condition}. These can form solutions that interpolate between the $\FMtwo$ and BN phases:
\begin{equation}
    \zeta^\mathrm{FM_2-BN} = 
    \frac{1}{\sqrt2}
    \left(
        e^{i\chi_2}\sqrt{1 + \eta} ,
        0 ,
        0 ,
        0 ,
        e^{i\chi_{-2}}\sqrt{1 - \eta}
    \right)^\text{T},
    \label{eq: FM-BN-interpolating-spinor}
\end{equation}
where $\eta = \langle \hat{F}_z\rangle/2$, providing the BN phase at $\eta=0$, and $\FMtwo$ at $\eta =\pm 1$. The uniform mean-field energy of Eq.~\eqref{eq: FM-BN-interpolating-spinor} reads
\begin{equation}
    \mathcal{E}^\text{FM$_2$-BN}  = 4q -  2p \eta + \left(2\tilde{c}_1-\frac{\tilde{c}_2}{10}\right) \eta^2  + \frac{\tilde{c}_2}{10},
    \label{eq: FM_BN_energy}
\end{equation}
and is minimized by $\eta = p/(2\tilde{c}_1-\tilde{c}_2/10)$, provided that the spin-dependent and spin-singlet interactions satisfy $\tilde{c}_1\geq \tilde{c}_2/20$ at fixed $p,q$. 
We construct a topological interface between the $\FMtwo$ and BN phases in Eq.~\eqref{eq: FM-BN-interpolating-spinor} with a linear Zeeman shift varying spatially between $p=\pm (2\tilde{c}_1-\tilde{c}_2/10$) (FM$_2$), and $p=0$ (BN). Additionally, in the absence of Zeeman shifts, Eq.~\eqref{eq: FM_BN_energy} shows that the BN phase is energetically favoured for $\tilde{c}_1>\tilde{c}_2/20$, and the FM$_2$ limit otherwise. Thus, FM-BN interfaces are also obtained with spatially varying $\tilde{c}_1$.  These are numerically illustrated in Sec.~\ref{sec: FM-BN_numeric}. 

\textit{Connections among phase, spin and half-quantum vortices}---The spin-gauge symmetry of the $\FMtwo$ limits in Eq.~\eqref{eq: FM-BN-interpolating-spinor} allows further varieties of vortex connections, identified by combinations of winding of the phase coefficients $\chi_{\pm 2}$. 
The cases of interest here read
\begin{align}
    &\zeta^\mathrm{FM_2-BN}_\text{pv-sv} \!=\! \frac{1}{\sqrt{2}}
    \left(
         e^{-ik\varphi}D_+ ,
         0 ,
         0 ,
         0 ,
         e^{ik\varphi}D_-
    \right)^\text{T},  \label{eq: FM-BN-sqv-sv}\\
    &\zeta^\mathrm{FM_2-BN}_\text{pv-hqv} = \frac{1}{\sqrt{2}}
    \left(
         D_+ ,
         0 ,
         0 ,
         0 ,
         e^{i\varphi}D_- 
    \right)^\text{T},
    \label{eq: FM-BN-sqv-hqv}
\end{align}
corresponding to $\chi_{\pm 2} = \mp k\varphi$, and $\chi_2 = 0, \chi_{-2}=\varphi$, where $D_\pm$ is defined in Eq.~\eqref{eq: D_pm}. Eq.~\eqref{eq: FM-BN-sqv-sv} yields phase vortices in the $\FMtwo$ limits ($D_+ = 0$ or $D_- = 0$), connecting to a spin vortex in the BN ($D_\pm=1$), identified by $\gamma=\varphi/2$ only. In Eq.~\eqref{eq: FM-BN-sqv-hqv}, fractional $(1/2,1/4)$ BN vortex connects to a vortex-free state or phase-vortex in the $\FMtwo$. 
The vortex connections identified through winding of $\chi_{\pm 2}$ are shown in Table~\ref{tab: FM-BN-vortices}.
\begin{table}
    \centering
    \begin{tabular}{cccc}
        \toprule
        \multicolumn{4}{c}{FM-BN: Vortices from spinor-component phase winding} \\
        \midrule
        FM$_{2}$ limit & BN limit &  $\chi_2/\varphi$ & $\chi_{-2}/\varphi$  \\
        \midrule
         Phase vortex & Phase vortex & $k$ & $k$\\ 
         Phase vortex & Spin vortex  & $-k$ & $k$\\
         Phase vortex & Half-quantum vortex & 1 & 0 \\
         Vortex-free & Half-quantum vortex & 0 & 1 \\
        \bottomrule
    \end{tabular}
    \caption{Singular vortex connections across a FM$_2$-BN interface, characterized by the phase windings $\chi_m$ in Eq.~\eqref{eq: FM-BN-interpolating-spinor}.}
    \label{tab: FM-BN-vortices}
\end{table}

\textit{Nonsingular textures and monopoles}---Applying Eq.~\eqref{eq: euler_angles} to Eq.~\eqref{eq: FM-BN-interpolating-spinor} yields
\begin{equation}
    \begin{gathered}
        \zeta^\mathrm{FM_2-BN} =  \frac{1}{\sqrt{2}}\left(D_+\zeta^{\text{FM}_2^+} +  D_-\zeta^{\text{FM}_2^-}\right),
    \end{gathered}
    \label{eq: FM-BN-orbit}
\end{equation}
 where $\zeta^{\text{FM}_2^+}$ is defined in Eq.~\eqref{eq: FM2_orbit}, and $\zeta^{\text{FM}_{2}^-}$ is similarly obtained by applying Eq.~\eqref{eq: euler_angles} to $\zeta=(0,0,0,0,1)^\text{T}$.
Following the procedure outlined in Sec.~\ref{sec: C-FM}, 
we can construct spinors representing FM nonsingular vortices connecting to singular BN vortices, and Dirac monopoles that form the termination point of vortices at the FM-BN  interface. 
By choosing the condensate phase and Euler angles as in Eq.~\eqref{eq: FM2-coreless-condition}, we obtain states of the form~\eqref{eq: FM-BN-orbit} with the $\FMtwo$ limits
\begin{align}
    &\zeta^{\text{FM}_2^+} = 
    \begin{pNiceMatrix}
        e^{-4i\varphi} \cos^4 \frac{\beta}{2} \\
        2e^{-3i\varphi} \cos^3 \frac{\beta}{2} \sin \frac{\beta}{2} \\
        \sqrt{6}e^{-2i\varphi}\cos^2\frac{\beta}{2}\sin^2\frac{\beta}{2}\\
        2e^{-i\varphi} \sin^3 \frac{\beta}{2} \cos \frac{\beta}{2} \\
        \sin^4\frac{\beta}{2}
    \end{pNiceMatrix}, \label{eq: FM-BN-dirac2a}\\[1ex]
    &\zeta^{\text{FM}_{2}^-} = 
    \begin{pNiceMatrix}
        e^{-4i\varphi} \sin^4 \frac{\beta}{2} \\
        -2e^{-3i\varphi} \sin^3 \frac{\beta}{2} \cos \frac{\beta}{2} \\
        \sqrt{6}e^{-2i\varphi}\cos^2\frac{\beta}{2}\sin^2\frac{\beta}{2}\\
        -2e^{-i\varphi} \cos^3 \frac{\beta}{2} \sin \frac{\beta}{2} \\
        \cos^4\frac{\beta}{2}
    \end{pNiceMatrix}. \label{eq: FM-BN-dirac-2b}
\end{align}
In the BN limit, $D_\pm =1$ in Eq.~\eqref{eq: FM-BN-orbit}, the spinor yields phase-, spin- and HQVs, depending on the choice of winding $\tau=2, 0, 1/2$ respectively.
For example, when $\beta=\theta$, $\zeta^{\text{FM}_2^+}$ 
and $\zeta^{\text{FM}_2^-}$ represent a $\FMtwo$ monopole with an associated line singularity along $z>0$ and $z<0$, respectively. The interpolating spinor then connects the monopole to a singular vortex in the BN limit.
We summarize the interpolating states obtained in this manner in Table~\ref{tab: FM-BN-other}.
\begin{table}
    \centering
    \begin{tabular}{cccccc}
        \toprule
        \multicolumn{6}{c}{FM-BN: Vortices, textures and monopoles from Euler angles} \\
        \midrule
        FM$_{2}$ limit & BN limit & $\tau/\varphi$ & $\alpha/\varphi$ & $\gamma/\varphi$ & $\beta$ \\
        \midrule
        Nonsingular vortex & Phase vortex & $2$ & 1 & $0$ & $\beta(\rho)$ \\ 
        Nonsingular spin vortex & Spin vortex  & $0$ & 1 & $\pm 1$ & $\beta(\rho)$ \\ 
        Nonsingular vortex & Half-quantum vortex & 1/2 & 1 & -3/4 & $\beta(\rho)$ \\ 
        Dirac monopole & Phase vortex & 2 & 1  & 0 & $\theta$\\
        Dirac monopole & Spin vortex & 0 & 1  & $\pm 1$ & $\theta$\\
        Dirac monopole & Half-quantum vortex & 1/2 & 1  & -3/4 & $\theta$\\
        \bottomrule
    \end{tabular}
    \caption{Singular and nonsingular vortices and monopoles connecting across a FM$_2$-BN interface, constructed through spatial dependent $\tau$, $\alpha$,
    and $\gamma$ (given as multiples of the azimuthal angle $\varphi$), and $\beta$ (given as a multiple of the polar angle $\theta$ or, for nonsingular vortices, as a monotonically increasing function of the transverse radius $\rho$).}
    \label{tab: FM-BN-other}
\end{table}

\textit{Non-Abelian vortex pair at the interface}---The BN and C magnetic phases support non-Abelian vortices. The BN $(1/2,1/4)$ and $(1/2,1/2)$ vortices belong to different conjugacy classes and do not commute~\cite{Borgh2016b}. A spinor representing parallel $(1/2,1/4)$ and $(1/2,1/2)$ vortices terminating on the $\FMtwo$-BN interface can be constructed starting from Eq.~\eqref{eq: FM-BN-sqv-hqv}, where the azimuthal angle $\varphi=\varphi_1$ determines the $(1/2,1/4)$ vortex line.
When the $(1/2,1/2)$ vortex is added, the transformations of the BN order parameter that correspond to the spin-rotation charges of the vortex lines combine nontrivially. For a pure BN condensate, this vortex combination was constructed in Ref.~\cite{Borgh2016b} (with technical details in its Supplemental Material). Applying the same construction to Eq.~\eqref{eq: FM-BN-sqv-hqv} results in an interface spinor on the form~\eqref{eq: FM-BN-orbit}, with the $\FMtwo^\pm$ limits given by 
\begin{align}
    &\zeta^\text{FM$_2^+$} = e^{i\frac{\varphi_2}{2}}
    \begin{pNiceMatrix}
        \cos^4\frac{\varphi_2}{4} \\
        \frac12 e^{i\frac{\pi}{4}}e^{i\frac{\varphi_1}{4}}\sin\frac{\varphi_2}{2}\left(1+\cos\frac{\varphi_2}{2}\right) \\
        i\sqrt{\frac38}e^{i\frac{\varphi_1}{2}}\sin^2\frac{\varphi_2}{2} \\
        \frac12 e^{i\frac{3\pi}{4}}e^{i\frac{3\varphi_1}{4}}\sin\frac{\varphi_2}{2}\left(1-\cos\frac{\varphi_2}{2}\right) \\
        -e^{i\varphi_1}\sin^4\frac{\varphi_2}{4}
    \end{pNiceMatrix}, \label{eq: FM-BN-pair2}\\[1ex]
    &\zeta^\text{FM$_2^-$} = e^{i\frac{\varphi_2}{2}}
    \begin{pNiceMatrix}
        -\sin^4\frac{\varphi_2}{4} \\
        \frac12 e^{i\frac{\pi}{4}}e^{i\frac{\varphi_1}{4}}\sin\frac{\varphi_2}{2}\left(1-\cos\frac{\varphi_2}{2}\right) \\
        -i\sqrt{\frac38}e^{i\frac{\varphi_1}{2}}\sin^2\frac{\varphi_2}{2} \\
        \frac12 e^{i\frac{3\pi}{4}}e^{i\frac{3\varphi_1}{4}}\sin\frac{\varphi_2}{2}\left(1+\cos\frac{\varphi_2}{2}\right) \\
        e^{i\varphi_1}\cos^4\frac{\varphi_2}{4}
    \end{pNiceMatrix}, \label{eq: FM-BN-pair-2}
\end{align}
where $\varphi_2$ is the azimuthal angle determining the $(1/2,1/2)$ vortex line. Note, however, that this construction yields a well-defined defect state only in the BN limit. The addition of the $(1/2,1/2)$ vortex results in a discontinuous semiplane at $\varphi_2 = 0$, where the $\FMtwo$ order parameter jumps from $F_z=2$ to $F_z= -2$. Despite this, Eqs.~\eqref{eq: FM-BN-pair2} and~\eqref{eq: FM-BN-pair-2} approximate the desired defect combination and the discontinuity corresponds only to a rapidly relaxing excitation, as illustrated by numerical simulation in Sec.~\ref{sec: FM-BN_numeric}.

\section{Numerical simulations of core structures \label{sec: numerics}}
We study the dynamics and energy relaxation for illustrative examples of interface-crossing defect states, as constructed in  Sec.~\ref{sec: theory}, by numerically propagating the coupled Gross-Pitaevskii equations derived from the Hamiltonian density~\eqref{eq: energy-functional} using a split-step method~\cite{Javanainen2006}.
Simulations are performed on a $128^3$-point grid, and we choose $^{87}$Rb interaction parameters~\cite{Klausen2001} with $Nc_0 = 1.32\times10^4\,\hbar\omega\ell^3$ (corresponding to $N = 2\times10^5$ atoms in an $\omega = 2\pi\times130~\mathrm{Hz}$ trap).
The UN-BN and C-BN topological interfaces are stabilized through spatial variations of level shifts, while the C-FM and the FM$_2$-BN interfaces by spatially varying the interaction strength $c_1$ along the $z$-direction.
A weak phenomenological damping, $t \rightarrow (1-i\nu)\,t$ with $\nu\approx 10^{-2}$, accounts for dissipation in time-evolution simulations. Energy relaxation is determined using imaginary-time propagation.

\subsection{Phase vortex crossing the UN-BN interface \label{sec: un-bn_numeric}} 

As our first example of evolution of interface-crossing defects, we consider a singly quantized phase vortex that perforates the UN-BN topological interface. The continuously interpolating initial state is given by Eq.~\eqref{eq: UN-BN-SQV-SQV}, with $k=1$ and $\eta = \eta(z)$ reaching $D_- = 0$ for $z>\xi_a$, and $D_+ = 0$ for $z<-\xi_a$. The interface is stabilized by choosing the quadratic level shift such that $q = \pm |q_\text{max}|$, with $q_\text{max}=0.1\,\hbar \omega$, away from the interface at $z=0$, around which $q$ interpolates smoothly over the region smaller than the singlet healing length $\xi_a$.

\begin{figure}
    \centering
    \includegraphics[width=1.02\columnwidth]{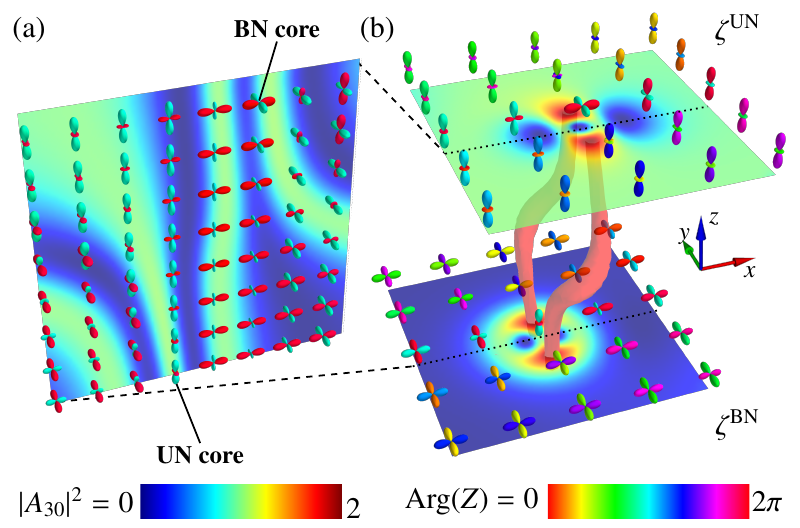}
    \caption{Complex-time evolution of singly quantized phase vortices connecting across the UN-BN interface. (a) Longitudinal cut of $|A_{30}|^2$ and spherical harmonics in the $z,x$-plane, showing the vortex core structure of two vortices terminating at the interface that originate from an initial line at $\rho=0$.
    The locations of the two cores are indicated.
    (b)  Transverse cuts of $|A_{30}|^2$ and spherical harmonics at both sides of the interface, where the filled core structures lead to a mixture of phases including UN ($|A_{30}|^2=1$), C ($|A_{30}|^2=2$) and BN ($|A_{30}|^2=0$). The C regions penetrate the interface as illustrated by red isosurfaces at $|A_{30}|^2 = 2$.
}
    \label{fig: UN-BN-SQV-SQV}
\end{figure}

Figure~\ref{fig: UN-BN-SQV-SQV}(a) shows the resulting vortex state after time propagation to $t = 100\,\omega^{-1}$. An azimuthal instability at the interface results in a local separation of the vortex lines, which terminate at displaced points on the interface within an extended core region whose size is determined by $\xi_a$ and where the UN, BN and C phases mix [Fig.~\ref{fig: UN-BN-SQV-SQV}(b)]. On the line singularity itself, the vortex on the BN side of the interface fills with the UN phase, while on the UN side, the singularity is accommodated by excitation to the BN superfluid.

\begin{figure}
    \centering
    \vspace*{-.2cm}
    \hspace*{-.13cm}\includegraphics[width=1.03\columnwidth]{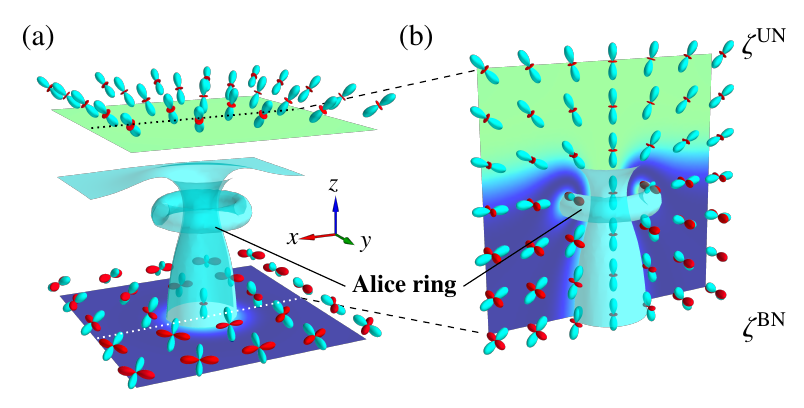}\vspace*{-.5cm}
    \hspace*{-.13cm}\includegraphics[width=1.03\columnwidth]{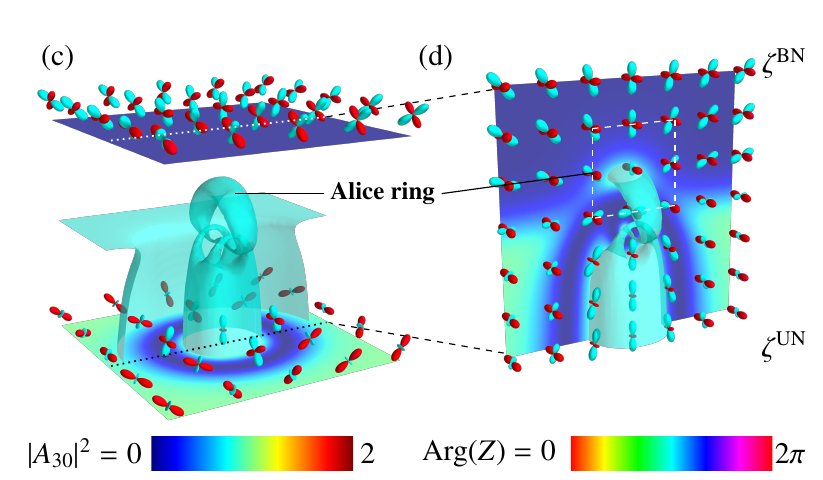}
    \caption{ Energy relaxation of spin vortices terminating as monopoles on the UN-BN interface. (a, b) A BN spin vortex terminating as a UN monopole. Cuts of $|A_{30}|^2$ and spherical harmonics show the UN monopole for $z>0$ attached to a vortex for $z<0$ in the BN phase. (c, d) Analogous representation for an initial singular UN spin vortex in $z<0$ terminating as a BN monopole in $z>0$. 
    Energy relaxation leads to Alice rings at the interface, illustrated by isosurfaces at $|A_{30}|^2=1/2$ and longitudinal cuts of  $|A_{30}|^2$. In both (b) and (d) the spherical harmonics show the nematic hedgehog and the continuous winding of the order parameter around the spin-HQV rings.}
   \label{fig: UN-BN-monopole}
\end{figure}

\subsection{Singular vortices terminating as monopoles on the UN-BN interface \label{sec: un-bn_numeric2}}

In Sec.~\ref{sec: UN-BN}, we constructed spinors representing singular vortices terminating as monopoles on the UN-BN interface. Here we numerically simulate their energy relaxation at zero Zeeman shifts to reveal the dissipative deformation of the defect core. In the analogous spin-1 case, an isolated nematic point-defect core is energetically unstable against deformation into a singular HQV-ring~\cite{Ruostekoski2003,Blinova2023}, called an Alice ring in analogy with similar objects in high-energy physics. 

We first consider a $(0,1)$ spin vortex in the BN phase along the negative $z$-axis terminating as a UN monopole on the interface at the origin. The initial state is given by Eqs.~\eqref{eq: UN-ncv}--\eqref{eq: UN-BN-orbit} with $\beta=\theta$ and the parameter $D_\pm (z)$ chosen such that $D_- = 0$ for $z>\xi_a$, and $D_+ = 0$ for $z<-\xi_a$. The spin vortex rapidly develops a UN superfluid core, reducing its energy. However, the point defect cannot do the same without deforming to a vortex ring due to the ``hairy ball'' theorem~\cite{Ruostekoski2003}. This happens within one trap period of imaginary-time evolution, shown in Fig.~\ref{fig: UN-BN-monopole}(a) and (b). The ring appears parallel to the interface and encircles the BN spin vortex around its termination point. It is readily identified as a $(0,1/2)$ BN spin-HQV (a ``spin-Alice ring''~\cite{Borgh2016b}). The BEC away from the defect core retains the topological asymptotics of the point defect.

As outlined in Sec.~\ref{sec: UN-BN}, we may also construct a spinor where the roles of the BN and UN regions are switched, such that a UN spin vortex terminates as a BN monopole on the interface, forming a different boojum. The initial state has the form~\eqref{eq: UN-BN-orbit}, with the single-phase limits Eqs.~\eqref{eq: UN-monopole2} and \eqref{eq: BN-monopole2}.  The numerics in this case leads to a more complex geometry of an Alice-ring threaded by a pair of spin-HQVs, shown in Fig.~\ref{fig: UN-BN-monopole}(c). Energy relaxation causes the core of the UN spin vortex (here appearing in $z<0$), initially connected to the monopole, to develop a composite-defect structure with a BN outer core, shown in Fig.~\ref{fig: UN-BN-monopole}(d).

\subsection{Spin vortices crossing the C-BN interface \label{sec: c-bn_numeric}} 

We can also form interfaces using interpolating solutions for energetically excited states that decay due to energy relaxation when they exhibit sufficient dynamical stability. A topological interface between C and BN phases for the parameters of $^{87}$Rb clearly represents such a case. Similarly to recent experimental observations~\cite{Xiao2022}, the C-BN interface relaxes into a uniform BN state over timescales sufficiently slower than the relevant vortex core dynamics, allowing for characterizations of vortex stability properties. We consider the C-BN interpolating spinor in Eq.~\eqref{eq: C-BN-interpolating-spinor} when $c_2 < 0$. 
In the numerics we take $q(z)$ such that $q=-0.1\,\hbar\omega$ for $z<-\xi_a$, and $q=0$ for $z>\xi_a$, continuously interpolating between the two values.

\begin{figure}
    \centering
    \hspace*{-.15cm}\includegraphics[width=1.04\columnwidth]{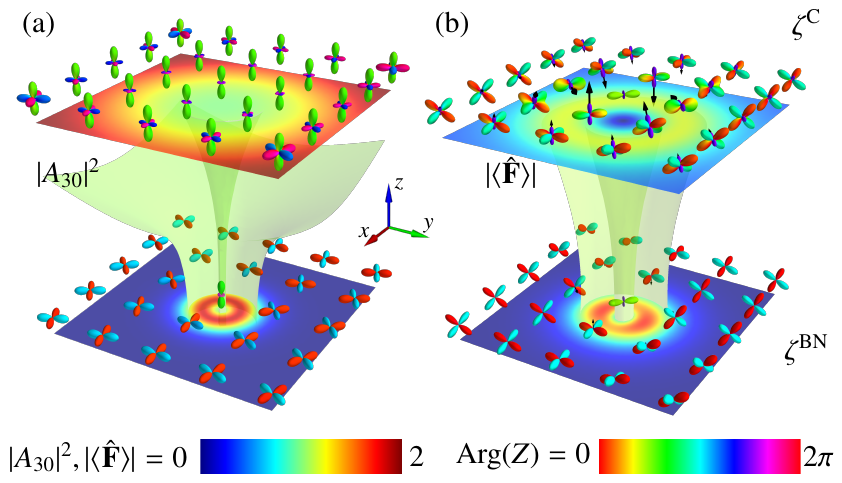}
    \caption{Complex-time evolution of spin-vortices penetrating a C-BN interface. (a) Spin-HQV crossing the interface. Transverse cuts of  $|A_{30}|^2$ on both sides of the interface together with the spherical-harmonic representation of the order parameter show a UN core, highlighted by the isosurface $|A_{30}|^2=1$. (b) Singly quantized spin vortices connecting across the interface developing a composite-defect structure with FM outer core, shown by $|\langle \hat{\mathbf{F}}\rangle|$ on transverse cuts on both C and BN sides and isosurface $|\langle \hat{\mathbf{F}}\rangle|=1$. The spherical-harmonics representation shows the spin rotation of the order parameter, defined in Eq.~(\ref{eq: C-BN_sv_rot}), about the  $\hat{\mathbf{e}}_{(1,0,0)}$ axis.}
    \label{fig: C-BN-SV-SV}
\end{figure}

The BN and C order parameters allow for different spin vortices to connect across the topological interface, as shown in Sec.~\ref{sec: C-BN} and summarised in Tabs.~\ref{tab: C-BN-vortices} and~\ref{tab: C-BN-other}.
We examine a spin-HQV of Eq.~\eqref{eq: C-BN-SV-SV} that penetrates the interface, represented by the spatially dependent  $D_\pm(z)$, such that $D_+ = 0$ for $z<-\xi_a$, and $D_\pm = 1$ for $z>\xi_a$. The dynamics is shown in Fig.~\ref{fig: C-BN-SV-SV}(a), where the spherical-harmonics representation after two trap periods highlights the characteristic spin winding around  $z$ on both sides of the interface. 

For singly quantized spin vortices given by Eq.~(\ref{eq: C-BN-SV2-SV2}), dissipation rapidly develops a composite vortex core, characterized by a FM cylindrical structure across the interface and illustrated by an isosurface at $|\langle\hat{\mathbf{F}}\rangle|=1$ in Fig.~\ref{fig: C-BN-SV-SV}(b). This configuration is unstable and eventually decays into vortices with $\FMtwo$ cores. Interestingly, in a pure BN phase, the same vortex instead develops a $\FMone$ core, highlighting how the presence of the interface can strongly influence core dissipation.

\subsection{Fractional and nonsingular vortices at the C-FM interface \label{sec: c_fm_numeric}}

The interface between C and $\FMtwo$ phases allows the smooth connection of fractionally quantized vortices with singular and nonsingular vortices as well as vortex-free states on the $\FMtwo$ side (Sec.~\ref{sec: C-FM} and Tabs.~\ref{tab: C-FM-vortices} and~\ref{tab: C-FM-other}). 
We simulate the dynamics of two example defect states: a C $(1/3,-1/3)$ vortex connecting to a singly quantized $\FMtwo$ vortex, and a C $(2/3,1/3)$ vortex connecting to a vortex-free $\FMtwo$ state. 

The initial spinor wave functions are given by
Eq.~\eqref{eq: C-FM-third} in the former example, and \eqref{eq: C-FM-2third} in the latter, with spatially dependent $D_{\pm m}(z)$ interpolating between $D_2=1, D_{-1}=\sqrt2$ for $z<-\xi_F$, and $D_{-1}=0$ for $z>\xi_F$. For our numerical simulation, we stabilize the interface by introducing a sign-changing spin-dependent interaction $c_1$, such that $c_1 > 0$ on the C side, assumed for $z < -\xi_F$, and $c_1 <0$ on the FM side for $z > \xi_F$.
\begin{figure}
    \centering
    \vspace*{-.14cm}
    \hspace*{-.15cm}\includegraphics[width =1.04\columnwidth]{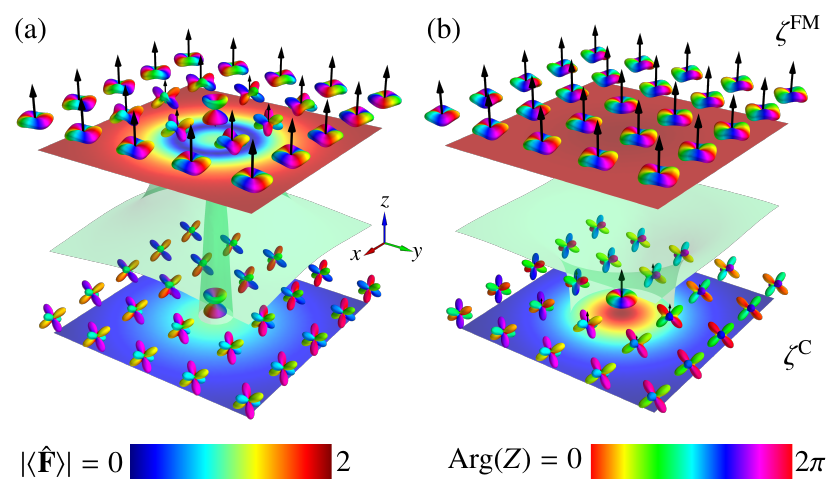}
    \caption{Defect-core structures after complex-time evolution of initial configurations containing a fractional vortex on the C side of a C-$\FMtwo$ interface. Transverse cuts of $|\langle \hat{\mathbf{F}}\rangle|$ on both sides of the interface are shown together with the spherical-harmonic representation of the order parameter.
    (a) $(1/3,-1/3)$ vortex in the C phase connecting to a singly quantized $\FMtwo$ phase vortex, forming a $\FMone$ core that penetrates the interface, as highlighted by the isosurface $|\langle \hat{\mathbf{F}}\rangle|=1$.
    (b) $(2/3,1/3)$ vortex in the C phase terminating at the interface. The $\FMtwo$ phase penetrates the interface, forming the core of the C vortex.
    }
    \label{fig: C-FM=2_third-SQV}
\end{figure}

Figure~\ref{fig: C-FM=2_third-SQV} shows the core structures emerging after approximately ten trap periods of complex-time evolution. The spin magnitude as the overlaid spherical-harmonics representation of the order parameter show that the core of the $(1/3,-1/3)$ vortex [Fig.~\ref{fig: C-FM=2_third-SQV}(a)] fills with the \FMone~phase, extending also throughout the FM$_2$ region to form the outer core of a composite defect. By contrast, the $(2/3,1/3)$ vortex forms a FM$_2$ core that smoothly connects to a vortex-free state on the $\FMtwo$ side of the interface [Fig.~\ref{fig: C-FM=2_third-SQV}(b)]. 
\begin{figure}
    \centering
    \vspace*{-.1cm}
    \hspace*{-.12cm}\includegraphics[width=1.04\columnwidth]{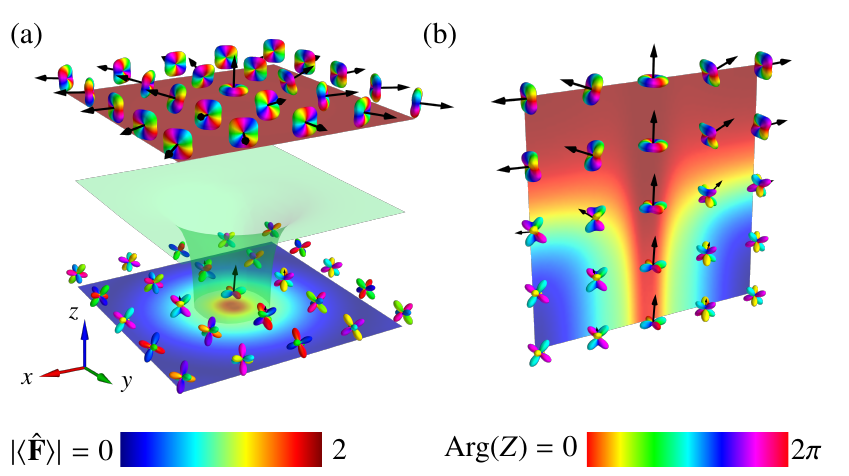}
    \caption{Energy relaxation of a doubly quantized phase vortex in the C phase connecting to a $\FMtwo$ nonsingular vortex. (a) Transverse cuts of $|\langle \hat{\mathbf{F}}\rangle|$ on both sides of the interface together with the spherical-harmonic representation of the order parameter. The C vortex develops a $\FMtwo$ core that forms an interface-penetrating continuation of the nonsingular $\FMtwo$ vortex, as shown by the isosurface at $|\langle \hat{\mathbf{F}} \rangle|=1$. (b) Longitudinal cut of $|\langle \hat{\mathbf{F}}\rangle|$ and spherical-harmonic representation, showing the fountain-like spin texture on the $\FMtwo$ side.}
    \label{fig: C-FM=2_coreless}
\end{figure}

As an additional illustrative example, we also consider a singular, doubly quantized C phase vortex connecting to a nonsingular vortex in the $\FMtwo$ limit. This state was constructed in Eq.~(\ref{eq: C-FM-orbit}), choosing $\beta = \beta(\rho)$ to be a function of the radial distance such that $\beta(0) = 0$ on the vortex line and $\beta(\rho)=\pi/2$ far away, forming a Mermin-Ho texture~\cite{Mermin1976} on the FM side.
Figure~\ref{fig: C-FM=2_coreless} shows the rapidly forming defect core (half a trap period of energy relaxation). The $\FMtwo$ phase quickly intrudes on the C side to fill the core of the singular C vortex. The spherical-harmonics representation shows the $4\pi$-winding of the condensate phase about the C vortex, coupled to a spin rotation.
The doubly quantized vortex line is, however, not stable and quickly decays via an azimuthal instability under further energy relaxation.

\subsection{Non-Abelian vortex pair terminating on a FM-BN interface \label{sec: FM-BN_numeric}}

Intriguingly, the order-parameters with discrete polytope symmetries in spin-2 BECs exhibit non-Abelian vortices~\cite{Semenoff2007,Kobayashi2009,Borgh2016b}. As our final example, we examine a pair of non-commuting BN $(1/2,1/4)$ and $(1/2,1/2)$ vortices, a distance 2$\ell$ apart, terminating on a BN-$\FMtwo$ interface. The initial spinor was constructed in Eqs.~\eqref{eq: FM-BN-pair2} and~\eqref{eq: FM-BN-pair-2} to approximate Eq.~\eqref{eq: FM-BN-orbit}, with the discontinuity in the $\FMtwo$ limit rapidly disappearing after short energy relaxation. Similarly to the C-FM case  (Sec.~\ref{sec: c_fm_numeric}), the interface is stabilized in the numerical simulation by spatially varying the $c_1$ interaction strength along the $z$ axis, such that $c_1>0$ on the BN side at $z < -\xi_F$ and $c_1>0$ on the FM side at $z > \xi_F$, at fixed $c_2$.
\begin{figure}
    \centering
    \hspace*{-.15cm}\includegraphics[width=1.03\columnwidth]{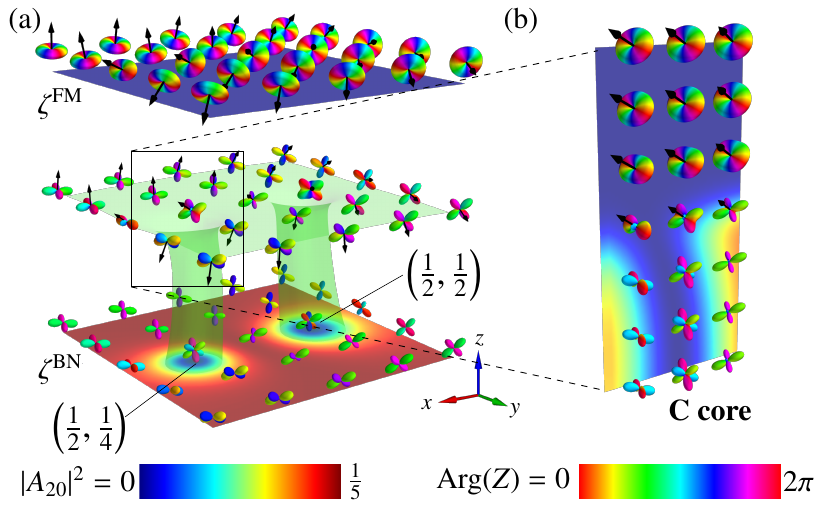}
    \caption{Energy relaxation of a pair of non-commuting BN $(1/2,1/4)$ and $(1/2,1/2)$ HQVs terminating on the C-$\FMtwo$ interface. (a) Transverse cuts of $|A_{20}|^2$ at the interface and both sides, together with the spherical harmonics representation of the order parameter. Both initial HQVs develop C cores, as shown by the isosurface at $|A_{20}|^2=1/10$ and order-parameter representation. (b) Longitudinal cut of $|A_{20}|^2$ together with spherical harmonics, showing the continuous $\FMtwo$-C transition occurring at the termination point of the $(1/2,1/4)$ vortex (similar transition in the $(1/2,1/2)$ vortex core not shown).}
    \label{fig: FM-BN_pair}
\end{figure}

Figure~\ref{fig: FM-BN_pair} shows the defect core relaxation after one trap period of imaginary-time propagation. Both non-commuting BN HQVs terminate to a nonsingular $\FMtwo$ spin texture. For the chosen interaction strengths $c_{1,2}$, the HQV cores surprisingly rapidly relax to the C phase (consistent with the core structure of a single BN $(1/2,1/4)$ vortex~\cite{Borgh2016b}), resulting in a spontaneous C-$\FMtwo$ interface between the vortex cores and the $\FMtwo$ phase, as illustrated in Fig.~\ref{fig: FM-BN_pair}(b).

\section{Concluding remarks and experimental prospects \label{sec: conclusions}}

To summarise, we have demonstrated the potential of spin-2 BECs as rich laboratories for topological-interface physics by systematically constructing spinor wave functions for defects and textures connecting across the interfaces between UN, BN, C and $\FMtwo$ phases. These wave functions are derived from continuously interpolating steady-state solutions to the spin-2 Gross-Pitaevskii equations, and include connecting singular vortices carrying mass and/or spin circulation as well as vortices that terminate at the interface to a vortex-free state or as monopoles. For a selection of examples, we have simulated their time evolution and energy relaxation, demonstrating the appearance of non-trivial core structures that include the formation of composite defects as well as the deformation of point defects into Alice-ring structures at the interface. The discrete, polytope order-parameter symmetries of the spin-2 BN and C phases also open the possibility of using spin-2 BECs to explore non-Abelian interface physics both theoretically and experimentally. We have demonstrated an example by numerically simulating a pair of non-commuting HQVs that terminates on the $\FMtwo$-BN interface.

Defects and textures with a non-zero topological charge are topologically stable, meaning that they cannot dissolve through purely local changes to the order parameter. In a trapped BEC, however, they will eventually escape through the edge of the cloud due to the density gradient, unless a stabilising mechanism, such as rotation or a shallow density minimum that inverts the density gradient~\cite{Savage2003a}, is applied. 
However, even without such measures, the time scale for a vortex leaving can be slow even compared with experimental timescales. Defects can also decay into other defects while preserving the total topological charge, such as the deformation of a point defect into an Alice ring~\cite{Ruostekoski2003} or splitting of a singly quantised vortex into HQVs~\cite{Lovegrove2012}. With the exception of monopole deformation, also these decays are generally slow for the states in Tabs.~\ref{tab: UN-BN-vortices}--\ref{tab: C-FM-vortices}, and \ref{tab: FM-BN-vortices}, while those in Tabs.~\ref{tab: C-FM-other} and \ref{tab: FM-BN-other} decay more rapidly. In addition, e.g., Fig.~\ref{fig: UN-BN-SQV-SQV}, connecting UN, BN or C phase vortices may separate at the interface but remain intact as terminating vortices.

The experimental creation of magnetic-phase interfaces in BECs has thus far been restricted to vortex cores~\cite{Weiss2019,Xiao2021,Xiao2022}. Creating extended interfaces and populating them with the desired topological excitations poses significant additional technical challenges. Nevertheless, we can sketch possible creation methods by combining relatively straightforward extensions of existing technologies, which have already been shown to adjust ground state magnetic phases and controllably generate vortex excitations.

Topological interfaces can be created by engineering spatial variation in interaction strengths, as discussed earlier. This could in principle be achieved by controlling microwave Fesh\-bach~\cite{Papoular2010} resonance conditions with optical ac Stark shifts, or by using optical Feshbach resonances~\cite{Theis2004}. However, here we have focused on interfaces that are obtained through spatially varying Zeeman shifts $p,q$ that could be generated by ac Stark shifts of lasers or microwaves~\cite{Gerbier2006}.

For example, the value of $q$ in the system Hamiltonian~\eqref{eq: energy-functional} can be modified by applying a linearly polarized microwave field detuned from the ground state hyperfine transition~\cite{Gerbier2006}, as realized for both $^{87}$Rb~\cite{Leslie2009a} and $^{23}$Na~\cite{Seo2015} in the spin-1 manifold. 
The spin-2 case is conceptually straightforward where the presence of the additional levels provides additional useful degrees of freedom that are conventionally expressed in terms of dynamical scalar, vector, and tensor polarizabilities~\cite{Manakov86}, extending the description beyond the single parameter $q$. 
Another intriguing possibility is to mimic microwave-induced level shifts with stimulated Raman dressing between the two ground-state hyperfine levels. 
Such transitions are highly state-selective and, with proper choice of polarization, intensity, and frequency, can differentially dress selected Zeeman sublevels in a spatially dependent fashion.

For example, UN-BN interface could be achieved by illuminating half of the condensate with light that depresses the energy of the $m=0$ state with respect to the others, and the other half with light that depresses the energy of the $m=\pm 2$ states. It remains to populate these two regions with atoms in the appropriate magnetic phases, which could be achieved by magnetic phase exchange~\cite{Xiao2021} in a spatially selective fashion. Ideally, the resulting condensate would change smoothly from its ground state UN phase to the ground state BN phase at the boundary between the beams.

Vortices can currently be introduced through a variety of methods, including direct phase imprinting through optical~\cite{Andersen2006,Wright2009} or magnetic~\cite{Leanhardt2002} means, as well as by stirring~\cite{Madison2000,Hodby2001}. These techniques have been employed in recent spinor experiments to create some of the defects and textures that form the basis for our considerations, e.g., HQVs~\cite{Seo2015}, non-singular vortices~\cite{Choi2012,Choi2012b}, singular vortices through controlling the instabilities of nonsingular vortices~\cite{Weiss2019}, and monopoles~\cite{Ray2014,Ray2015}. One possible approach is to use a combination of magnetic phase imprinting~\cite{Leanhardt2002,Weiss2019} followed by magnetic phase conversion~\cite{Xiao2021,Xiao2022}, using optical fields instead of microwaves to achieve the requisite spatial selectivity. For example, a phase vortex crossing a UN-BN interface could be generated from a phase vortex in a spin-1 polar condensate ($F=1$, $m=0$) followed by optical pulse sequences to convert the spin-1 polar phase into spin-2 BN and UN phases~\cite{Xiao2022} in the two regions. Similarly, a vortex-free to phase vortex interface could be created by using magnetic phase imprinting techniques on a spin-1 mixed polar and FM condensate ($F=1$, $m=0$ and $m=1$), where the phase imprinting yields two vortices in the $m=-1$ spinor component and no vortices in the $m=0$ spinor component. Initially displaced from one another spatially by a magnetic field gradient, one of these components could subsequently be optically converted to the UN phase ($F=2$, $m=0$) and the other to the BN phase ($F=2$, $m=\pm 2$), leading to the desired topological configuration in the milieu of ground states described previously. Approaches such as these can be envisioned for creating spin vortices and HQVs.

Data used in this publication available at~\cite{dataset}.

\section*{Acknowledgments}
 G.B. and M.O.B. as well as J.R. acknowledge financial support from the Engineering and Physical Sciences Research Council, Grants No.\ EP/V03832X/1 and No.\ EP/S002952/1, respectively. D.S.H. acknowledges support from the National Science Foundation Grant No.\ PHY-2207631. The numerical results presented in this paper were carried out on the High Performance Computing Cluster supported by the Research and Specialist Computing Support service at the University of East Anglia. 

\bibliography{references}

\begin{thebibliography}{122}%
\makeatletter
\providecommand \@ifxundefined [1]{%
 \@ifx{#1\undefined}
}%
\providecommand \@ifnum [1]{%
 \ifnum #1\expandafter \@firstoftwo
 \else \expandafter \@secondoftwo
 \fi
}%
\providecommand \@ifx [1]{%
 \ifx #1\expandafter \@firstoftwo
 \else \expandafter \@secondoftwo
 \fi
}%
\providecommand \natexlab [1]{#1}%
\providecommand \enquote  [1]{``#1''}%
\providecommand \bibnamefont  [1]{#1}%
\providecommand \bibfnamefont [1]{#1}%
\providecommand \citenamefont [1]{#1}%
\providecommand \href@noop [0]{\@secondoftwo}%
\providecommand \href [0]{\begingroup \@sanitize@url \@href}%
\providecommand \@href[1]{\@@startlink{#1}\@@href}%
\providecommand \@@href[1]{\endgroup#1\@@endlink}%
\providecommand \@sanitize@url [0]{\catcode `\\12\catcode `\$12\catcode
  `\&12\catcode `\#12\catcode `\^12\catcode `\_12\catcode `\%12\relax}%
\providecommand \@@startlink[1]{}%
\providecommand \@@endlink[0]{}%
\providecommand \url  [0]{\begingroup\@sanitize@url \@url }%
\providecommand \@url [1]{\endgroup\@href {#1}{\urlprefix }}%
\providecommand \urlprefix  [0]{URL }%
\providecommand \Eprint [0]{\href }%
\providecommand \doibase [0]{https://doi.org/}%
\providecommand \selectlanguage [0]{\@gobble}%
\providecommand \bibinfo  [0]{\@secondoftwo}%
\providecommand \bibfield  [0]{\@secondoftwo}%
\providecommand \translation [1]{[#1]}%
\providecommand \BibitemOpen [0]{}%
\providecommand \bibitemStop [0]{}%
\providecommand \bibitemNoStop [0]{.\EOS\space}%
\providecommand \EOS [0]{\spacefactor3000\relax}%
\providecommand \BibitemShut  [1]{\csname bibitem#1\endcsname}%
\let\auto@bib@innerbib\@empty
\bibitem [{\citenamefont {Zel'dovich}\ \emph {et~al.}(1974)\citenamefont
  {Zel'dovich}, \citenamefont {Kobzarev},\ and\ \citenamefont
  {Okun'}}]{zeldovich1975}%
  \BibitemOpen
  \bibfield  {author} {\bibinfo {author} {\bibfnamefont {Y.~B.}\ \bibnamefont
  {Zel'dovich}}, \bibinfo {author} {\bibfnamefont {I.~Y.}\ \bibnamefont
  {Kobzarev}},\ and\ \bibinfo {author} {\bibfnamefont {L.~B.}\ \bibnamefont
  {Okun'}},\ }\bibfield  {title} {\bibinfo {title} {Cosmological consequences
  of a spontaneous breakdown of a discrete symmetry},\ }\href
  {http://jetp.ras.ru/cgi-bin/e/index/r/67/1/p3?a=list} {\bibfield  {journal}
  {\bibinfo  {journal} {Zh. Eksp. Teor. Fiz.}\ }\textbf {\bibinfo {volume}
  {67}},\ \bibinfo {pages} {3} (\bibinfo {year} {1974})},\ \bibinfo {note}
  {[Sov. Phys. JETP \textbf{40}, 1 (1975)]}\BibitemShut {NoStop}%
\bibitem [{\citenamefont {Kibble}(1976)}]{Kibble1976}%
  \BibitemOpen
  \bibfield  {author} {\bibinfo {author} {\bibfnamefont {T.~W.}\ \bibnamefont
  {Kibble}},\ }\bibfield  {title} {\bibinfo {title} {Topology of cosmic domains
  and strings},\ }\href {https://doi.org/10.1088/0305-4470/9/8/029} {\bibfield
  {journal} {\bibinfo  {journal} {J. Phys A: Math. Gen}\ }\textbf {\bibinfo
  {volume} {9}},\ \bibinfo {pages} {1387} (\bibinfo {year} {1976})}\BibitemShut
  {NoStop}%
\bibitem [{\citenamefont {Kibble}(1980)}]{Kibble1980}%
  \BibitemOpen
  \bibfield  {author} {\bibinfo {author} {\bibfnamefont {T.~W.}\ \bibnamefont
  {Kibble}},\ }\bibfield  {title} {\bibinfo {title} {Some implications of a
  cosmological phase transition},\ }\href
  {https://doi.org/10.1016/0370-1573(80)90091-5} {\bibfield  {journal}
  {\bibinfo  {journal} {Phys. Rep.}\ }\textbf {\bibinfo {volume} {67}},\
  \bibinfo {pages} {183} (\bibinfo {year} {1980})}\BibitemShut {NoStop}%
\bibitem [{\citenamefont {Vilenkin}(1985)}]{Vilenkin1985}%
  \BibitemOpen
  \bibfield  {author} {\bibinfo {author} {\bibfnamefont {A.}~\bibnamefont
  {Vilenkin}},\ }\bibfield  {title} {\bibinfo {title} {Cosmic strings and
  domain walls},\ }\href
  {https://doi.org/https://doi.org/10.1016/0370-1573(85)90033-X} {\bibfield
  {journal} {\bibinfo  {journal} {Phys. Rep.}\ }\textbf {\bibinfo {volume}
  {121}},\ \bibinfo {pages} {263} (\bibinfo {year} {1985})}\BibitemShut
  {NoStop}%
\bibitem [{\citenamefont {Dvali}\ and\ \citenamefont {Tye}(1999)}]{Dvali1999}%
  \BibitemOpen
  \bibfield  {author} {\bibinfo {author} {\bibfnamefont {G.}~\bibnamefont
  {Dvali}}\ and\ \bibinfo {author} {\bibfnamefont {S.~H.}\ \bibnamefont
  {Tye}},\ }\bibfield  {title} {\bibinfo {title} {Brane inflation},\ }\href
  {https://doi.org/10.1016/S0370-2693(99)00132-X} {\bibfield  {journal}
  {\bibinfo  {journal} {Phys. Lett. B}\ }\textbf {\bibinfo {volume} {450}},\
  \bibinfo {pages} {72} (\bibinfo {year} {1999})}\BibitemShut {NoStop}%
\bibitem [{\citenamefont {Sarangi}\ and\ \citenamefont
  {Tye}(2002)}]{Sarangi2002}%
  \BibitemOpen
  \bibfield  {author} {\bibinfo {author} {\bibfnamefont {S.}~\bibnamefont
  {Sarangi}}\ and\ \bibinfo {author} {\bibfnamefont {S.~H.}\ \bibnamefont
  {Tye}},\ }\bibfield  {title} {\bibinfo {title} {Cosmic string production
  towards the end of brane inflation},\ }\href
  {https://doi.org/10.1016/S0370-2693(02)01824-5} {\bibfield  {journal}
  {\bibinfo  {journal} {Phys. Lett. B}\ }\textbf {\bibinfo {volume} {536}},\
  \bibinfo {pages} {185} (\bibinfo {year} {2002})}\BibitemShut {NoStop}%
\bibitem [{\citenamefont {Gudnason}\ and\ \citenamefont
  {Nitta}(2015)}]{Gudnason2015}%
  \BibitemOpen
  \bibfield  {author} {\bibinfo {author} {\bibfnamefont {S.~B.}\ \bibnamefont
  {Gudnason}}\ and\ \bibinfo {author} {\bibfnamefont {M.}~\bibnamefont
  {Nitta}},\ }\bibfield  {title} {\bibinfo {title} {D-brane solitons in various
  dimensions},\ }\href {https://doi.org/10.1103/PhysRevD.91.045018} {\bibfield
  {journal} {\bibinfo  {journal} {Phys. Rev. D}\ }\textbf {\bibinfo {volume}
  {91}},\ \bibinfo {pages} {045018} (\bibinfo {year} {2015})}\BibitemShut
  {NoStop}%
\bibitem [{\citenamefont {Osheroff}\ and\ \citenamefont
  {Cross}(1977)}]{Osheroff1977}%
  \BibitemOpen
  \bibfield  {author} {\bibinfo {author} {\bibfnamefont {D.~D.}\ \bibnamefont
  {Osheroff}}\ and\ \bibinfo {author} {\bibfnamefont {M.~C.}\ \bibnamefont
  {Cross}},\ }\bibfield  {title} {\bibinfo {title} {{Interfacial Surface Energy
  between the Superfluid Phases of ${\mathrm{{H}e}}^{3}$}},\ }\href
  {https://doi.org/10.1103/PhysRevLett.38.905} {\bibfield  {journal} {\bibinfo
  {journal} {Phys. Rev. Lett.}\ }\textbf {\bibinfo {volume} {38}},\ \bibinfo
  {pages} {905} (\bibinfo {year} {1977})}\BibitemShut {NoStop}%
\bibitem [{\citenamefont {Yip}\ and\ \citenamefont {Leggett}(1986)}]{Yip1986}%
  \BibitemOpen
  \bibfield  {author} {\bibinfo {author} {\bibfnamefont {S.}~\bibnamefont
  {Yip}}\ and\ \bibinfo {author} {\bibfnamefont {A.~J.}\ \bibnamefont
  {Leggett}},\ }\bibfield  {title} {\bibinfo {title} {Dynamics of the
  $^{3}\mathrm{He} \ {A}\ensuremath{-}{B}$ phase boundary},\ }\href
  {https://doi.org/10.1103/PhysRevLett.57.345} {\bibfield  {journal} {\bibinfo
  {journal} {Phys. Rev. Lett.}\ }\textbf {\bibinfo {volume} {57}},\ \bibinfo
  {pages} {345} (\bibinfo {year} {1986})}\BibitemShut {NoStop}%
\bibitem [{\citenamefont {Salomaa}(1987)}]{Salomaa1987}%
  \BibitemOpen
  \bibfield  {author} {\bibinfo {author} {\bibfnamefont {M.~M.}\ \bibnamefont
  {Salomaa}},\ }\bibfield  {title} {\bibinfo {title} {Monopoles in the rotating
  superfluid helium-3 {A}–{B} interface},\ }\href
  {https://doi.org/10.1038/326367a0} {\bibfield  {journal} {\bibinfo  {journal}
  {Nature}\ }\textbf {\bibinfo {volume} {326}},\ \bibinfo {pages} {367}
  (\bibinfo {year} {1987})}\BibitemShut {NoStop}%
\bibitem [{\citenamefont {Finne}\ \emph {et~al.}(2006)\citenamefont {Finne},
  \citenamefont {Eltsov}, \citenamefont {Hänninen}, \citenamefont {Kopnin},
  \citenamefont {Kopu}, \citenamefont {Krusius}, \citenamefont {Tsubota},\ and\
  \citenamefont {Volovik}}]{Finne2006}%
  \BibitemOpen
  \bibfield  {author} {\bibinfo {author} {\bibfnamefont {A.~P.}\ \bibnamefont
  {Finne}}, \bibinfo {author} {\bibfnamefont {V.~B.}\ \bibnamefont {Eltsov}},
  \bibinfo {author} {\bibfnamefont {R.}~\bibnamefont {Hänninen}}, \bibinfo
  {author} {\bibfnamefont {N.~B.}\ \bibnamefont {Kopnin}}, \bibinfo {author}
  {\bibfnamefont {J.}~\bibnamefont {Kopu}}, \bibinfo {author} {\bibfnamefont
  {M.}~\bibnamefont {Krusius}}, \bibinfo {author} {\bibfnamefont
  {M.}~\bibnamefont {Tsubota}},\ and\ \bibinfo {author} {\bibfnamefont {G.~E.}\
  \bibnamefont {Volovik}},\ }\bibfield  {title} {\bibinfo {title} {Dynamics of
  vortices and interfaces in superfluid $^3\mathrm{He}$},\ }\href
  {https://doi.org/10.1088/0034-4885/69/12/R03} {\bibfield  {journal} {\bibinfo
   {journal} {Rep. Prog. Phys.}\ }\textbf {\bibinfo {volume} {69}},\ \bibinfo
  {pages} {3157} (\bibinfo {year} {2006})}\BibitemShut {NoStop}%
\bibitem [{\citenamefont {Bradley}\ \emph {et~al.}(2008)\citenamefont
  {Bradley}, \citenamefont {Fisher}, \citenamefont {Gu{\'e}nault},
  \citenamefont {Haley}, \citenamefont {Kopu}, \citenamefont {Martin},
  \citenamefont {Pickett}, \citenamefont {Roberts},\ and\ \citenamefont
  {Tsepelin}}]{Bradley2008}%
  \BibitemOpen
  \bibfield  {author} {\bibinfo {author} {\bibfnamefont {D.~I.}\ \bibnamefont
  {Bradley}}, \bibinfo {author} {\bibfnamefont {S.~N.}\ \bibnamefont {Fisher}},
  \bibinfo {author} {\bibfnamefont {A.~M.}\ \bibnamefont {Gu{\'e}nault}},
  \bibinfo {author} {\bibfnamefont {R.~P.}\ \bibnamefont {Haley}}, \bibinfo
  {author} {\bibfnamefont {J.}~\bibnamefont {Kopu}}, \bibinfo {author}
  {\bibfnamefont {H.}~\bibnamefont {Martin}}, \bibinfo {author} {\bibfnamefont
  {G.~R.}\ \bibnamefont {Pickett}}, \bibinfo {author} {\bibfnamefont {J.~E.}\
  \bibnamefont {Roberts}},\ and\ \bibinfo {author} {\bibfnamefont
  {V.}~\bibnamefont {Tsepelin}},\ }\bibfield  {title} {\bibinfo {title} {Relic
  topological defects from brane annihilation simulated in superfluid
  $^3${{He}}},\ }\href {https://doi.org/10.1038/nphys815} {\bibfield  {journal}
  {\bibinfo  {journal} {Nat. Phys.}\ }\textbf {\bibinfo {volume} {4}},\
  \bibinfo {pages} {46} (\bibinfo {year} {2008})}\BibitemShut {NoStop}%
\bibitem [{\citenamefont {Volovik}(2009)}]{Volovik2009}%
  \BibitemOpen
  \bibfield  {author} {\bibinfo {author} {\bibfnamefont {G.~E.}\ \bibnamefont
  {Volovik}},\ }\href
  {https://doi.org/10.1093/acprof:oso/9780199564842.001.0001} {\emph {\bibinfo
  {title} {The Universe in a Helium Droplet}}}\ (\bibinfo  {publisher} {Oxford
  University Press},\ \bibinfo {year} {2009})\BibitemShut {NoStop}%
\bibitem [{\citenamefont {Takeuchi}\ and\ \citenamefont
  {Tsubota}(2006)}]{Takeuchi2006}%
  \BibitemOpen
  \bibfield  {author} {\bibinfo {author} {\bibfnamefont {H.}~\bibnamefont
  {Takeuchi}}\ and\ \bibinfo {author} {\bibfnamefont {M.}~\bibnamefont
  {Tsubota}},\ }\bibfield  {title} {\bibinfo {title} {Boojums in rotating
  two-component {B}ose–{E}instein condensates},\ }\href
  {https://doi.org/10.1143/JPSJ.75.063601} {\bibfield  {journal} {\bibinfo
  {journal} {J. Phys. Soc. Jpn.}\ }\textbf {\bibinfo {volume} {75}},\ \bibinfo
  {pages} {063601} (\bibinfo {year} {2006})}\BibitemShut {NoStop}%
\bibitem [{\citenamefont {Kasamatsu}\ \emph {et~al.}(2010)\citenamefont
  {Kasamatsu}, \citenamefont {Takeuchi}, \citenamefont {Nitta},\ and\
  \citenamefont {Tsubota}}]{Kasamatsu2010}%
  \BibitemOpen
  \bibfield  {author} {\bibinfo {author} {\bibfnamefont {K.}~\bibnamefont
  {Kasamatsu}}, \bibinfo {author} {\bibfnamefont {H.}~\bibnamefont {Takeuchi}},
  \bibinfo {author} {\bibfnamefont {M.}~\bibnamefont {Nitta}},\ and\ \bibinfo
  {author} {\bibfnamefont {M.}~\bibnamefont {Tsubota}},\ }\bibfield  {title}
  {\bibinfo {title} {Analogues of {D}-branes in bose-einstein condensates},\
  }\href {https://doi.org/10.1007/JHEP11(2010)068} {\bibfield  {journal}
  {\bibinfo  {journal} {J. High Energy Phys.}\ }\textbf {\bibinfo {volume}
  {2010}}\bibinfo  {number} { (11)},\ \bibinfo {pages} {68}}\BibitemShut
  {NoStop}%
\bibitem [{\citenamefont {Borgh}\ and\ \citenamefont
  {Ruostekoski}(2012)}]{Borgh2012}%
  \BibitemOpen
\bibfield  {number} {  }\bibfield  {author} {\bibinfo {author} {\bibfnamefont
  {M.~O.}\ \bibnamefont {Borgh}}\ and\ \bibinfo {author} {\bibfnamefont
  {J.}~\bibnamefont {Ruostekoski}},\ }\bibfield  {title} {\bibinfo {title}
  {Topological interface engineering and defect crossing in ultracold atomic
  gases},\ }\href {https://doi.org/10.1103/PHYSREVLETT.109.015302} {\bibfield
  {journal} {\bibinfo  {journal} {Phys. Rev. Lett.}\ }\textbf {\bibinfo
  {volume} {109}},\ \bibinfo {pages} {015302} (\bibinfo {year}
  {2012})}\BibitemShut {NoStop}%
\bibitem [{\citenamefont {Borgh}\ and\ \citenamefont
  {Ruostekoski}(2013)}]{Borgh2013}%
  \BibitemOpen
  \bibfield  {author} {\bibinfo {author} {\bibfnamefont {M.~O.}\ \bibnamefont
  {Borgh}}\ and\ \bibinfo {author} {\bibfnamefont {J.}~\bibnamefont
  {Ruostekoski}},\ }\bibfield  {title} {\bibinfo {title} {Topological interface
  physics of defects and textures in spinor {B}ose-{E}instein condensates},\
  }\href {https://doi.org/10.1103/PHYSREVA.87.033617} {\bibfield  {journal}
  {\bibinfo  {journal} {Phys. Rev. A}\ }\textbf {\bibinfo {volume} {87}},\
  \bibinfo {pages} {033617} (\bibinfo {year} {2013})}\BibitemShut {NoStop}%
\bibitem [{\citenamefont {Borgh}\ \emph {et~al.}(2014)\citenamefont {Borgh},
  \citenamefont {Lovegrove},\ and\ \citenamefont {Ruostekoski}}]{Borgh2014}%
  \BibitemOpen
  \bibfield  {author} {\bibinfo {author} {\bibfnamefont {M.~O.}\ \bibnamefont
  {Borgh}}, \bibinfo {author} {\bibfnamefont {J.}~\bibnamefont {Lovegrove}},\
  and\ \bibinfo {author} {\bibfnamefont {J.}~\bibnamefont {Ruostekoski}},\
  }\bibfield  {title} {\bibinfo {title} {Imprinting a topological interface
  using {Z}eeman shifts in an atomic spinor {B}ose-{E}instein condensate},\
  }\href {https://doi.org/10.1088/1367-2630/16/5/053046} {\bibfield  {journal}
  {\bibinfo  {journal} {New. J. Phys.}\ }\textbf {\bibinfo {volume} {16}},\
  \bibinfo {pages} {053046} (\bibinfo {year} {2014})}\BibitemShut {NoStop}%
\bibitem [{\citenamefont {Kaneda}\ and\ \citenamefont
  {Saito}(2014)}]{Kaneda2014}%
  \BibitemOpen
  \bibfield  {author} {\bibinfo {author} {\bibfnamefont {T.}~\bibnamefont
  {Kaneda}}\ and\ \bibinfo {author} {\bibfnamefont {H.}~\bibnamefont {Saito}},\
  }\bibfield  {title} {\bibinfo {title} {Dynamics of a vortex dipole across a
  magnetic phase boundary in a spinor {B}ose-{E}instein condensate},\ }\href
  {https://doi.org/10.1103/PhysRevA.90.053632} {\bibfield  {journal} {\bibinfo
  {journal} {Phys. Rev. A}\ }\textbf {\bibinfo {volume} {90}},\ \bibinfo
  {pages} {053632} (\bibinfo {year} {2014})}\BibitemShut {NoStop}%
\bibitem [{\citenamefont {Alford}\ \emph {et~al.}(2001)\citenamefont {Alford},
  \citenamefont {Rajagopal}, \citenamefont {Reddy},\ and\ \citenamefont
  {Wilczek}}]{Alford2001}%
  \BibitemOpen
  \bibfield  {author} {\bibinfo {author} {\bibfnamefont {M.}~\bibnamefont
  {Alford}}, \bibinfo {author} {\bibfnamefont {K.}~\bibnamefont {Rajagopal}},
  \bibinfo {author} {\bibfnamefont {S.}~\bibnamefont {Reddy}},\ and\ \bibinfo
  {author} {\bibfnamefont {F.}~\bibnamefont {Wilczek}},\ }\bibfield  {title}
  {\bibinfo {title} {Minimal color-flavor-locked--nuclear interface},\ }\href
  {https://doi.org/10.1103/PhysRevD.64.074017} {\bibfield  {journal} {\bibinfo
  {journal} {Phys. Rev. D}\ }\textbf {\bibinfo {volume} {64}},\ \bibinfo
  {pages} {074017} (\bibinfo {year} {2001})}\BibitemShut {NoStop}%
\bibitem [{\citenamefont {Cipriani}\ \emph {et~al.}(2012)\citenamefont
  {Cipriani}, \citenamefont {Vinci},\ and\ \citenamefont
  {Nitta}}]{Cipriani2012}%
  \BibitemOpen
  \bibfield  {author} {\bibinfo {author} {\bibfnamefont {M.}~\bibnamefont
  {Cipriani}}, \bibinfo {author} {\bibfnamefont {W.}~\bibnamefont {Vinci}},\
  and\ \bibinfo {author} {\bibfnamefont {M.}~\bibnamefont {Nitta}},\ }\bibfield
   {title} {\bibinfo {title} {Colorful boojums at the interface of a color
  superconductor},\ }\href {https://doi.org/10.1103/PhysRevD.86.121704}
  {\bibfield  {journal} {\bibinfo  {journal} {Phys. Rev. D}\ }\textbf {\bibinfo
  {volume} {86}},\ \bibinfo {pages} {121704(R)} (\bibinfo {year}
  {2012})}\BibitemShut {NoStop}%
\bibitem [{\citenamefont {Eto}\ \emph {et~al.}(2014)\citenamefont {Eto},
  \citenamefont {Hirono}, \citenamefont {Nitta},\ and\ \citenamefont
  {Yasui}}]{Eto2014}%
  \BibitemOpen
  \bibfield  {author} {\bibinfo {author} {\bibfnamefont {M.}~\bibnamefont
  {Eto}}, \bibinfo {author} {\bibfnamefont {Y.}~\bibnamefont {Hirono}},
  \bibinfo {author} {\bibfnamefont {M.}~\bibnamefont {Nitta}},\ and\ \bibinfo
  {author} {\bibfnamefont {S.}~\bibnamefont {Yasui}},\ }\bibfield  {title}
  {\bibinfo {title} {{Vortices and other topological solitons in dense quark
  matter}},\ }\href {https://doi.org/10.1093/ptep/ptt095} {\bibfield  {journal}
  {\bibinfo  {journal} {Prog. Theor. Exp. Phys.}\ }\textbf {\bibinfo {volume}
  {2014}} (\bibinfo {year} {2014})}\BibitemShut {NoStop}%
\bibitem [{\citenamefont {Xiao}\ \emph {et~al.}(2022)\citenamefont {Xiao},
  \citenamefont {Borgh}, \citenamefont {Blinova}, \citenamefont {Ollikainen},
  \citenamefont {Ruostekoski},\ and\ \citenamefont {Hall}}]{Xiao2022}%
  \BibitemOpen
  \bibfield  {author} {\bibinfo {author} {\bibfnamefont {Y.}~\bibnamefont
  {Xiao}}, \bibinfo {author} {\bibfnamefont {M.~O.}\ \bibnamefont {Borgh}},
  \bibinfo {author} {\bibfnamefont {A.}~\bibnamefont {Blinova}}, \bibinfo
  {author} {\bibfnamefont {T.}~\bibnamefont {Ollikainen}}, \bibinfo {author}
  {\bibfnamefont {J.}~\bibnamefont {Ruostekoski}},\ and\ \bibinfo {author}
  {\bibfnamefont {D.~S.}\ \bibnamefont {Hall}},\ }\bibfield  {title} {\bibinfo
  {title} {Topological superfluid defects with discrete point group
  symmetries},\ }\href {https://doi.org/10.1038/s41467-022-32362-5} {\bibfield
  {journal} {\bibinfo  {journal} {Nat. Commun.}\ }\textbf {\bibinfo {volume}
  {13}},\ \bibinfo {pages} {4635} (\bibinfo {year} {2022})}\BibitemShut
  {NoStop}%
\bibitem [{\citenamefont {Poenaru}\ and\ \citenamefont
  {Toulouse}(1977)}]{Poenaru1977}%
  \BibitemOpen
  \bibfield  {author} {\bibinfo {author} {\bibfnamefont {V.}~\bibnamefont
  {Poenaru}}\ and\ \bibinfo {author} {\bibfnamefont {G.}~\bibnamefont
  {Toulouse}},\ }\bibfield  {title} {\bibinfo {title} {The crossing of defects
  in ordered media and the topology of 3-manifolds},\ }\href
  {https://doi.org/10.1051/jphys:01977003808088700} {\bibfield  {journal}
  {\bibinfo  {journal} {J. Phys. (Paris)}\ }\textbf {\bibinfo {volume} {38}},\
  \bibinfo {pages} {887} (\bibinfo {year} {1977})}\BibitemShut {NoStop}%
\bibitem [{\citenamefont {Mermin}(1979)}]{Mermin1979}%
  \BibitemOpen
  \bibfield  {author} {\bibinfo {author} {\bibfnamefont {N.~D.}\ \bibnamefont
  {Mermin}},\ }\bibfield  {title} {\bibinfo {title} {The topological theory of
  defects in ordered media},\ }\href
  {https://doi.org/10.1103/RevModPhys.51.591} {\bibfield  {journal} {\bibinfo
  {journal} {Rev. Mod. Phys.}\ }\textbf {\bibinfo {volume} {51}},\ \bibinfo
  {pages} {591} (\bibinfo {year} {1979})}\BibitemShut {NoStop}%
\bibitem [{\citenamefont {Barnett}\ \emph {et~al.}(2006)\citenamefont
  {Barnett}, \citenamefont {Turner},\ and\ \citenamefont
  {Demler}}]{Barnett2006}%
  \BibitemOpen
  \bibfield  {author} {\bibinfo {author} {\bibfnamefont {R.}~\bibnamefont
  {Barnett}}, \bibinfo {author} {\bibfnamefont {A.}~\bibnamefont {Turner}},\
  and\ \bibinfo {author} {\bibfnamefont {E.}~\bibnamefont {Demler}},\
  }\bibfield  {title} {\bibinfo {title} {Classifying novel phases of spinor
  atoms},\ }\href {https://doi.org/10.1103/PhysRevLett.97.180412} {\bibfield
  {journal} {\bibinfo  {journal} {Phys. Rev. Lett.}\ }\textbf {\bibinfo
  {volume} {97}},\ \bibinfo {pages} {180412} (\bibinfo {year}
  {2006})}\BibitemShut {NoStop}%
\bibitem [{\citenamefont {Santos}\ and\ \citenamefont
  {Pfau}(2006)}]{Santos2006}%
  \BibitemOpen
  \bibfield  {author} {\bibinfo {author} {\bibfnamefont {L.}~\bibnamefont
  {Santos}}\ and\ \bibinfo {author} {\bibfnamefont {T.}~\bibnamefont {Pfau}},\
  }\bibfield  {title} {\bibinfo {title} {Spin-3 chromium {B}ose-{E}instein
  condensates},\ }\href {https://doi.org/10.1103/PhysRevLett.96.190404}
  {\bibfield  {journal} {\bibinfo  {journal} {Phys. Rev. Lett.}\ }\textbf
  {\bibinfo {volume} {96}},\ \bibinfo {pages} {190404} (\bibinfo {year}
  {2006})}\BibitemShut {NoStop}%
\bibitem [{\citenamefont {Diener}\ and\ \citenamefont {Ho}(2006)}]{Diener2006}%
  \BibitemOpen
  \bibfield  {author} {\bibinfo {author} {\bibfnamefont {R.~B.}\ \bibnamefont
  {Diener}}\ and\ \bibinfo {author} {\bibfnamefont {T.-L.}\ \bibnamefont
  {Ho}},\ }\bibfield  {title} {\bibinfo {title} {$^{52}\mathrm{Cr}$ spinor
  condensate: A biaxial or uniaxial spin nematic},\ }\href
  {https://doi.org/10.1103/PhysRevLett.96.190405} {\bibfield  {journal}
  {\bibinfo  {journal} {Phys. Rev. Lett.}\ }\textbf {\bibinfo {volume} {96}},\
  \bibinfo {pages} {190405} (\bibinfo {year} {2006})}\BibitemShut {NoStop}%
\bibitem [{\citenamefont {Barnett}\ \emph {et~al.}(2007)\citenamefont
  {Barnett}, \citenamefont {Turner},\ and\ \citenamefont
  {Demler}}]{Barnett2007}%
  \BibitemOpen
  \bibfield  {author} {\bibinfo {author} {\bibfnamefont {R.}~\bibnamefont
  {Barnett}}, \bibinfo {author} {\bibfnamefont {A.}~\bibnamefont {Turner}},\
  and\ \bibinfo {author} {\bibfnamefont {E.}~\bibnamefont {Demler}},\
  }\bibfield  {title} {\bibinfo {title} {Classifying vortices in ${S}=3$
  {B}ose--{E}instein condensates},\ }\href
  {https://doi.org/10.1103/PhysRevA.76.013605} {\bibfield  {journal} {\bibinfo
  {journal} {Phys. Rev. A}\ }\textbf {\bibinfo {volume} {76}},\ \bibinfo
  {pages} {013605} (\bibinfo {year} {2007})}\BibitemShut {NoStop}%
\bibitem [{\citenamefont {Semenoff}\ and\ \citenamefont
  {Zhou}(2007)}]{Semenoff2007}%
  \BibitemOpen
  \bibfield  {author} {\bibinfo {author} {\bibfnamefont {G.~W.}\ \bibnamefont
  {Semenoff}}\ and\ \bibinfo {author} {\bibfnamefont {F.}~\bibnamefont
  {Zhou}},\ }\bibfield  {title} {\bibinfo {title} {Discrete symmetries and
  1/3-quantum vortices in condensates of {F}=2 cold atoms},\ }\href
  {https://doi.org/10.1103/PHYSREVLETT.98.100401} {\bibfield  {journal}
  {\bibinfo  {journal} {Phys. Rev. Lett.}\ }\textbf {\bibinfo {volume} {98}},\
  \bibinfo {pages} {100401} (\bibinfo {year} {2007})}\BibitemShut {NoStop}%
\bibitem [{\citenamefont {M\"akel\"a}\ and\ \citenamefont
  {Suominen}(2007)}]{Makela2007}%
  \BibitemOpen
  \bibfield  {author} {\bibinfo {author} {\bibfnamefont {H.}~\bibnamefont
  {M\"akel\"a}}\ and\ \bibinfo {author} {\bibfnamefont {K.-A.}\ \bibnamefont
  {Suominen}},\ }\bibfield  {title} {\bibinfo {title} {Ground states of spin-3
  {B}ose-{E}instein condensates for conserved magnetization},\ }\href
  {https://doi.org/10.1103/PhysRevA.75.033610} {\bibfield  {journal} {\bibinfo
  {journal} {Phys. Rev. A}\ }\textbf {\bibinfo {volume} {75}},\ \bibinfo
  {pages} {033610} (\bibinfo {year} {2007})}\BibitemShut {NoStop}%
\bibitem [{\citenamefont {Yip}(2007)}]{Yip2007}%
  \BibitemOpen
  \bibfield  {author} {\bibinfo {author} {\bibfnamefont {S.-K.}\ \bibnamefont
  {Yip}},\ }\bibfield  {title} {\bibinfo {title} {Symmetry and inert states of
  spin {B}ose-{E}instein condensates},\ }\href
  {https://doi.org/10.1103/PhysRevA.75.023625} {\bibfield  {journal} {\bibinfo
  {journal} {Phys. Rev. A}\ }\textbf {\bibinfo {volume} {75}},\ \bibinfo
  {pages} {023625} (\bibinfo {year} {2007})}\BibitemShut {NoStop}%
\bibitem [{\citenamefont {Kawaguchi}\ and\ \citenamefont
  {Ueda}(2011)}]{Kawaguchi2011}%
  \BibitemOpen
  \bibfield  {author} {\bibinfo {author} {\bibfnamefont {Y.}~\bibnamefont
  {Kawaguchi}}\ and\ \bibinfo {author} {\bibfnamefont {M.}~\bibnamefont
  {Ueda}},\ }\bibfield  {title} {\bibinfo {title} {Symmetry classification of
  spinor {B}ose-{E}instein condensates},\ }\href
  {https://doi.org/10.1103/PhysRevA.84.053616} {\bibfield  {journal} {\bibinfo
  {journal} {Phys. Rev. A}\ }\textbf {\bibinfo {volume} {84}},\ \bibinfo
  {pages} {053616} (\bibinfo {year} {2011})}\BibitemShut {NoStop}%
\bibitem [{\citenamefont {Kobayashi}\ \emph {et~al.}(2009)\citenamefont
  {Kobayashi}, \citenamefont {Kawaguchi}, \citenamefont {Nitta},\ and\
  \citenamefont {Ueda}}]{Kobayashi2009}%
  \BibitemOpen
  \bibfield  {author} {\bibinfo {author} {\bibfnamefont {M.}~\bibnamefont
  {Kobayashi}}, \bibinfo {author} {\bibfnamefont {Y.}~\bibnamefont
  {Kawaguchi}}, \bibinfo {author} {\bibfnamefont {M.}~\bibnamefont {Nitta}},\
  and\ \bibinfo {author} {\bibfnamefont {M.}~\bibnamefont {Ueda}},\ }\bibfield
  {title} {\bibinfo {title} {{Collision dynamics and rung formation of
  non-Abelian vortices}},\ }\href
  {https://doi.org/10.1103/PHYSREVLETT.103.115301} {\bibfield  {journal}
  {\bibinfo  {journal} {Phys. Rev. Lett.}\ }\textbf {\bibinfo {volume} {103}},\
  \bibinfo {pages} {115301} (\bibinfo {year} {2009})}\BibitemShut {NoStop}%
\bibitem [{\citenamefont {Borgh}\ and\ \citenamefont
  {Ruostekoski}(2016)}]{Borgh2016b}%
  \BibitemOpen
  \bibfield  {author} {\bibinfo {author} {\bibfnamefont {M.~O.}\ \bibnamefont
  {Borgh}}\ and\ \bibinfo {author} {\bibfnamefont {J.}~\bibnamefont
  {Ruostekoski}},\ }\bibfield  {title} {\bibinfo {title} {{Core Structure and
  Non-Abelian Reconnection of Defects in a Biaxial Nematic Spin-2
  {B}ose-{E}instein Condensate}},\ }\href
  {https://doi.org/10.1103/PhysRevLett.117.275302} {\bibfield  {journal}
  {\bibinfo  {journal} {Phys. Rev. Lett.}\ }\textbf {\bibinfo {volume} {117}},\
  \bibinfo {pages} {275302} (\bibinfo {year} {2016})}\BibitemShut {NoStop}%
\bibitem [{\citenamefont {Borgh}\ \emph {et~al.}(2017)\citenamefont {Borgh},
  \citenamefont {Lovegrove},\ and\ \citenamefont {Ruostekoski}}]{Borgh2017}%
  \BibitemOpen
  \bibfield  {author} {\bibinfo {author} {\bibfnamefont {M.~O.}\ \bibnamefont
  {Borgh}}, \bibinfo {author} {\bibfnamefont {J.}~\bibnamefont {Lovegrove}},\
  and\ \bibinfo {author} {\bibfnamefont {J.}~\bibnamefont {Ruostekoski}},\
  }\bibfield  {title} {\bibinfo {title} {Internal structure and stability of
  vortices in a dipolar spinor {B}ose-{E}instein condensate},\ }\href
  {https://doi.org/10.1103/PhysRevA.95.053601} {\bibfield  {journal} {\bibinfo
  {journal} {Phys. Rev. A}\ }\textbf {\bibinfo {volume} {95}},\ \bibinfo
  {pages} {053601} (\bibinfo {year} {2017})}\BibitemShut {NoStop}%
\bibitem [{\citenamefont {Mawson}\ \emph {et~al.}(2019)\citenamefont {Mawson},
  \citenamefont {Petersen}, \citenamefont {Slingerland},\ and\ \citenamefont
  {Simula}}]{Mawson2019}%
  \BibitemOpen
  \bibfield  {author} {\bibinfo {author} {\bibfnamefont {T.}~\bibnamefont
  {Mawson}}, \bibinfo {author} {\bibfnamefont {T.~C.}\ \bibnamefont
  {Petersen}}, \bibinfo {author} {\bibfnamefont {J.~K.}\ \bibnamefont
  {Slingerland}},\ and\ \bibinfo {author} {\bibfnamefont {T.~P.}\ \bibnamefont
  {Simula}},\ }\bibfield  {title} {\bibinfo {title} {{Braiding and Fusion of
  Non-Abelian Vortex Anyons}},\ }\href
  {https://doi.org/10.1103/PhysRevLett.123.140404} {\bibfield  {journal}
  {\bibinfo  {journal} {Phys. Rev. Lett.}\ }\textbf {\bibinfo {volume} {123}},\
  \bibinfo {pages} {140404} (\bibinfo {year} {2019})}\BibitemShut {NoStop}%
\bibitem [{\citenamefont {Kawaguchi}\ and\ \citenamefont
  {Ueda}(2012)}]{Kawaguchi2012}%
  \BibitemOpen
  \bibfield  {author} {\bibinfo {author} {\bibfnamefont {Y.}~\bibnamefont
  {Kawaguchi}}\ and\ \bibinfo {author} {\bibfnamefont {M.}~\bibnamefont
  {Ueda}},\ }\bibfield  {title} {\bibinfo {title} {Spinor {B}ose-{E}instein
  condensates},\ }\href {https://doi.org/10.1016/j.physrep.2012.07.005}
  {\bibfield  {journal} {\bibinfo  {journal} {Phys. Rep.}\ }\textbf {\bibinfo
  {volume} {520}},\ \bibinfo {pages} {253} (\bibinfo {year}
  {2012})}\BibitemShut {NoStop}%
\bibitem [{\citenamefont {Stamper-Kurn}\ and\ \citenamefont
  {Ueda}(2013)}]{StamperKurn2013}%
  \BibitemOpen
  \bibfield  {author} {\bibinfo {author} {\bibfnamefont {D.~M.}\ \bibnamefont
  {Stamper-Kurn}}\ and\ \bibinfo {author} {\bibfnamefont {M.}~\bibnamefont
  {Ueda}},\ }\bibfield  {title} {\bibinfo {title} {{Spinor Bose gases:
  Symmetries, magnetism, and quantum dynamics}},\ }\href
  {https://doi.org/10.1103/RevModPhys.85.1191} {\bibfield  {journal} {\bibinfo
  {journal} {Rev. Mod. Phys.}\ }\textbf {\bibinfo {volume} {85}},\ \bibinfo
  {pages} {1191} (\bibinfo {year} {2013})}\BibitemShut {NoStop}%
\bibitem [{\citenamefont {Stamper-Kurn}\ \emph {et~al.}(1998)\citenamefont
  {Stamper-Kurn}, \citenamefont {Andrews}, \citenamefont {Chikkatur},
  \citenamefont {Inouye}, \citenamefont {Miesner}, \citenamefont {Stenger},\
  and\ \citenamefont {Ketterle}}]{StamperKurn1998}%
  \BibitemOpen
  \bibfield  {author} {\bibinfo {author} {\bibfnamefont {D.~M.}\ \bibnamefont
  {Stamper-Kurn}}, \bibinfo {author} {\bibfnamefont {M.~R.}\ \bibnamefont
  {Andrews}}, \bibinfo {author} {\bibfnamefont {A.~P.}\ \bibnamefont
  {Chikkatur}}, \bibinfo {author} {\bibfnamefont {S.}~\bibnamefont {Inouye}},
  \bibinfo {author} {\bibfnamefont {H.~J.}\ \bibnamefont {Miesner}}, \bibinfo
  {author} {\bibfnamefont {J.}~\bibnamefont {Stenger}},\ and\ \bibinfo {author}
  {\bibfnamefont {W.}~\bibnamefont {Ketterle}},\ }\bibfield  {title} {\bibinfo
  {title} {Optical confinement of a {B}ose-{E}instein condensate},\ }\href
  {https://doi.org/10.1103/PhysRevLett.80.2027} {\bibfield  {journal} {\bibinfo
   {journal} {Phys. Rev. Lett.}\ }\textbf {\bibinfo {volume} {80}},\ \bibinfo
  {pages} {2027} (\bibinfo {year} {1998})}\BibitemShut {NoStop}%
\bibitem [{\citenamefont {Lovegrove}\ \emph {et~al.}(2016)\citenamefont
  {Lovegrove}, \citenamefont {Borgh},\ and\ \citenamefont
  {Ruostekoski}}]{Lovegrove2016}%
  \BibitemOpen
  \bibfield  {author} {\bibinfo {author} {\bibfnamefont {J.}~\bibnamefont
  {Lovegrove}}, \bibinfo {author} {\bibfnamefont {M.~O.}\ \bibnamefont
  {Borgh}},\ and\ \bibinfo {author} {\bibfnamefont {J.}~\bibnamefont
  {Ruostekoski}},\ }\bibfield  {title} {\bibinfo {title} {Stability and
  internal structure of vortices in spin-1 {B}ose-{E}instein condensates with
  conserved magnetization},\ }\href
  {https://journals.aps.org/pra/abstract/10.1103/PhysRevA.93.033633} {\bibfield
   {journal} {\bibinfo  {journal} {Phys. Rev. A}\ }\textbf {\bibinfo {volume}
  {93}},\ \bibinfo {pages} {033633} (\bibinfo {year} {2016})}\BibitemShut
  {NoStop}%
\bibitem [{\citenamefont {Weiss}\ \emph {et~al.}(2019)\citenamefont {Weiss},
  \citenamefont {Borgh}, \citenamefont {Blinova}, \citenamefont {Ollikainen},
  \citenamefont {Möttönen}, \citenamefont {Ruostekoski},\ and\ \citenamefont
  {Hall}}]{Weiss2019}%
  \BibitemOpen
  \bibfield  {author} {\bibinfo {author} {\bibfnamefont {L.~S.}\ \bibnamefont
  {Weiss}}, \bibinfo {author} {\bibfnamefont {M.~O.}\ \bibnamefont {Borgh}},
  \bibinfo {author} {\bibfnamefont {A.}~\bibnamefont {Blinova}}, \bibinfo
  {author} {\bibfnamefont {T.}~\bibnamefont {Ollikainen}}, \bibinfo {author}
  {\bibfnamefont {M.}~\bibnamefont {Möttönen}}, \bibinfo {author}
  {\bibfnamefont {J.}~\bibnamefont {Ruostekoski}},\ and\ \bibinfo {author}
  {\bibfnamefont {D.~S.}\ \bibnamefont {Hall}},\ }\bibfield  {title} {\bibinfo
  {title} {Controlled creation of a singular spinor vortex by circumventing the
  {D}irac belt trick},\ }\href {https://doi.org/10.1038/s41467-019-12787-1}
  {\bibfield  {journal} {\bibinfo  {journal} {Nat. Commun.}\ }\textbf {\bibinfo
  {volume} {10}},\ \bibinfo {pages} {1} (\bibinfo {year} {2019})}\BibitemShut
  {NoStop}%
\bibitem [{\citenamefont {Xiao}\ \emph {et~al.}(2021)\citenamefont {Xiao},
  \citenamefont {Borgh}, \citenamefont {Weiss}, \citenamefont {Blinova},
  \citenamefont {Ruostekoski},\ and\ \citenamefont {Hall}}]{Xiao2021}%
  \BibitemOpen
  \bibfield  {author} {\bibinfo {author} {\bibfnamefont {Y.}~\bibnamefont
  {Xiao}}, \bibinfo {author} {\bibfnamefont {M.~O.}\ \bibnamefont {Borgh}},
  \bibinfo {author} {\bibfnamefont {L.~S.}\ \bibnamefont {Weiss}}, \bibinfo
  {author} {\bibfnamefont {A.~A.}\ \bibnamefont {Blinova}}, \bibinfo {author}
  {\bibfnamefont {J.}~\bibnamefont {Ruostekoski}},\ and\ \bibinfo {author}
  {\bibfnamefont {D.~S.}\ \bibnamefont {Hall}},\ }\bibfield  {title} {\bibinfo
  {title} {Controlled creation and decay of singly-quantized vortices in a
  polar magnetic phase},\ }\href {https://doi.org/10.1038/s42005-021-00554-y}
  {\bibfield  {journal} {\bibinfo  {journal} {Commun. Phys.}\ }\textbf
  {\bibinfo {volume} {4}},\ \bibinfo {pages} {1} (\bibinfo {year}
  {2021})}\BibitemShut {NoStop}%
\bibitem [{\citenamefont {Ueda}(2014)}]{Ueda2014}%
  \BibitemOpen
  \bibfield  {author} {\bibinfo {author} {\bibfnamefont {M.}~\bibnamefont
  {Ueda}},\ }\bibfield  {title} {\bibinfo {title} {Topological aspects in
  spinor {Bose–Einstein} condensates},\ }\href
  {https://doi.org/10.1088/0034-4885/77/12/122401} {\bibfield  {journal}
  {\bibinfo  {journal} {Rep. Prog. Phys.}\ }\textbf {\bibinfo {volume} {77}},\
  \bibinfo {pages} {122401} (\bibinfo {year} {2014})}\BibitemShut {NoStop}%
\bibitem [{\citenamefont {Yip}(1999)}]{Yip1999}%
  \BibitemOpen
  \bibfield  {author} {\bibinfo {author} {\bibfnamefont {S.~K.}\ \bibnamefont
  {Yip}},\ }\bibfield  {title} {\bibinfo {title} {Internal vortex structure of
  a trapped spinor {B}ose-{E}instein condensate},\ }\href
  {https://doi.org/10.1103/PhysRevLett.83.4677} {\bibfield  {journal} {\bibinfo
   {journal} {Phys. Rev. Lett.}\ }\textbf {\bibinfo {volume} {83}},\ \bibinfo
  {pages} {4677} (\bibinfo {year} {1999})}\BibitemShut {NoStop}%
\bibitem [{\citenamefont {Isoshima}\ and\ \citenamefont
  {Machida}(2002)}]{Isoshima2002}%
  \BibitemOpen
  \bibfield  {author} {\bibinfo {author} {\bibfnamefont {T.}~\bibnamefont
  {Isoshima}}\ and\ \bibinfo {author} {\bibfnamefont {K.}~\bibnamefont
  {Machida}},\ }\bibfield  {title} {\bibinfo {title} {Axisymmetric vortices in
  spinor {B}ose-{E}instein condensates under rotation},\ }\href
  {https://doi.org/10.1103/PhysRevA.66.023602} {\bibfield  {journal} {\bibinfo
  {journal} {Phys. Rev. A}\ }\textbf {\bibinfo {volume} {66}},\ \bibinfo
  {pages} {023602} (\bibinfo {year} {2002})}\BibitemShut {NoStop}%
\bibitem [{\citenamefont {Mizushima}\ \emph
  {et~al.}(2002{\natexlab{a}})\citenamefont {Mizushima}, \citenamefont
  {Machida},\ and\ \citenamefont {Kita}}]{MizushimaPRA2002}%
  \BibitemOpen
  \bibfield  {author} {\bibinfo {author} {\bibfnamefont {T.}~\bibnamefont
  {Mizushima}}, \bibinfo {author} {\bibfnamefont {K.}~\bibnamefont {Machida}},\
  and\ \bibinfo {author} {\bibfnamefont {T.}~\bibnamefont {Kita}},\ }\bibfield
  {title} {\bibinfo {title} {Axisymmetric versus nonaxisymmetric vortices in
  spinor {B}ose-{E}instein condensates},\ }\href
  {https://doi.org/10.1103/PhysRevA.66.053610} {\bibfield  {journal} {\bibinfo
  {journal} {Phys. Rev. A}\ }\textbf {\bibinfo {volume} {66}},\ \bibinfo
  {pages} {053610} (\bibinfo {year} {2002}{\natexlab{a}})}\BibitemShut
  {NoStop}%
\bibitem [{\citenamefont {Sadler}\ \emph {et~al.}(2006)\citenamefont {Sadler},
  \citenamefont {Higbie}, \citenamefont {Leslie}, \citenamefont
  {Vengalattore},\ and\ \citenamefont {Stamper-Kurn}}]{Sadler2006}%
  \BibitemOpen
  \bibfield  {author} {\bibinfo {author} {\bibfnamefont {L.~E.}\ \bibnamefont
  {Sadler}}, \bibinfo {author} {\bibfnamefont {J.~M.}\ \bibnamefont {Higbie}},
  \bibinfo {author} {\bibfnamefont {S.~R.}\ \bibnamefont {Leslie}}, \bibinfo
  {author} {\bibfnamefont {M.}~\bibnamefont {Vengalattore}},\ and\ \bibinfo
  {author} {\bibfnamefont {D.~M.}\ \bibnamefont {Stamper-Kurn}},\ }\bibfield
  {title} {\bibinfo {title} {Spontaneous symmetry breaking in a quenched
  ferromagnetic spinor {Bose–Einstein} condensate},\ }\href
  {https://doi.org/10.1038/nature05094} {\bibfield  {journal} {\bibinfo
  {journal} {Nature}\ }\textbf {\bibinfo {volume} {443}},\ \bibinfo {pages}
  {312} (\bibinfo {year} {2006})}\BibitemShut {NoStop}%
\bibitem [{\citenamefont {Lovegrove}\ \emph {et~al.}(2012)\citenamefont
  {Lovegrove}, \citenamefont {Borgh},\ and\ \citenamefont
  {Ruostekoski}}]{Lovegrove2012}%
  \BibitemOpen
  \bibfield  {author} {\bibinfo {author} {\bibfnamefont {J.}~\bibnamefont
  {Lovegrove}}, \bibinfo {author} {\bibfnamefont {M.~O.}\ \bibnamefont
  {Borgh}},\ and\ \bibinfo {author} {\bibfnamefont {J.}~\bibnamefont
  {Ruostekoski}},\ }\bibfield  {title} {\bibinfo {title} {Energetically stable
  singular vortex cores in an atomic spin-1 {B}ose-{E}instein condensate},\
  }\href {https://doi.org/10.1103/PHYSREVA.86.013613} {\bibfield  {journal}
  {\bibinfo  {journal} {Phys.\ Rev.\ A}\ }\textbf {\bibinfo {volume} {86}},\
  \bibinfo {pages} {013613} (\bibinfo {year} {2012})}\BibitemShut {NoStop}%
\bibitem [{\citenamefont {Leonhardt}\ and\ \citenamefont
  {Volovik}(2000)}]{Leonhardt2000}%
  \BibitemOpen
  \bibfield  {author} {\bibinfo {author} {\bibfnamefont {U.}~\bibnamefont
  {Leonhardt}}\ and\ \bibinfo {author} {\bibfnamefont {G.~E.}\ \bibnamefont
  {Volovik}},\ }\bibfield  {title} {\bibinfo {title} {How to create an {A}lice
  string (half-quantum vortex) in a vector {B}ose-{E}instein condensate},\
  }\href {https://doi.org/10.1134/1.1312008} {\bibfield  {journal} {\bibinfo
  {journal} {JETP Lett.}\ }\textbf {\bibinfo {volume} {72}},\ \bibinfo {pages}
  {46} (\bibinfo {year} {2000})}\BibitemShut {NoStop}%
\bibitem [{\citenamefont {Zhou}(2003)}]{Zhou2003}%
  \BibitemOpen
  \bibfield  {author} {\bibinfo {author} {\bibfnamefont {F.}~\bibnamefont
  {Zhou}},\ }\bibfield  {title} {\bibinfo {title} {Quantum spin nematic states
  in {B}ose–{E}instein condensates},\ }\href
  {https://doi.org/10.1142/S0217979203018399} {\bibfield  {journal} {\bibinfo
  {journal} {Int. J. Mod. Phys. B}\ }\textbf {\bibinfo {volume} {17}},\
  \bibinfo {pages} {2643} (\bibinfo {year} {2003})}\BibitemShut {NoStop}%
\bibitem [{\citenamefont {Ji}\ \emph {et~al.}(2008)\citenamefont {Ji},
  \citenamefont {Liu}, \citenamefont {Song},\ and\ \citenamefont
  {Zhou}}]{Ji2008}%
  \BibitemOpen
  \bibfield  {author} {\bibinfo {author} {\bibfnamefont {A.~C.}\ \bibnamefont
  {Ji}}, \bibinfo {author} {\bibfnamefont {W.~M.}\ \bibnamefont {Liu}},
  \bibinfo {author} {\bibfnamefont {J.~L.}\ \bibnamefont {Song}},\ and\
  \bibinfo {author} {\bibfnamefont {F.}~\bibnamefont {Zhou}},\ }\bibfield
  {title} {\bibinfo {title} {Dynamical creation of fractionalized vortices and
  vortex lattices},\ }\href {https://doi.org/10.1103/PHYSREVLETT.101.010402}
  {\bibfield  {journal} {\bibinfo  {journal} {Phys. Rev. Lett.}\ }\textbf
  {\bibinfo {volume} {101}},\ \bibinfo {pages} {010402} (\bibinfo {year}
  {2008})}\BibitemShut {NoStop}%
\bibitem [{\citenamefont {Seo}\ \emph {et~al.}(2015)\citenamefont {Seo},
  \citenamefont {Kang}, \citenamefont {Kwon},\ and\ \citenamefont
  {Shin}}]{Seo2015}%
  \BibitemOpen
  \bibfield  {author} {\bibinfo {author} {\bibfnamefont {S.~W.}\ \bibnamefont
  {Seo}}, \bibinfo {author} {\bibfnamefont {S.}~\bibnamefont {Kang}}, \bibinfo
  {author} {\bibfnamefont {W.~J.}\ \bibnamefont {Kwon}},\ and\ \bibinfo
  {author} {\bibfnamefont {Y.~I.}\ \bibnamefont {Shin}},\ }\bibfield  {title}
  {\bibinfo {title} {Half-quantum vortices in an antiferromagnetic spinor
  {B}ose-{E}instein condensate},\ }\href
  {https://doi.org/10.1103/PHYSREVLETT.115.015301} {\bibfield  {journal}
  {\bibinfo  {journal} {Phys. Rev. Lett.}\ }\textbf {\bibinfo {volume} {115}},\
  \bibinfo {pages} {015301} (\bibinfo {year} {2015})}\BibitemShut {NoStop}%
\bibitem [{\citenamefont {Ohmi}\ and\ \citenamefont
  {Machida}(1998)}]{Ohmi1998}%
  \BibitemOpen
  \bibfield  {author} {\bibinfo {author} {\bibfnamefont {T.}~\bibnamefont
  {Ohmi}}\ and\ \bibinfo {author} {\bibfnamefont {K.}~\bibnamefont {Machida}},\
  }\bibfield  {title} {\bibinfo {title} {{B}ose--{E}instein condensation with
  internal degrees of freedom in alkali atom gases},\ }\href
  {https://doi.org/10.1143/jpsj.67.1822} {\bibfield  {journal} {\bibinfo
  {journal} {J. Phys. Soc. Jpn.}\ }\textbf {\bibinfo {volume} {67}},\ \bibinfo
  {pages} {1822} (\bibinfo {year} {1998})}\BibitemShut {NoStop}%
\bibitem [{\citenamefont {Ho}(1998)}]{Ho1998}%
  \BibitemOpen
  \bibfield  {author} {\bibinfo {author} {\bibfnamefont {T.~L.}\ \bibnamefont
  {Ho}},\ }\bibfield  {title} {\bibinfo {title} {{Spinor Bose Condensates in
  Optical Traps}},\ }\href {https://doi.org/10.1103/PhysRevLett.81.742}
  {\bibfield  {journal} {\bibinfo  {journal} {Phys. Rev. Lett.}\ }\textbf
  {\bibinfo {volume} {81}},\ \bibinfo {pages} {742} (\bibinfo {year}
  {1998})}\BibitemShut {NoStop}%
\bibitem [{\citenamefont {Mizushima}\ \emph
  {et~al.}(2002{\natexlab{b}})\citenamefont {Mizushima}, \citenamefont
  {Machida},\ and\ \citenamefont {Kita}}]{MizushimaPRL2002}%
  \BibitemOpen
  \bibfield  {author} {\bibinfo {author} {\bibfnamefont {T.}~\bibnamefont
  {Mizushima}}, \bibinfo {author} {\bibfnamefont {K.}~\bibnamefont {Machida}},\
  and\ \bibinfo {author} {\bibfnamefont {T.}~\bibnamefont {Kita}},\ }\bibfield
  {title} {\bibinfo {title} {Mermin-{H}o vortex in ferromagnetic spinor
  {B}ose-{E}instein condensates},\ }\href
  {https://doi.org/10.1103/PhysRevLett.89.030401} {\bibfield  {journal}
  {\bibinfo  {journal} {Phys. Rev. Lett.}\ }\textbf {\bibinfo {volume} {89}},\
  \bibinfo {pages} {030401} (\bibinfo {year} {2002}{\natexlab{b}})}\BibitemShut
  {NoStop}%
\bibitem [{\citenamefont {Martikainen}\ \emph {et~al.}(2002)\citenamefont
  {Martikainen}, \citenamefont {Collin},\ and\ \citenamefont
  {Suominen}}]{Martikainen2002}%
  \BibitemOpen
  \bibfield  {author} {\bibinfo {author} {\bibfnamefont {J.-P.}\ \bibnamefont
  {Martikainen}}, \bibinfo {author} {\bibfnamefont {A.}~\bibnamefont
  {Collin}},\ and\ \bibinfo {author} {\bibfnamefont {K.-A.}\ \bibnamefont
  {Suominen}},\ }\bibfield  {title} {\bibinfo {title} {Coreless vortex ground
  state of the rotating spinor condensate},\ }\href
  {https://doi.org/10.1103/PhysRevA.66.053604} {\bibfield  {journal} {\bibinfo
  {journal} {Phys. Rev. A}\ }\textbf {\bibinfo {volume} {66}},\ \bibinfo
  {pages} {053604} (\bibinfo {year} {2002})}\BibitemShut {NoStop}%
\bibitem [{\citenamefont {Leanhardt}\ \emph {et~al.}(2003)\citenamefont
  {Leanhardt}, \citenamefont {Shin}, \citenamefont {Kielpinski}, \citenamefont
  {Pritchard},\ and\ \citenamefont {Ketterle}}]{Leanhardt2003}%
  \BibitemOpen
  \bibfield  {author} {\bibinfo {author} {\bibfnamefont {A.~E.}\ \bibnamefont
  {Leanhardt}}, \bibinfo {author} {\bibfnamefont {Y.}~\bibnamefont {Shin}},
  \bibinfo {author} {\bibfnamefont {D.}~\bibnamefont {Kielpinski}}, \bibinfo
  {author} {\bibfnamefont {D.~E.}\ \bibnamefont {Pritchard}},\ and\ \bibinfo
  {author} {\bibfnamefont {W.}~\bibnamefont {Ketterle}},\ }\bibfield  {title}
  {\bibinfo {title} {Coreless vortex formation in a spinor {B}ose-{E}instein
  condensate},\ }\href {https://doi.org/10.1103/PHYSREVLETT.90.140403}
  {\bibfield  {journal} {\bibinfo  {journal} {Phys. Rev. Lett.}\ }\textbf
  {\bibinfo {volume} {90}},\ \bibinfo {pages} {140403} (\bibinfo {year}
  {2003})}\BibitemShut {NoStop}%
\bibitem [{\citenamefont {Mizushima}\ \emph {et~al.}(2004)\citenamefont
  {Mizushima}, \citenamefont {Kobayashi},\ and\ \citenamefont
  {Machida}}]{Mizushima2004}%
  \BibitemOpen
  \bibfield  {author} {\bibinfo {author} {\bibfnamefont {T.}~\bibnamefont
  {Mizushima}}, \bibinfo {author} {\bibfnamefont {N.}~\bibnamefont
  {Kobayashi}},\ and\ \bibinfo {author} {\bibfnamefont {K.}~\bibnamefont
  {Machida}},\ }\bibfield  {title} {\bibinfo {title} {Coreless and singular
  vortex lattices in rotating spinor {B}ose-{E}instein condensates},\ }\href
  {https://doi.org/10.1103/PHYSREVA.70.043613} {\bibfield  {journal} {\bibinfo
  {journal} {Phys. Rev. A}\ }\textbf {\bibinfo {volume} {70}},\ \bibinfo
  {pages} {043613} (\bibinfo {year} {2004})}\BibitemShut {NoStop}%
\bibitem [{\citenamefont {Choi}\ \emph
  {et~al.}(2012{\natexlab{a}})\citenamefont {Choi}, \citenamefont {Kwon},\ and\
  \citenamefont {Shin}}]{Choi2012}%
  \BibitemOpen
  \bibfield  {author} {\bibinfo {author} {\bibfnamefont {J.-Y.}\ \bibnamefont
  {Choi}}, \bibinfo {author} {\bibfnamefont {W.~J.}\ \bibnamefont {Kwon}},\
  and\ \bibinfo {author} {\bibfnamefont {Y.-I.}\ \bibnamefont {Shin}},\
  }\bibfield  {title} {\bibinfo {title} {Observation of topologically stable
  2{D} {S}kyrmions in an antiferromagnetic spinor {B}ose-{E}instein
  condensate},\ }\href {https://doi.org/10.1103/PHYSREVLETT.108.035301}
  {\bibfield  {journal} {\bibinfo  {journal} {Phys. Rev. Lett.}\ }\textbf
  {\bibinfo {volume} {108}},\ \bibinfo {pages} {035301} (\bibinfo {year}
  {2012}{\natexlab{a}})}\BibitemShut {NoStop}%
\bibitem [{\citenamefont {Choi}\ \emph
  {et~al.}(2012{\natexlab{b}})\citenamefont {Choi}, \citenamefont {Kwon},
  \citenamefont {Lee}, \citenamefont {Jeong}, \citenamefont {An},\ and\
  \citenamefont {Shin}}]{Choi2012b}%
  \BibitemOpen
  \bibfield  {author} {\bibinfo {author} {\bibfnamefont {J.-Y.}\ \bibnamefont
  {Choi}}, \bibinfo {author} {\bibfnamefont {W.~J.}\ \bibnamefont {Kwon}},
  \bibinfo {author} {\bibfnamefont {M.}~\bibnamefont {Lee}}, \bibinfo {author}
  {\bibfnamefont {H.}~\bibnamefont {Jeong}}, \bibinfo {author} {\bibfnamefont
  {K.}~\bibnamefont {An}},\ and\ \bibinfo {author} {\bibfnamefont {Y.-I.}\
  \bibnamefont {Shin}},\ }\bibfield  {title} {\bibinfo {title} {Imprinting
  {S}kyrmion spin textures in {B}ose--{E}instein condensates},\ }\href
  {http://stacks.iop.org/1367-2630/14/i=5/a=053013} {\bibfield  {journal}
  {\bibinfo  {journal} {New J. Phys.}\ }\textbf {\bibinfo {volume} {14}},\
  \bibinfo {pages} {053013} (\bibinfo {year} {2012}{\natexlab{b}})}\BibitemShut
  {NoStop}%
\bibitem [{\citenamefont {Lovegrove}\ \emph {et~al.}(2014)\citenamefont
  {Lovegrove}, \citenamefont {Borgh},\ and\ \citenamefont
  {Ruostekoski}}]{Lovegrove2014}%
  \BibitemOpen
  \bibfield  {author} {\bibinfo {author} {\bibfnamefont {J.}~\bibnamefont
  {Lovegrove}}, \bibinfo {author} {\bibfnamefont {M.~O.}\ \bibnamefont
  {Borgh}},\ and\ \bibinfo {author} {\bibfnamefont {J.}~\bibnamefont
  {Ruostekoski}},\ }\bibfield  {title} {\bibinfo {title} {Energetic stability
  of coreless vortices in spin-1 {B}ose-{E}instein condensates with conserved
  magnetization},\ }\href
  {https://journals.aps.org/prl/abstract/10.1103/PhysRevLett.112.075301}
  {\bibfield  {journal} {\bibinfo  {journal} {Phys. Rev. Lett.}\ }\textbf
  {\bibinfo {volume} {112}},\ \bibinfo {pages} {075301} (\bibinfo {year}
  {2014})}\BibitemShut {NoStop}%
\bibitem [{\citenamefont {Mühlbauer}\ \emph {et~al.}(2009)\citenamefont
  {Mühlbauer}, \citenamefont {Binz}, \citenamefont {Jonietz}, \citenamefont
  {Pfleiderer}, \citenamefont {Rosch}, \citenamefont {Neubauer}, \citenamefont
  {Georgii},\ and\ \citenamefont {Böni}}]{Muhlbauer2009}%
  \BibitemOpen
  \bibfield  {author} {\bibinfo {author} {\bibfnamefont {S.}~\bibnamefont
  {Mühlbauer}}, \bibinfo {author} {\bibfnamefont {B.}~\bibnamefont {Binz}},
  \bibinfo {author} {\bibfnamefont {F.}~\bibnamefont {Jonietz}}, \bibinfo
  {author} {\bibfnamefont {C.}~\bibnamefont {Pfleiderer}}, \bibinfo {author}
  {\bibfnamefont {A.}~\bibnamefont {Rosch}}, \bibinfo {author} {\bibfnamefont
  {A.}~\bibnamefont {Neubauer}}, \bibinfo {author} {\bibfnamefont
  {R.}~\bibnamefont {Georgii}},\ and\ \bibinfo {author} {\bibfnamefont
  {P.}~\bibnamefont {Böni}},\ }\bibfield  {title} {\bibinfo {title}
  {{S}kyrmion lattice in a chiral magnet},\ }\href
  {https://doi.org/10.1126/science.1166767} {\bibfield  {journal} {\bibinfo
  {journal} {Science}\ }\textbf {\bibinfo {volume} {323}},\ \bibinfo {pages}
  {915} (\bibinfo {year} {2009})}\BibitemShut {NoStop}%
\bibitem [{\citenamefont {Nagaosa}\ and\ \citenamefont
  {Tokura}(2013)}]{Nagaosa2013}%
  \BibitemOpen
  \bibfield  {author} {\bibinfo {author} {\bibfnamefont {N.}~\bibnamefont
  {Nagaosa}}\ and\ \bibinfo {author} {\bibfnamefont {Y.}~\bibnamefont
  {Tokura}},\ }\bibfield  {title} {\bibinfo {title} {Topological properties and
  dynamics of magnetic {S}kyrmions},\ }\href
  {https://doi.org/10.1038/nnano.2013.243} {\bibfield  {journal} {\bibinfo
  {journal} {Nat. Nanotech.}\ }\textbf {\bibinfo {volume} {8}},\ \bibinfo
  {pages} {899} (\bibinfo {year} {2013})}\BibitemShut {NoStop}%
\bibitem [{\citenamefont {Kang}\ \emph {et~al.}(2019)\citenamefont {Kang},
  \citenamefont {Seo}, \citenamefont {Takeuchi},\ and\ \citenamefont
  {Shin}}]{Kang2019}%
  \BibitemOpen
  \bibfield  {author} {\bibinfo {author} {\bibfnamefont {S.}~\bibnamefont
  {Kang}}, \bibinfo {author} {\bibfnamefont {S.~W.}\ \bibnamefont {Seo}},
  \bibinfo {author} {\bibfnamefont {H.}~\bibnamefont {Takeuchi}},\ and\
  \bibinfo {author} {\bibfnamefont {Y.}~\bibnamefont {Shin}},\ }\bibfield
  {title} {\bibinfo {title} {Observation of wall-vortex composite defects in a
  spinor {B}ose-{E}instein condensate},\ }\href
  {https://doi.org/10.1103/PhysRevLett.122.095301} {\bibfield  {journal}
  {\bibinfo  {journal} {Phys. Rev. Lett.}\ }\textbf {\bibinfo {volume} {122}},\
  \bibinfo {pages} {095301} (\bibinfo {year} {2019})}\BibitemShut {NoStop}%
\bibitem [{\citenamefont {Takeuchi}(2021)}]{Takeuchi2021}%
  \BibitemOpen
  \bibfield  {author} {\bibinfo {author} {\bibfnamefont {H.}~\bibnamefont
  {Takeuchi}},\ }\bibfield  {title} {\bibinfo {title} {Quantum elliptic vortex
  in a nematic-spin {B}ose-{E}instein condensate},\ }\href
  {https://doi.org/10.1103/PhysRevLett.126.195302} {\bibfield  {journal}
  {\bibinfo  {journal} {Phys. Rev. Lett.}\ }\textbf {\bibinfo {volume} {126}},\
  \bibinfo {pages} {195302} (\bibinfo {year} {2021})}\BibitemShut {NoStop}%
\bibitem [{\citenamefont {Stoof}\ \emph {et~al.}(2001)\citenamefont {Stoof},
  \citenamefont {Vliegen},\ and\ \citenamefont {Al~Khawaja}}]{Stoof2001}%
  \BibitemOpen
  \bibfield  {author} {\bibinfo {author} {\bibfnamefont {H.~T.~C.}\
  \bibnamefont {Stoof}}, \bibinfo {author} {\bibfnamefont {E.}~\bibnamefont
  {Vliegen}},\ and\ \bibinfo {author} {\bibfnamefont {U.}~\bibnamefont
  {Al~Khawaja}},\ }\bibfield  {title} {\bibinfo {title} {Monopoles in an
  antiferromagnetic {B}ose-{E}instein condensate},\ }\href
  {https://doi.org/10.1103/PhysRevLett.87.120407} {\bibfield  {journal}
  {\bibinfo  {journal} {Phys. Rev. Lett.}\ }\textbf {\bibinfo {volume} {87}},\
  \bibinfo {pages} {120407} (\bibinfo {year} {2001})}\BibitemShut {NoStop}%
\bibitem [{\citenamefont {Savage}\ and\ \citenamefont
  {Ruostekoski}(2003{\natexlab{a}})}]{Savage2003b}%
  \BibitemOpen
  \bibfield  {author} {\bibinfo {author} {\bibfnamefont {C.~M.}\ \bibnamefont
  {Savage}}\ and\ \bibinfo {author} {\bibfnamefont {J.}~\bibnamefont
  {Ruostekoski}},\ }\bibfield  {title} {\bibinfo {title} {{D}irac monopoles and
  dipoles in ferromagnetic spinor {B}ose-{E}instein condensates},\ }\href
  {https://doi.org/10.1103/PhysRevA.68.043604} {\bibfield  {journal} {\bibinfo
  {journal} {Phys. Rev. A}\ }\textbf {\bibinfo {volume} {68}},\ \bibinfo
  {pages} {043604} (\bibinfo {year} {2003}{\natexlab{a}})}\BibitemShut
  {NoStop}%
\bibitem [{\citenamefont {Ruostekoski}\ and\ \citenamefont
  {Anglin}(2003)}]{Ruostekoski2003}%
  \BibitemOpen
  \bibfield  {author} {\bibinfo {author} {\bibfnamefont {J.}~\bibnamefont
  {Ruostekoski}}\ and\ \bibinfo {author} {\bibfnamefont {J.~R.}\ \bibnamefont
  {Anglin}},\ }\bibfield  {title} {\bibinfo {title} {Monopole core instability
  and {A}lice rings in spinor {B}ose-{E}instein condensates},\ }\href
  {https://doi.org/10.1103/PHYSREVLETT.91.190402} {\bibfield  {journal}
  {\bibinfo  {journal} {Phys. Rev. Lett.}\ }\textbf {\bibinfo {volume} {91}},\
  \bibinfo {pages} {190402} (\bibinfo {year} {2003})}\BibitemShut {NoStop}%
\bibitem [{\citenamefont {Pietil\"a}\ and\ \citenamefont
  {M\"ott\"onen}(2009)}]{Pietila2009}%
  \BibitemOpen
  \bibfield  {author} {\bibinfo {author} {\bibfnamefont {V.}~\bibnamefont
  {Pietil\"a}}\ and\ \bibinfo {author} {\bibfnamefont {M.}~\bibnamefont
  {M\"ott\"onen}},\ }\bibfield  {title} {\bibinfo {title} {Creation of {D}irac
  monopoles in spinor {B}ose-{E}instein condensates},\ }\href
  {https://doi.org/10.1103/PHYSREVLETT.103.030401} {\bibfield  {journal}
  {\bibinfo  {journal} {Phys. Rev. Lett.}\ }\textbf {\bibinfo {volume} {103}},\
  \bibinfo {pages} {030401} (\bibinfo {year} {2009})}\BibitemShut {NoStop}%
\bibitem [{\citenamefont {Ray}\ \emph {et~al.}(2014)\citenamefont {Ray},
  \citenamefont {Ruokokoski}, \citenamefont {Kandel}, \citenamefont
  {Möttönen},\ and\ \citenamefont {Hall}}]{Ray2014}%
  \BibitemOpen
  \bibfield  {author} {\bibinfo {author} {\bibfnamefont {M.~W.}\ \bibnamefont
  {Ray}}, \bibinfo {author} {\bibfnamefont {E.}~\bibnamefont {Ruokokoski}},
  \bibinfo {author} {\bibfnamefont {S.}~\bibnamefont {Kandel}}, \bibinfo
  {author} {\bibfnamefont {M.}~\bibnamefont {Möttönen}},\ and\ \bibinfo
  {author} {\bibfnamefont {D.~S.}\ \bibnamefont {Hall}},\ }\bibfield  {title}
  {\bibinfo {title} {Observation of {D}irac monopoles in a synthetic magnetic
  field},\ }\href {https://doi.org/10.1038/nature12954} {\bibfield  {journal}
  {\bibinfo  {journal} {Nature}\ }\textbf {\bibinfo {volume} {505}},\ \bibinfo
  {pages} {657} (\bibinfo {year} {2014})}\BibitemShut {NoStop}%
\bibitem [{\citenamefont {Ray}\ \emph {et~al.}(2015)\citenamefont {Ray},
  \citenamefont {Ruokokoski}, \citenamefont {Tiurev}, \citenamefont
  {Möttönen},\ and\ \citenamefont {Hall}}]{Ray2015}%
  \BibitemOpen
  \bibfield  {author} {\bibinfo {author} {\bibfnamefont {M.~W.}\ \bibnamefont
  {Ray}}, \bibinfo {author} {\bibfnamefont {E.}~\bibnamefont {Ruokokoski}},
  \bibinfo {author} {\bibfnamefont {K.}~\bibnamefont {Tiurev}}, \bibinfo
  {author} {\bibfnamefont {M.}~\bibnamefont {Möttönen}},\ and\ \bibinfo
  {author} {\bibfnamefont {D.~S.}\ \bibnamefont {Hall}},\ }\bibfield  {title}
  {\bibinfo {title} {Observation of isolated monopoles in a quantum field},\
  }\href {https://doi.org/10.1126/science.1258289} {\bibfield  {journal}
  {\bibinfo  {journal} {Science}\ }\textbf {\bibinfo {volume} {348}},\ \bibinfo
  {pages} {544} (\bibinfo {year} {2015})}\BibitemShut {NoStop}%
\bibitem [{\citenamefont {Ollikainen}\ \emph {et~al.}(2017)\citenamefont
  {Ollikainen}, \citenamefont {Tiurev}, \citenamefont {Blinova}, \citenamefont
  {Lee}, \citenamefont {Hall},\ and\ \citenamefont
  {M\"ott\"onen}}]{Ollikainen2017}%
  \BibitemOpen
  \bibfield  {author} {\bibinfo {author} {\bibfnamefont {T.}~\bibnamefont
  {Ollikainen}}, \bibinfo {author} {\bibfnamefont {K.}~\bibnamefont {Tiurev}},
  \bibinfo {author} {\bibfnamefont {A.}~\bibnamefont {Blinova}}, \bibinfo
  {author} {\bibfnamefont {W.}~\bibnamefont {Lee}}, \bibinfo {author}
  {\bibfnamefont {D.~S.}\ \bibnamefont {Hall}},\ and\ \bibinfo {author}
  {\bibfnamefont {M.}~\bibnamefont {M\"ott\"onen}},\ }\bibfield  {title}
  {\bibinfo {title} {Experimental realization of a {D}irac monopole through the
  decay of an isolated monopole},\ }\href
  {https://doi.org/10.1103/PHYSREVX.7.021023} {\bibfield  {journal} {\bibinfo
  {journal} {Phys. Rev. X}\ }\textbf {\bibinfo {volume} {7}},\ \bibinfo {pages}
  {021023} (\bibinfo {year} {2017})}\BibitemShut {NoStop}%
\bibitem [{\citenamefont {Mithun}\ \emph {et~al.}(2022)\citenamefont {Mithun},
  \citenamefont {Carretero-Gonz\'alez}, \citenamefont {Charalampidis},
  \citenamefont {Hall},\ and\ \citenamefont {Kevrekidis}}]{Mithun2022}%
  \BibitemOpen
  \bibfield  {author} {\bibinfo {author} {\bibfnamefont {T.}~\bibnamefont
  {Mithun}}, \bibinfo {author} {\bibfnamefont {R.}~\bibnamefont
  {Carretero-Gonz\'alez}}, \bibinfo {author} {\bibfnamefont {E.~G.}\
  \bibnamefont {Charalampidis}}, \bibinfo {author} {\bibfnamefont {D.~S.}\
  \bibnamefont {Hall}},\ and\ \bibinfo {author} {\bibfnamefont {P.~G.}\
  \bibnamefont {Kevrekidis}},\ }\bibfield  {title} {\bibinfo {title}
  {Existence, stability, and dynamics of monopole and {A}lice ring solutions in
  antiferromagnetic spinor condensates},\ }\href
  {https://doi.org/10.1103/PhysRevA.105.053303} {\bibfield  {journal} {\bibinfo
   {journal} {Phys. Rev. A}\ }\textbf {\bibinfo {volume} {105}},\ \bibinfo
  {pages} {053303} (\bibinfo {year} {2022})}\BibitemShut {NoStop}%
\bibitem [{\citenamefont {Blinova}\ \emph {et~al.}(2023)\citenamefont
  {Blinova}, \citenamefont {Zamora-Zamora}, \citenamefont {Ollikainen},
  \citenamefont {Kivioja}, \citenamefont {M{\"o}tt{\"o}nen},\ and\
  \citenamefont {Hall}}]{Blinova2023}%
  \BibitemOpen
  \bibfield  {author} {\bibinfo {author} {\bibfnamefont {A.}~\bibnamefont
  {Blinova}}, \bibinfo {author} {\bibfnamefont {R.}~\bibnamefont
  {Zamora-Zamora}}, \bibinfo {author} {\bibfnamefont {T.}~\bibnamefont
  {Ollikainen}}, \bibinfo {author} {\bibfnamefont {M.}~\bibnamefont {Kivioja}},
  \bibinfo {author} {\bibfnamefont {M.}~\bibnamefont {M{\"o}tt{\"o}nen}},\ and\
  \bibinfo {author} {\bibfnamefont {D.~S.}\ \bibnamefont {Hall}},\ }\bibfield
  {title} {\bibinfo {title} {Observation of an {A}lice ring in a
  {B}ose--{E}instein condensate},\ }\href
  {https://doi.org/10.1038/s41467-023-40710-2} {\bibfield  {journal} {\bibinfo
  {journal} {Nat. Commun.}\ }\textbf {\bibinfo {volume} {14}},\ \bibinfo
  {pages} {5100} (\bibinfo {year} {2023})}\BibitemShut {NoStop}%
\bibitem [{\citenamefont {Al~Khawaja}\ and\ \citenamefont
  {Stoof}(2001)}]{Al_Khawaja2001}%
  \BibitemOpen
  \bibfield  {author} {\bibinfo {author} {\bibfnamefont {U.}~\bibnamefont
  {Al~Khawaja}}\ and\ \bibinfo {author} {\bibfnamefont {H.}~\bibnamefont
  {Stoof}},\ }\bibfield  {title} {\bibinfo {title} {{S}kyrmions in a
  ferromagnetic {B}ose-{E}instein condensate},\ }\href
  {https://doi.org/10.1038/35082010} {\bibfield  {journal} {\bibinfo  {journal}
  {Nature}\ }\textbf {\bibinfo {volume} {411}},\ \bibinfo {pages} {918}
  (\bibinfo {year} {2001})}\BibitemShut {NoStop}%
\bibitem [{\citenamefont {Ruostekoski}\ and\ \citenamefont
  {Anglin}(2001)}]{Ruostekoski2001}%
  \BibitemOpen
  \bibfield  {author} {\bibinfo {author} {\bibfnamefont {J.}~\bibnamefont
  {Ruostekoski}}\ and\ \bibinfo {author} {\bibfnamefont {J.~R.}\ \bibnamefont
  {Anglin}},\ }\bibfield  {title} {\bibinfo {title} {Creating vortex rings and
  three-dimensional {S}kyrmions in {B}ose-{E}instein condensates},\ }\href
  {https://doi.org/10.1103/PhysRevLett.86.3934} {\bibfield  {journal} {\bibinfo
   {journal} {Phys. Rev. Lett.}\ }\textbf {\bibinfo {volume} {86}},\ \bibinfo
  {pages} {3934} (\bibinfo {year} {2001})}\BibitemShut {NoStop}%
\bibitem [{\citenamefont {Battye}\ \emph {et~al.}(2002)\citenamefont {Battye},
  \citenamefont {Cooper},\ and\ \citenamefont {Sutcliffe}}]{Battye2002}%
  \BibitemOpen
  \bibfield  {author} {\bibinfo {author} {\bibfnamefont {R.~A.}\ \bibnamefont
  {Battye}}, \bibinfo {author} {\bibfnamefont {N.~R.}\ \bibnamefont {Cooper}},\
  and\ \bibinfo {author} {\bibfnamefont {P.~M.}\ \bibnamefont {Sutcliffe}},\
  }\bibfield  {title} {\bibinfo {title} {Stable {S}kyrmions in two-component
  {B}ose-{E}instein condensates},\ }\href
  {https://doi.org/10.1103/PhysRevLett.88.080401} {\bibfield  {journal}
  {\bibinfo  {journal} {Phys. Rev. Lett.}\ }\textbf {\bibinfo {volume} {88}},\
  \bibinfo {pages} {080401} (\bibinfo {year} {2002})}\BibitemShut {NoStop}%
\bibitem [{\citenamefont {Savage}\ and\ \citenamefont
  {Ruostekoski}(2003{\natexlab{b}})}]{Savage2003a}%
  \BibitemOpen
  \bibfield  {author} {\bibinfo {author} {\bibfnamefont {C.~M.}\ \bibnamefont
  {Savage}}\ and\ \bibinfo {author} {\bibfnamefont {J.}~\bibnamefont
  {Ruostekoski}},\ }\bibfield  {title} {\bibinfo {title} {Energetically stable
  particlelike {S}kyrmions in a trapped {B}ose-{E}instein condensate},\ }\href
  {https://doi.org/10.1103/PhysRevLett.91.010403} {\bibfield  {journal}
  {\bibinfo  {journal} {Phys. Rev. Lett.}\ }\textbf {\bibinfo {volume} {91}},\
  \bibinfo {pages} {010403} (\bibinfo {year} {2003}{\natexlab{b}})}\BibitemShut
  {NoStop}%
\bibitem [{\citenamefont {Ruostekoski}(2004)}]{Ruostekoski2004}%
  \BibitemOpen
  \bibfield  {author} {\bibinfo {author} {\bibfnamefont {J.}~\bibnamefont
  {Ruostekoski}},\ }\bibfield  {title} {\bibinfo {title} {Stable particlelike
  solitons with multiply quantized vortex lines in {B}ose-{E}instein
  condensates},\ }\href {https://doi.org/10.1103/PhysRevA.70.041601} {\bibfield
   {journal} {\bibinfo  {journal} {Phys. Rev. A}\ }\textbf {\bibinfo {volume}
  {70}},\ \bibinfo {pages} {041601(R)} (\bibinfo {year} {2004})}\BibitemShut
  {NoStop}%
\bibitem [{\citenamefont {Kawakami}\ \emph {et~al.}(2012)\citenamefont
  {Kawakami}, \citenamefont {Mizushima}, \citenamefont {Nitta},\ and\
  \citenamefont {Machida}}]{Kawakami2012}%
  \BibitemOpen
  \bibfield  {author} {\bibinfo {author} {\bibfnamefont {T.}~\bibnamefont
  {Kawakami}}, \bibinfo {author} {\bibfnamefont {T.}~\bibnamefont {Mizushima}},
  \bibinfo {author} {\bibfnamefont {M.}~\bibnamefont {Nitta}},\ and\ \bibinfo
  {author} {\bibfnamefont {K.}~\bibnamefont {Machida}},\ }\bibfield  {title}
  {\bibinfo {title} {Stable {S}kyrmions in {$\text{SU}(2)$} gauged
  {B}ose-{E}instein condensates},\ }\href
  {https://doi.org/10.1103/PhysRevLett.109.015301} {\bibfield  {journal}
  {\bibinfo  {journal} {Phys. Rev. Lett.}\ }\textbf {\bibinfo {volume} {109}},\
  \bibinfo {pages} {015301} (\bibinfo {year} {2012})}\BibitemShut {NoStop}%
\bibitem [{\citenamefont {Tiurev}\ \emph {et~al.}(2018)\citenamefont {Tiurev},
  \citenamefont {Ollikainen}, \citenamefont {Kuopanportti}, \citenamefont
  {Nakahara}, \citenamefont {Hall},\ and\ \citenamefont
  {Möttönen}}]{Tiurev2018}%
  \BibitemOpen
  \bibfield  {author} {\bibinfo {author} {\bibfnamefont {K.}~\bibnamefont
  {Tiurev}}, \bibinfo {author} {\bibfnamefont {T.}~\bibnamefont {Ollikainen}},
  \bibinfo {author} {\bibfnamefont {P.}~\bibnamefont {Kuopanportti}}, \bibinfo
  {author} {\bibfnamefont {M.}~\bibnamefont {Nakahara}}, \bibinfo {author}
  {\bibfnamefont {D.~S.}\ \bibnamefont {Hall}},\ and\ \bibinfo {author}
  {\bibfnamefont {M.}~\bibnamefont {Möttönen}},\ }\bibfield  {title}
  {\bibinfo {title} {Three-dimensional {S}kyrmions in spin-2
  {B}ose–{E}instein condensates},\ }\href
  {https://doi.org/10.1088/1367-2630/AAC2A8} {\bibfield  {journal} {\bibinfo
  {journal} {New. J. Phys.}\ }\textbf {\bibinfo {volume} {20}},\ \bibinfo
  {pages} {055011} (\bibinfo {year} {2018})}\BibitemShut {NoStop}%
\bibitem [{\citenamefont {Lee}\ \emph {et~al.}(2018)\citenamefont {Lee},
  \citenamefont {Gheorghe}, \citenamefont {Tiurev}, \citenamefont {Ollikainen},
  \citenamefont {M\"ott\"onen},\ and\ \citenamefont {Hall}}]{Lee2018}%
  \BibitemOpen
  \bibfield  {author} {\bibinfo {author} {\bibfnamefont {W.}~\bibnamefont
  {Lee}}, \bibinfo {author} {\bibfnamefont {A.~H.}\ \bibnamefont {Gheorghe}},
  \bibinfo {author} {\bibfnamefont {K.}~\bibnamefont {Tiurev}}, \bibinfo
  {author} {\bibfnamefont {T.}~\bibnamefont {Ollikainen}}, \bibinfo {author}
  {\bibfnamefont {M.}~\bibnamefont {M\"ott\"onen}},\ and\ \bibinfo {author}
  {\bibfnamefont {D.~S.}\ \bibnamefont {Hall}},\ }\bibfield  {title} {\bibinfo
  {title} {Synthetic electromagnetic knot in a three-dimensional {S}kyrmion},\
  }\href {https://doi.org/10.1126/sciadv.aao3820} {\bibfield  {journal}
  {\bibinfo  {journal} {Sci. Adv.}\ }\textbf {\bibinfo {volume} {4}},\ \bibinfo
  {pages} {eaao3820} (\bibinfo {year} {2018})}\BibitemShut {NoStop}%
\bibitem [{\citenamefont {Skyrme}(1961)}]{Skyrme1961}%
  \BibitemOpen
  \bibfield  {author} {\bibinfo {author} {\bibfnamefont {T.~H.~R.}\
  \bibnamefont {Skyrme}},\ }\bibfield  {title} {\bibinfo {title} {A non-linear
  field theory},\ }\href {https://doi.org/10.1098/rspa.1961.0018} {\bibfield
  {journal} {\bibinfo  {journal} {Proc. R. Soc. Lond. A}\ }\textbf {\bibinfo
  {volume} {260}},\ \bibinfo {pages} {127} (\bibinfo {year}
  {1961})}\BibitemShut {NoStop}%
\bibitem [{\citenamefont {Kawaguchi}\ \emph {et~al.}(2008)\citenamefont
  {Kawaguchi}, \citenamefont {Nitta},\ and\ \citenamefont
  {Ueda}}]{Kawaguchi2008}%
  \BibitemOpen
  \bibfield  {author} {\bibinfo {author} {\bibfnamefont {Y.}~\bibnamefont
  {Kawaguchi}}, \bibinfo {author} {\bibfnamefont {M.}~\bibnamefont {Nitta}},\
  and\ \bibinfo {author} {\bibfnamefont {M.}~\bibnamefont {Ueda}},\ }\bibfield
  {title} {\bibinfo {title} {Knots in a spinor {B}ose-{E}instein condensate},\
  }\href {https://doi.org/10.1103/PhysRevLett.100.180403} {\bibfield  {journal}
  {\bibinfo  {journal} {Phys. Rev. Lett.}\ }\textbf {\bibinfo {volume} {100}},\
  \bibinfo {pages} {180403} (\bibinfo {year} {2008})}\BibitemShut {NoStop}%
\bibitem [{\citenamefont {Hall}\ \emph {et~al.}(2016)\citenamefont {Hall},
  \citenamefont {Ray}, \citenamefont {Tiurev}, \citenamefont {Ruokokoski},
  \citenamefont {Gheorghe},\ and\ \citenamefont {M{\"o}tt{\"o}nen}}]{Hall2016}%
  \BibitemOpen
  \bibfield  {author} {\bibinfo {author} {\bibfnamefont {D.~S.}\ \bibnamefont
  {Hall}}, \bibinfo {author} {\bibfnamefont {M.~W.}\ \bibnamefont {Ray}},
  \bibinfo {author} {\bibfnamefont {K.}~\bibnamefont {Tiurev}}, \bibinfo
  {author} {\bibfnamefont {E.}~\bibnamefont {Ruokokoski}}, \bibinfo {author}
  {\bibfnamefont {A.~H.}\ \bibnamefont {Gheorghe}},\ and\ \bibinfo {author}
  {\bibfnamefont {M.}~\bibnamefont {M{\"o}tt{\"o}nen}},\ }\bibfield  {title}
  {\bibinfo {title} {Tying quantum knots},\ }\href
  {https://doi.org/10.1038/nphys3624} {\bibfield  {journal} {\bibinfo
  {journal} {Nat. Phys.}\ }\textbf {\bibinfo {volume} {12}},\ \bibinfo {pages}
  {478} (\bibinfo {year} {2016})}\BibitemShut {NoStop}%
\bibitem [{\citenamefont {Faddeev}\ and\ \citenamefont
  {Niemi}(1997)}]{Faddeev1997}%
  \BibitemOpen
  \bibfield  {author} {\bibinfo {author} {\bibfnamefont {L.}~\bibnamefont
  {Faddeev}}\ and\ \bibinfo {author} {\bibfnamefont {A.~J.}\ \bibnamefont
  {Niemi}},\ }\bibfield  {title} {\bibinfo {title} {Stable knot-like structures
  in classical field theory},\ }\href {https://doi.org/10.1038/387058a0}
  {\bibfield  {journal} {\bibinfo  {journal} {Nature}\ }\textbf {\bibinfo
  {volume} {387}},\ \bibinfo {pages} {58} (\bibinfo {year} {1997})}\BibitemShut
  {NoStop}%
\bibitem [{\citenamefont {Battye}\ and\ \citenamefont
  {Sutcliffe}(1998)}]{Battye1998}%
  \BibitemOpen
  \bibfield  {author} {\bibinfo {author} {\bibfnamefont {R.~A.}\ \bibnamefont
  {Battye}}\ and\ \bibinfo {author} {\bibfnamefont {P.~M.}\ \bibnamefont
  {Sutcliffe}},\ }\bibfield  {title} {\bibinfo {title} {Knots as stable soliton
  solutions in a three-dimensional classical field theory},\ }\href
  {https://doi.org/10.1103/PhysRevLett.81.4798} {\bibfield  {journal} {\bibinfo
   {journal} {Phys. Rev. Lett.}\ }\textbf {\bibinfo {volume} {81}},\ \bibinfo
  {pages} {4798} (\bibinfo {year} {1998})}\BibitemShut {NoStop}%
\bibitem [{\citenamefont {Hietarinta}\ and\ \citenamefont
  {Salo}(1999)}]{Hietarinta1999}%
  \BibitemOpen
  \bibfield  {author} {\bibinfo {author} {\bibfnamefont {J.}~\bibnamefont
  {Hietarinta}}\ and\ \bibinfo {author} {\bibfnamefont {P.}~\bibnamefont
  {Salo}},\ }\bibfield  {title} {\bibinfo {title} {Faddeev-hopf knots: dynamics
  of linked un-knots},\ }\href
  {https://doi.org/https://doi.org/10.1016/S0370-2693(99)00054-4} {\bibfield
  {journal} {\bibinfo  {journal} {Phys. Lett. B}\ }\textbf {\bibinfo {volume}
  {451}},\ \bibinfo {pages} {60} (\bibinfo {year} {1999})}\BibitemShut
  {NoStop}%
\bibitem [{\citenamefont {Sutcliffe}(2017)}]{Sutcliffe2017}%
  \BibitemOpen
  \bibfield  {author} {\bibinfo {author} {\bibfnamefont {P.}~\bibnamefont
  {Sutcliffe}},\ }\bibfield  {title} {\bibinfo {title} {{S}kyrmion knots in
  frustrated magnets},\ }\href {https://doi.org/10.1103/PhysRevLett.118.247203}
  {\bibfield  {journal} {\bibinfo  {journal} {Phys. Rev. Lett.}\ }\textbf
  {\bibinfo {volume} {118}},\ \bibinfo {pages} {247203} (\bibinfo {year}
  {2017})}\BibitemShut {NoStop}%
\bibitem [{\citenamefont {Parmee}\ \emph {et~al.}(2022)\citenamefont {Parmee},
  \citenamefont {Dennis},\ and\ \citenamefont {Ruostekoski}}]{Parmee2022}%
  \BibitemOpen
  \bibfield  {author} {\bibinfo {author} {\bibfnamefont {C.~D.}\ \bibnamefont
  {Parmee}}, \bibinfo {author} {\bibfnamefont {M.~R.}\ \bibnamefont {Dennis}},\
  and\ \bibinfo {author} {\bibfnamefont {J.}~\bibnamefont {Ruostekoski}},\
  }\bibfield  {title} {\bibinfo {title} {Optical excitations of {S}kyrmions,
  knotted solitons, and defects in atoms},\ }\href
  {https://doi.org/10.1038/s42005-022-00829-y} {\bibfield  {journal} {\bibinfo
  {journal} {Commun. Phys.}\ }\textbf {\bibinfo {volume} {5}},\ \bibinfo
  {pages} {54} (\bibinfo {year} {2022})}\BibitemShut {NoStop}%
\bibitem [{\citenamefont {Blaauwgeers}\ \emph {et~al.}(2002)\citenamefont
  {Blaauwgeers}, \citenamefont {Eltsov}, \citenamefont {Eska}, \citenamefont
  {Finne}, \citenamefont {Haley}, \citenamefont {Krusius}, \citenamefont
  {Ruohio}, \citenamefont {Skrbek},\ and\ \citenamefont
  {Volovik}}]{Blaauwgeers2002}%
  \BibitemOpen
  \bibfield  {author} {\bibinfo {author} {\bibfnamefont {R.}~\bibnamefont
  {Blaauwgeers}}, \bibinfo {author} {\bibfnamefont {V.~B.}\ \bibnamefont
  {Eltsov}}, \bibinfo {author} {\bibfnamefont {G.}~\bibnamefont {Eska}},
  \bibinfo {author} {\bibfnamefont {A.~P.}\ \bibnamefont {Finne}}, \bibinfo
  {author} {\bibfnamefont {R.~P.}\ \bibnamefont {Haley}}, \bibinfo {author}
  {\bibfnamefont {M.}~\bibnamefont {Krusius}}, \bibinfo {author} {\bibfnamefont
  {J.~J.}\ \bibnamefont {Ruohio}}, \bibinfo {author} {\bibfnamefont
  {L.}~\bibnamefont {Skrbek}},\ and\ \bibinfo {author} {\bibfnamefont {G.~E.}\
  \bibnamefont {Volovik}},\ }\bibfield  {title} {\bibinfo {title} {{Shear Flow
  and Kelvin-Helmholtz Instability in Superfluids}},\ }\href
  {https://doi.org/10.1103/PhysRevLett.89.155301} {\bibfield  {journal}
  {\bibinfo  {journal} {Phys. Rev. Lett.}\ }\textbf {\bibinfo {volume} {89}},\
  \bibinfo {pages} {155301} (\bibinfo {year} {2002})}\BibitemShut {NoStop}%
\bibitem [{\citenamefont {Ueda}\ and\ \citenamefont {Koashi}(2002)}]{Ueda2002}%
  \BibitemOpen
  \bibfield  {author} {\bibinfo {author} {\bibfnamefont {M.}~\bibnamefont
  {Ueda}}\ and\ \bibinfo {author} {\bibfnamefont {M.}~\bibnamefont {Koashi}},\
  }\bibfield  {title} {\bibinfo {title} {Theory of spin-2 {{B}ose-{E}instein}
  condensates: Spin correlations, magnetic response, and excitation spectra},\
  }\href {https://journals.aps.org/pra/abstract/10.1103/PhysRevA.65.063602}
  {\bibfield  {journal} {\bibinfo  {journal} {Phys. Rev. A}\ }\textbf {\bibinfo
  {volume} {65}},\ \bibinfo {pages} {063602} (\bibinfo {year}
  {2002})}\BibitemShut {NoStop}%
\bibitem [{\citenamefont {Corney}(1978)}]{corney1978atomic}%
  \BibitemOpen
  \bibfield  {author} {\bibinfo {author} {\bibfnamefont {A.}~\bibnamefont
  {Corney}},\ }\href@noop {} {\emph {\bibinfo {title} {Atomic and laser
  spectroscopy}}}\ (\bibinfo  {publisher} {Clarendon Press, Oxford},\ \bibinfo
  {year} {1978})\BibitemShut {NoStop}%
\bibitem [{\citenamefont {Gerbier}\ \emph {et~al.}(2006)\citenamefont
  {Gerbier}, \citenamefont {Widera}, \citenamefont {F\"olling}, \citenamefont
  {Mandel},\ and\ \citenamefont {Bloch}}]{Gerbier2006}%
  \BibitemOpen
  \bibfield  {author} {\bibinfo {author} {\bibfnamefont {F.}~\bibnamefont
  {Gerbier}}, \bibinfo {author} {\bibfnamefont {A.}~\bibnamefont {Widera}},
  \bibinfo {author} {\bibfnamefont {S.}~\bibnamefont {F\"olling}}, \bibinfo
  {author} {\bibfnamefont {O.}~\bibnamefont {Mandel}},\ and\ \bibinfo {author}
  {\bibfnamefont {I.}~\bibnamefont {Bloch}},\ }\bibfield  {title} {\bibinfo
  {title} {Resonant control of spin dynamics in ultracold quantum gases by
  microwave dressing},\ }\href {https://doi.org/10.1103/PhysRevA.73.041602}
  {\bibfield  {journal} {\bibinfo  {journal} {Phys. Rev. A}\ }\textbf {\bibinfo
  {volume} {73}},\ \bibinfo {pages} {041602(R)} (\bibinfo {year}
  {2006})}\BibitemShut {NoStop}%
\bibitem [{\citenamefont {Santos}\ \emph {et~al.}(2007)\citenamefont {Santos},
  \citenamefont {Fattori}, \citenamefont {Stuhler},\ and\ \citenamefont
  {Pfau}}]{Santos2007}%
  \BibitemOpen
  \bibfield  {author} {\bibinfo {author} {\bibfnamefont {L.}~\bibnamefont
  {Santos}}, \bibinfo {author} {\bibfnamefont {M.}~\bibnamefont {Fattori}},
  \bibinfo {author} {\bibfnamefont {J.}~\bibnamefont {Stuhler}},\ and\ \bibinfo
  {author} {\bibfnamefont {T.}~\bibnamefont {Pfau}},\ }\bibfield  {title}
  {\bibinfo {title} {Spinor condensates with a laser-induced quadratic {Z}eeman
  effect},\ }\href {https://doi.org/10.1103/PhysRevA.75.053606} {\bibfield
  {journal} {\bibinfo  {journal} {Phys. Rev. A}\ }\textbf {\bibinfo {volume}
  {75}},\ \bibinfo {pages} {053606} (\bibinfo {year} {2007})}\BibitemShut
  {NoStop}%
\bibitem [{\citenamefont {Ciobanu}\ \emph {et~al.}(2000)\citenamefont
  {Ciobanu}, \citenamefont {Yip},\ and\ \citenamefont {Ho}}]{Ciobanu2000}%
  \BibitemOpen
  \bibfield  {author} {\bibinfo {author} {\bibfnamefont {C.~V.}\ \bibnamefont
  {Ciobanu}}, \bibinfo {author} {\bibfnamefont {S.-K.}\ \bibnamefont {Yip}},\
  and\ \bibinfo {author} {\bibfnamefont {T.-L.}\ \bibnamefont {Ho}},\
  }\bibfield  {title} {\bibinfo {title} {Phase diagrams of {F}=2 spinor
  {B}ose-{E}instein condensates},\ }\href
  {https://doi.org/10.1103/PhysRevA.61.033607} {\bibfield  {journal} {\bibinfo
  {journal} {Phys. Rev. A}\ }\textbf {\bibinfo {volume} {61}},\ \bibinfo
  {pages} {033607} (\bibinfo {year} {2000})}\BibitemShut {NoStop}%
\bibitem [{\citenamefont {M\"{a}kel\"{a}}\ \emph {et~al.}(2003)\citenamefont
  {M\"{a}kel\"{a}}, \citenamefont {Zhang},\ and\ \citenamefont
  {Suominen}}]{Makela2003}%
  \BibitemOpen
  \bibfield  {author} {\bibinfo {author} {\bibfnamefont {H.}~\bibnamefont
  {M\"{a}kel\"{a}}}, \bibinfo {author} {\bibfnamefont {Y.}~\bibnamefont
  {Zhang}},\ and\ \bibinfo {author} {\bibfnamefont {K.-A.}\ \bibnamefont
  {Suominen}},\ }\bibfield  {title} {\bibinfo {title} {Topological defects in
  spinor condensates},\ }\href {https://doi.org/10.1088/0305-4470/36/32/302}
  {\bibfield  {journal} {\bibinfo  {journal} {J. Phys. A: Math. Gen.}\ }\textbf
  {\bibinfo {volume} {36}},\ \bibinfo {pages} {8555} (\bibinfo {year}
  {2003})}\BibitemShut {NoStop}%
\bibitem [{\citenamefont {Kobayashi}\ \emph {et~al.}(2012)\citenamefont
  {Kobayashi}, \citenamefont {Kobayashi}, \citenamefont {Kawaguchi},
  \citenamefont {Nitta},\ and\ \citenamefont {Ueda}}]{Kobayashi2012}%
  \BibitemOpen
  \bibfield  {author} {\bibinfo {author} {\bibfnamefont {S.}~\bibnamefont
  {Kobayashi}}, \bibinfo {author} {\bibfnamefont {M.}~\bibnamefont
  {Kobayashi}}, \bibinfo {author} {\bibfnamefont {Y.}~\bibnamefont
  {Kawaguchi}}, \bibinfo {author} {\bibfnamefont {M.}~\bibnamefont {Nitta}},\
  and\ \bibinfo {author} {\bibfnamefont {M.}~\bibnamefont {Ueda}},\ }\bibfield
  {title} {\bibinfo {title} {{Abe homotopy classification of topological
  excitations under the topological influence of vortices}},\ }\href
  {https://doi.org/https://doi.org/10.1016/j.nuclphysb.2011.11.003} {\bibfield
  {journal} {\bibinfo  {journal} {Nucl. Phys. B}\ }\textbf {\bibinfo {volume}
  {856}},\ \bibinfo {pages} {577} (\bibinfo {year} {2012})}\BibitemShut
  {NoStop}%
\bibitem [{\citenamefont {Song}\ \emph {et~al.}(2007)\citenamefont {Song},
  \citenamefont {Semenoff},\ and\ \citenamefont {Zhou}}]{Song2007}%
  \BibitemOpen
  \bibfield  {author} {\bibinfo {author} {\bibfnamefont {J.~L.}\ \bibnamefont
  {Song}}, \bibinfo {author} {\bibfnamefont {G.~W.}\ \bibnamefont {Semenoff}},\
  and\ \bibinfo {author} {\bibfnamefont {F.}~\bibnamefont {Zhou}},\ }\bibfield
  {title} {\bibinfo {title} {{Uniaxial and Biaxial Spin Nematic Phases Induced
  by Quantum Fluctuations}},\ }\href
  {https://doi.org/10.1103/PhysRevLett.98.160408} {\bibfield  {journal}
  {\bibinfo  {journal} {Phys. Rev. Lett.}\ }\textbf {\bibinfo {volume} {98}},\
  \bibinfo {pages} {160408} (\bibinfo {year} {2007})}\BibitemShut {NoStop}%
\bibitem [{\citenamefont {Klausen}\ \emph {et~al.}(2001)\citenamefont
  {Klausen}, \citenamefont {Bohn},\ and\ \citenamefont {Greene}}]{Klausen2001}%
  \BibitemOpen
  \bibfield  {author} {\bibinfo {author} {\bibfnamefont {N.~N.}\ \bibnamefont
  {Klausen}}, \bibinfo {author} {\bibfnamefont {J.~L.}\ \bibnamefont {Bohn}},\
  and\ \bibinfo {author} {\bibfnamefont {C.~H.}\ \bibnamefont {Greene}},\
  }\bibfield  {title} {\bibinfo {title} {Nature of spinor {B}ose-{E}instein
  condensates in rubidium},\ }\href
  {https://doi.org/10.1103/PhysRevA.64.053602} {\bibfield  {journal} {\bibinfo
  {journal} {Phys. Rev. A}\ }\textbf {\bibinfo {volume} {64}},\ \bibinfo
  {pages} {053602} (\bibinfo {year} {2001})}\BibitemShut {NoStop}%
\bibitem [{\citenamefont {Widera}\ \emph {et~al.}(2006)\citenamefont {Widera},
  \citenamefont {Gerbier}, \citenamefont {Fölling}, \citenamefont {Gericke},
  \citenamefont {Mandel},\ and\ \citenamefont {Bloch}}]{Widera2006}%
  \BibitemOpen
  \bibfield  {author} {\bibinfo {author} {\bibfnamefont {A.}~\bibnamefont
  {Widera}}, \bibinfo {author} {\bibfnamefont {F.}~\bibnamefont {Gerbier}},
  \bibinfo {author} {\bibfnamefont {S.}~\bibnamefont {Fölling}}, \bibinfo
  {author} {\bibfnamefont {T.}~\bibnamefont {Gericke}}, \bibinfo {author}
  {\bibfnamefont {O.}~\bibnamefont {Mandel}},\ and\ \bibinfo {author}
  {\bibfnamefont {I.}~\bibnamefont {Bloch}},\ }\bibfield  {title} {\bibinfo
  {title} {Precision measurement of spin-dependent interaction strengths for
  spin-1 and spin-2 $^{87}\text{Rb}$ atoms},\ }\href
  {https://doi.org/10.1088/1367-2630/8/8/152} {\bibfield  {journal} {\bibinfo
  {journal} {New J. Phys.}\ }\textbf {\bibinfo {volume} {8}},\ \bibinfo {pages}
  {152} (\bibinfo {year} {2006})}\BibitemShut {NoStop}%
\bibitem [{\citenamefont {Turner}\ \emph {et~al.}(2007)\citenamefont {Turner},
  \citenamefont {Barnett}, \citenamefont {Demler},\ and\ \citenamefont
  {Vishwanath}}]{Turner2007}%
  \BibitemOpen
  \bibfield  {author} {\bibinfo {author} {\bibfnamefont {A.~M.}\ \bibnamefont
  {Turner}}, \bibinfo {author} {\bibfnamefont {R.}~\bibnamefont {Barnett}},
  \bibinfo {author} {\bibfnamefont {E.}~\bibnamefont {Demler}},\ and\ \bibinfo
  {author} {\bibfnamefont {A.}~\bibnamefont {Vishwanath}},\ }\bibfield  {title}
  {\bibinfo {title} {Nematic order by disorder in spin-2 {B}ose-{E}instein
  condensates},\ }\href {https://doi.org/10.1103/PHYSREVLETT.98.190404}
  {\bibfield  {journal} {\bibinfo  {journal} {Phys. Rev. Lett.}\ }\textbf
  {\bibinfo {volume} {98}},\ \bibinfo {pages} {190404} (\bibinfo {year}
  {2007})}\BibitemShut {NoStop}%
\bibitem [{\citenamefont {Koashi}\ and\ \citenamefont
  {Ueda}(2000)}]{koashi_prl_2000}%
  \BibitemOpen
  \bibfield  {author} {\bibinfo {author} {\bibfnamefont {M.}~\bibnamefont
  {Koashi}}\ and\ \bibinfo {author} {\bibfnamefont {M.}~\bibnamefont {Ueda}},\
  }\bibfield  {title} {\bibinfo {title} {Exact eigenstates and magnetic
  response of spin-1 and spin-2 {Bose-Einstein} condensates},\ }\href
  {https://doi.org/https://doi.org/10.1103/PhysRevLett.84.1066} {\bibfield
  {journal} {\bibinfo  {journal} {Phys. Rev. Lett.}\ }\textbf {\bibinfo
  {volume} {84}},\ \bibinfo {pages} {1066} (\bibinfo {year}
  {2000})}\BibitemShut {NoStop}%
\bibitem [{\citenamefont {Zhang}\ \emph {et~al.}(2003)\citenamefont {Zhang},
  \citenamefont {Yi},\ and\ \citenamefont {You}}]{zhang_njp_2003}%
  \BibitemOpen
  \bibfield  {author} {\bibinfo {author} {\bibfnamefont {W.}~\bibnamefont
  {Zhang}}, \bibinfo {author} {\bibfnamefont {S.}~\bibnamefont {Yi}},\ and\
  \bibinfo {author} {\bibfnamefont {L.}~\bibnamefont {You}},\ }\bibfield
  {title} {\bibinfo {title} {Mean field ground state of a spin-1 condensate in
  a magnetic field},\ }\href {http://stacks.iop.org/1367-2630/5/i=1/a=377}
  {\bibfield  {journal} {\bibinfo  {journal} {New J. Phys.}\ }\textbf {\bibinfo
  {volume} {5}},\ \bibinfo {pages} {77} (\bibinfo {year} {2003})}\BibitemShut
  {NoStop}%
\bibitem [{\citenamefont {Murata}\ \emph {et~al.}(2007)\citenamefont {Murata},
  \citenamefont {Saito},\ and\ \citenamefont {Ueda}}]{murata_pra_2007}%
  \BibitemOpen
  \bibfield  {author} {\bibinfo {author} {\bibfnamefont {K.}~\bibnamefont
  {Murata}}, \bibinfo {author} {\bibfnamefont {H.}~\bibnamefont {Saito}},\ and\
  \bibinfo {author} {\bibfnamefont {M.}~\bibnamefont {Ueda}},\ }\bibfield
  {title} {\bibinfo {title} {Broken-axisymmetry phase of a spin-1 ferromagnetic
  {Bose-Einstein} condensate},\ }\href
  {https://doi.org/10.1103/PhysRevA.75.013607} {\bibfield  {journal} {\bibinfo
  {journal} {Phys. Rev. A}\ }\textbf {\bibinfo {volume} {75}},\ \bibinfo
  {pages} {013607} (\bibinfo {year} {2007})}\BibitemShut {NoStop}%
\bibitem [{\citenamefont {Ruostekoski}\ and\ \citenamefont
  {Dutton}(2007)}]{ruostekoski_pra_2007}%
  \BibitemOpen
  \bibfield  {author} {\bibinfo {author} {\bibfnamefont {J.}~\bibnamefont
  {Ruostekoski}}\ and\ \bibinfo {author} {\bibfnamefont {Z.}~\bibnamefont
  {Dutton}},\ }\bibfield  {title} {\bibinfo {title} {Dynamical and energetic
  instabilities in multicomponent {Bose-Einstein} condensates in optical
  lattices},\ }\href {https://doi.org/10.1103/PhysRevA.76.063607} {\bibfield
  {journal} {\bibinfo  {journal} {Phys. Rev. A}\ }\textbf {\bibinfo {volume}
  {76}},\ \bibinfo {pages} {063607} (\bibinfo {year} {2007})}\BibitemShut
  {NoStop}%
\bibitem [{\citenamefont {Borgh}\ \emph {et~al.}(2016)\citenamefont {Borgh},
  \citenamefont {Nitta},\ and\ \citenamefont {Ruostekoski}}]{Borgh2016a}%
  \BibitemOpen
  \bibfield  {author} {\bibinfo {author} {\bibfnamefont {M.~O.}\ \bibnamefont
  {Borgh}}, \bibinfo {author} {\bibfnamefont {M.}~\bibnamefont {Nitta}},\ and\
  \bibinfo {author} {\bibfnamefont {J.}~\bibnamefont {Ruostekoski}},\
  }\bibfield  {title} {\bibinfo {title} {Stable core symmetries and confined
  textures for a vortex line in a spinor {B}ose-{E}instein condensate},\ }\href
  {https://journals.aps.org/prl/abstract/10.1103/PhysRevLett.116.085301}
  {\bibfield  {journal} {\bibinfo  {journal} {Phys. Rev. Lett.}\ }\textbf
  {\bibinfo {volume} {116}},\ \bibinfo {pages} {085301} (\bibinfo {year}
  {2016})}\BibitemShut {NoStop}%
\bibitem [{\citenamefont {Fatemi}\ \emph {et~al.}(2000)\citenamefont {Fatemi},
  \citenamefont {Jones},\ and\ \citenamefont {Lett}}]{Fatemi2000}%
  \BibitemOpen
  \bibfield  {author} {\bibinfo {author} {\bibfnamefont {F.~K.}\ \bibnamefont
  {Fatemi}}, \bibinfo {author} {\bibfnamefont {K.~M.}\ \bibnamefont {Jones}},\
  and\ \bibinfo {author} {\bibfnamefont {P.~D.}\ \bibnamefont {Lett}},\
  }\bibfield  {title} {\bibinfo {title} {{Observation of Optically Induced
  Feshbach Resonances in Collisions of Cold Atoms}},\ }\href
  {https://doi.org/10.1103/PhysRevLett.85.4462} {\bibfield  {journal} {\bibinfo
   {journal} {Phys. Rev. Lett.}\ }\textbf {\bibinfo {volume} {85}},\ \bibinfo
  {pages} {4462} (\bibinfo {year} {2000})}\BibitemShut {NoStop}%
\bibitem [{\citenamefont {Papoular}\ \emph {et~al.}(2010)\citenamefont
  {Papoular}, \citenamefont {Shlyapnikov},\ and\ \citenamefont
  {Dalibard}}]{Papoular2010}%
  \BibitemOpen
  \bibfield  {author} {\bibinfo {author} {\bibfnamefont {D.~J.}\ \bibnamefont
  {Papoular}}, \bibinfo {author} {\bibfnamefont {G.~V.}\ \bibnamefont
  {Shlyapnikov}},\ and\ \bibinfo {author} {\bibfnamefont {J.}~\bibnamefont
  {Dalibard}},\ }\bibfield  {title} {\bibinfo {title} {{Microwave-induced
  Fano-Feshbach resonances}},\ }\href
  {https://doi.org/10.1103/PhysRevA.81.041603} {\bibfield  {journal} {\bibinfo
  {journal} {Phys. Rev. A}\ }\textbf {\bibinfo {volume} {81}},\ \bibinfo
  {pages} {041603(R)} (\bibinfo {year} {2010})}\BibitemShut {NoStop}%
\bibitem [{\citenamefont {Huhtam\"aki}\ \emph {et~al.}(2009)\citenamefont
  {Huhtam\"aki}, \citenamefont {Simula}, \citenamefont {Kobayashi},\ and\
  \citenamefont {Machida}}]{Huhtamaki2009}%
  \BibitemOpen
  \bibfield  {author} {\bibinfo {author} {\bibfnamefont {J.~A.~M.}\
  \bibnamefont {Huhtam\"aki}}, \bibinfo {author} {\bibfnamefont {T.~P.}\
  \bibnamefont {Simula}}, \bibinfo {author} {\bibfnamefont {M.}~\bibnamefont
  {Kobayashi}},\ and\ \bibinfo {author} {\bibfnamefont {K.}~\bibnamefont
  {Machida}},\ }\bibfield  {title} {\bibinfo {title} {Stable fractional
  vortices in the cyclic states of {B}ose-{E}instein condensates},\ }\href
  {https://doi.org/10.1103/PHYSREVA.80.051601} {\bibfield  {journal} {\bibinfo
  {journal} {Phys. Rev. A}\ }\textbf {\bibinfo {volume} {80}},\ \bibinfo
  {pages} {051601(R)} (\bibinfo {year} {2009})}\BibitemShut {NoStop}%
\bibitem [{\citenamefont {Javanainen}\ and\ \citenamefont
  {Ruostekoski}(2006)}]{Javanainen2006}%
  \BibitemOpen
  \bibfield  {author} {\bibinfo {author} {\bibfnamefont {J.}~\bibnamefont
  {Javanainen}}\ and\ \bibinfo {author} {\bibfnamefont {J.}~\bibnamefont
  {Ruostekoski}},\ }\bibfield  {title} {\bibinfo {title} {Symbolic calculation
  in development of algorithms: split-step methods for the
  {G}ross–{P}itaevskii equation},\ }\href
  {https://doi.org/10.1088/0305-4470/39/12/L02} {\bibfield  {journal} {\bibinfo
   {journal} {J. Phys. A: Math. Gen.}\ }\textbf {\bibinfo {volume} {39}},\
  \bibinfo {pages} {L179} (\bibinfo {year} {2006})}\BibitemShut {NoStop}%
\bibitem [{\citenamefont {Mermin}\ and\ \citenamefont {Ho}(1976)}]{Mermin1976}%
  \BibitemOpen
  \bibfield  {author} {\bibinfo {author} {\bibfnamefont {N.~D.}\ \bibnamefont
  {Mermin}}\ and\ \bibinfo {author} {\bibfnamefont {T.-L.}\ \bibnamefont
  {Ho}},\ }\bibfield  {title} {\bibinfo {title} {{Circulation and Angular
  Momentum in the $A$ Phase of Superfluid Helium-3}},\ }\href
  {https://doi.org/10.1103/PhysRevLett.36.594} {\bibfield  {journal} {\bibinfo
  {journal} {Phys. Rev. Lett.}\ }\textbf {\bibinfo {volume} {36}},\ \bibinfo
  {pages} {594} (\bibinfo {year} {1976})}\BibitemShut {NoStop}%
\bibitem [{\citenamefont {Theis}\ \emph {et~al.}(2004)\citenamefont {Theis},
  \citenamefont {Thalhammer}, \citenamefont {Winkler}, \citenamefont {Hellwig},
  \citenamefont {Ruff}, \citenamefont {Grimm},\ and\ \citenamefont
  {Denschlag}}]{Theis2004}%
  \BibitemOpen
  \bibfield  {author} {\bibinfo {author} {\bibfnamefont {M.}~\bibnamefont
  {Theis}}, \bibinfo {author} {\bibfnamefont {G.}~\bibnamefont {Thalhammer}},
  \bibinfo {author} {\bibfnamefont {K.}~\bibnamefont {Winkler}}, \bibinfo
  {author} {\bibfnamefont {M.}~\bibnamefont {Hellwig}}, \bibinfo {author}
  {\bibfnamefont {G.}~\bibnamefont {Ruff}}, \bibinfo {author} {\bibfnamefont
  {R.}~\bibnamefont {Grimm}},\ and\ \bibinfo {author} {\bibfnamefont {J.~H.}\
  \bibnamefont {Denschlag}},\ }\bibfield  {title} {\bibinfo {title} {Tuning the
  scattering length with an optically induced {F}eshbach resonance},\ }\href
  {https://doi.org/10.1103/PhysRevLett.93.123001} {\bibfield  {journal}
  {\bibinfo  {journal} {Phys. Rev. Lett.}\ }\textbf {\bibinfo {volume} {93}},\
  \bibinfo {pages} {123001} (\bibinfo {year} {2004})}\BibitemShut {NoStop}%
\bibitem [{\citenamefont {Leslie}\ \emph {et~al.}(2009)\citenamefont {Leslie},
  \citenamefont {Guzman}, \citenamefont {Vengalattore}, \citenamefont {Sau},
  \citenamefont {Cohen},\ and\ \citenamefont {Stamper-Kurn}}]{Leslie2009a}%
  \BibitemOpen
  \bibfield  {author} {\bibinfo {author} {\bibfnamefont {S.~R.}\ \bibnamefont
  {Leslie}}, \bibinfo {author} {\bibfnamefont {J.}~\bibnamefont {Guzman}},
  \bibinfo {author} {\bibfnamefont {M.}~\bibnamefont {Vengalattore}}, \bibinfo
  {author} {\bibfnamefont {J.~D.}\ \bibnamefont {Sau}}, \bibinfo {author}
  {\bibfnamefont {M.~L.}\ \bibnamefont {Cohen}},\ and\ \bibinfo {author}
  {\bibfnamefont {D.~M.}\ \bibnamefont {Stamper-Kurn}},\ }\bibfield  {title}
  {\bibinfo {title} {Amplification of fluctuations in a spinor
  {B}ose--{E}instein condensate},\ }\href
  {https://doi.org/10.1103/PhysRevA.79.043631} {\bibfield  {journal} {\bibinfo
  {journal} {Phys. Rev. A}\ }\textbf {\bibinfo {volume} {79}},\ \bibinfo
  {pages} {043631} (\bibinfo {year} {2009})}\BibitemShut {NoStop}%
\bibitem [{\citenamefont {Manakov}\ \emph {et~al.}(1986)\citenamefont
  {Manakov}, \citenamefont {Ovsiannikov},\ and\ \citenamefont
  {Rapoport}}]{Manakov86}%
  \BibitemOpen
  \bibfield  {author} {\bibinfo {author} {\bibfnamefont {N.}~\bibnamefont
  {Manakov}}, \bibinfo {author} {\bibfnamefont {V.}~\bibnamefont
  {Ovsiannikov}},\ and\ \bibinfo {author} {\bibfnamefont {L.}~\bibnamefont
  {Rapoport}},\ }\bibfield  {title} {\bibinfo {title} {Atoms in a laser
  field},\ }\href
  {https://doi.org/https://doi.org/10.1016/S0370-1573(86)80001-1} {\bibfield
  {journal} {\bibinfo  {journal} {Phys. Rep.}\ }\textbf {\bibinfo {volume}
  {141}},\ \bibinfo {pages} {320} (\bibinfo {year} {1986})}\BibitemShut
  {NoStop}%
\bibitem [{\citenamefont {Andersen}\ \emph {et~al.}(2006)\citenamefont
  {Andersen}, \citenamefont {Ryu}, \citenamefont {Clad\'e}, \citenamefont
  {Natarajan}, \citenamefont {Vaziri}, \citenamefont {Helmerson},\ and\
  \citenamefont {Phillips}}]{Andersen2006}%
  \BibitemOpen
  \bibfield  {author} {\bibinfo {author} {\bibfnamefont {M.~F.}\ \bibnamefont
  {Andersen}}, \bibinfo {author} {\bibfnamefont {C.}~\bibnamefont {Ryu}},
  \bibinfo {author} {\bibfnamefont {P.}~\bibnamefont {Clad\'e}}, \bibinfo
  {author} {\bibfnamefont {V.}~\bibnamefont {Natarajan}}, \bibinfo {author}
  {\bibfnamefont {A.}~\bibnamefont {Vaziri}}, \bibinfo {author} {\bibfnamefont
  {K.}~\bibnamefont {Helmerson}},\ and\ \bibinfo {author} {\bibfnamefont
  {W.~D.}\ \bibnamefont {Phillips}},\ }\bibfield  {title} {\bibinfo {title}
  {Quantized rotation of atoms from photons with orbital angular momentum},\
  }\href {https://doi.org/10.1103/PhysRevLett.97.170406} {\bibfield  {journal}
  {\bibinfo  {journal} {Phys. Rev. Lett.}\ }\textbf {\bibinfo {volume} {97}},\
  \bibinfo {pages} {170406} (\bibinfo {year} {2006})}\BibitemShut {NoStop}%
\bibitem [{\citenamefont {Wright}\ \emph {et~al.}(2009)\citenamefont {Wright},
  \citenamefont {Leslie}, \citenamefont {Hansen},\ and\ \citenamefont
  {Bigelow}}]{Wright2009}%
  \BibitemOpen
  \bibfield  {author} {\bibinfo {author} {\bibfnamefont {K.~C.}\ \bibnamefont
  {Wright}}, \bibinfo {author} {\bibfnamefont {L.~S.}\ \bibnamefont {Leslie}},
  \bibinfo {author} {\bibfnamefont {A.}~\bibnamefont {Hansen}},\ and\ \bibinfo
  {author} {\bibfnamefont {N.~P.}\ \bibnamefont {Bigelow}},\ }\bibfield
  {title} {\bibinfo {title} {Sculpting the vortex state of a spinor {BEC}},\
  }\href {https://doi.org/10.1103/PhysRevLett.102.030405} {\bibfield  {journal}
  {\bibinfo  {journal} {Phys. Rev. Lett.}\ }\textbf {\bibinfo {volume} {102}},\
  \bibinfo {pages} {030405} (\bibinfo {year} {2009})}\BibitemShut {NoStop}%
\bibitem [{\citenamefont {Leanhardt}\ \emph {et~al.}(2002)\citenamefont
  {Leanhardt}, \citenamefont {G\"orlitz}, \citenamefont {Chikkatur},
  \citenamefont {Kielpinski}, \citenamefont {Shin}, \citenamefont {Pritchard},\
  and\ \citenamefont {Ketterle}}]{Leanhardt2002}%
  \BibitemOpen
  \bibfield  {author} {\bibinfo {author} {\bibfnamefont {A.~E.}\ \bibnamefont
  {Leanhardt}}, \bibinfo {author} {\bibfnamefont {A.}~\bibnamefont
  {G\"orlitz}}, \bibinfo {author} {\bibfnamefont {A.~P.}\ \bibnamefont
  {Chikkatur}}, \bibinfo {author} {\bibfnamefont {D.}~\bibnamefont
  {Kielpinski}}, \bibinfo {author} {\bibfnamefont {Y.}~\bibnamefont {Shin}},
  \bibinfo {author} {\bibfnamefont {D.~E.}\ \bibnamefont {Pritchard}},\ and\
  \bibinfo {author} {\bibfnamefont {W.}~\bibnamefont {Ketterle}},\ }\bibfield
  {title} {\bibinfo {title} {Imprinting vortices in a {B}ose--{E}instein
  condensate using topological phases},\ }\href
  {https://doi.org/10.1103/PhysRevLett.89.190403} {\bibfield  {journal}
  {\bibinfo  {journal} {Phys. Rev. Lett.}\ }\textbf {\bibinfo {volume} {89}},\
  \bibinfo {pages} {190403} (\bibinfo {year} {2002})}\BibitemShut {NoStop}%
\bibitem [{\citenamefont {Madison}\ \emph {et~al.}(2000)\citenamefont
  {Madison}, \citenamefont {Chevy}, \citenamefont {Wohlleben},\ and\
  \citenamefont {Dalibard}}]{Madison2000}%
  \BibitemOpen
  \bibfield  {author} {\bibinfo {author} {\bibfnamefont {K.~W.}\ \bibnamefont
  {Madison}}, \bibinfo {author} {\bibfnamefont {F.}~\bibnamefont {Chevy}},
  \bibinfo {author} {\bibfnamefont {W.}~\bibnamefont {Wohlleben}},\ and\
  \bibinfo {author} {\bibfnamefont {J.}~\bibnamefont {Dalibard}},\ }\bibfield
  {title} {\bibinfo {title} {Vortex formation in a stirred {B}ose--{E}instein
  condensate},\ }\href {https://doi.org/10.1103/PhysRevLett.84.806} {\bibfield
  {journal} {\bibinfo  {journal} {Phys. Rev. Lett.}\ }\textbf {\bibinfo
  {volume} {84}},\ \bibinfo {pages} {806} (\bibinfo {year} {2000})}\BibitemShut
  {NoStop}%
\bibitem [{\citenamefont {Hodby}\ \emph {et~al.}(2001)\citenamefont {Hodby},
  \citenamefont {Hechenblaikner}, \citenamefont {Hopkins}, \citenamefont
  {Marag\`o},\ and\ \citenamefont {Foot}}]{Hodby2001}%
  \BibitemOpen
  \bibfield  {author} {\bibinfo {author} {\bibfnamefont {E.}~\bibnamefont
  {Hodby}}, \bibinfo {author} {\bibfnamefont {G.}~\bibnamefont
  {Hechenblaikner}}, \bibinfo {author} {\bibfnamefont {S.~A.}\ \bibnamefont
  {Hopkins}}, \bibinfo {author} {\bibfnamefont {O.~M.}\ \bibnamefont
  {Marag\`o}},\ and\ \bibinfo {author} {\bibfnamefont {C.~J.}\ \bibnamefont
  {Foot}},\ }\bibfield  {title} {\bibinfo {title} {Vortex nucleation in
  {B}ose--{E}instein condensates in an oblate, purely magnetic potential},\
  }\href {https://doi.org/10.1103/PhysRevLett.88.010405} {\bibfield  {journal}
  {\bibinfo  {journal} {Phys. Rev. Lett.}\ }\textbf {\bibinfo {volume} {88}},\
  \bibinfo {pages} {010405} (\bibinfo {year} {2001})}\BibitemShut {NoStop}%
\bibitem [{\citenamefont {Baio}\ \emph {et~al.}()\citenamefont {Baio},
  \citenamefont {Wheeler}, \citenamefont {Hall}, \citenamefont {Ruostekoski},\
  and\ \citenamefont {Borgh}}]{dataset}%
  \BibitemOpen
  \bibfield  {author} {\bibinfo {author} {\bibfnamefont {G.}~\bibnamefont
  {Baio}}, \bibinfo {author} {\bibfnamefont {M.~T.}\ \bibnamefont {Wheeler}},
  \bibinfo {author} {\bibfnamefont {D.~S.}\ \bibnamefont {Hall}}, \bibinfo
  {author} {\bibfnamefont {J.}~\bibnamefont {Ruostekoski}},\ and\ \bibinfo
  {author} {\bibfnamefont {M.~O.}\ \bibnamefont {Borgh}},\ }\href
  {https://doi.org/10.5281/zenodo.10406562} {\bibinfo {title} {Dataset from
  'topological interfaces crossed by defects and textures of continuous and
  discrete point group symmetries in spin-2 bose-einstein condensates'}},\
  \Eprint {https://arxiv.org/abs/{D}OI: 10.5281/zenodo.10406562} {{D}OI:
  10.5281/zenodo.10406562} \BibitemShut {NoStop}%
\end{thebibliography}%
\end{document}